\documentclass[11pt,a4]{article}
\usepackage{enumitem}
\usepackage{amsthm,amsmath,latexsym,amssymb,amsfonts,amscd}
\usepackage{graphics,lscape,fancyhdr,array,stmaryrd,euscript}
\usepackage{mathabx}
\usepackage{tikz-cd}
\tikzcdset{arrow style=math font}
\usepackage{tikz}

\makeatletter
\newsavebox{\@brx}
\newcommand{\llangle}[1][]{\savebox{\@brx}{\(\m@th{#1\langle}\)}%
  \mathopen{\copy\@brx\kern-0.5\wd\@brx\usebox{\@brx}}}
\newcommand{\rrangle}[1][]{\savebox{\@brx}{\(\m@th{#1\rangle}\)}%
  \mathclose{\copy\@brx\kern-0.5\wd\@brx\usebox{\@brx}}}
\makeatother

\usepackage{skak}

\newcommand{\bep}{\begin{picture}}
\newcommand{\eep}{\end{picture}}

\newcounter{YoungHeight}\newcounter{YoungWidth}

\newcounter{Mul1}\newcounter{Mul2}\newcounter{Mul3}\newcounter{Mul4}
\newcounter{A0}\newcounter{A1}\newcounter{A2}
\newcounter{B3}
\newcounter{C3}\newcounter{C4}
\newcounter{D1}\newcounter{D2}\newcounter{D3}
\newcounter{T0}\newcounter{T1}

\newlength{\txtHShift}

\newlength{\txtWidth}

\newcommand{\HalfLength}[2]{\setcounter{Mul1}{#1}\setcounter{Mul2}{#1}\addtocounter{Mul1}{\value{Mul2}}\addtocounter{Mul1}{\value{Mul2}}%
\addtocounter{Mul1}{\value{Mul2}}\addtocounter{Mul1}{\value{Mul2}}\setcounter{#2}{\value{Mul1}}}

\newcommand{\Add}[3]{\setcounter{#1}{#2}\addtocounter{#1}{#3}}

\newcommand{\Length}[1]{#10}
\newcommand{\YoungScale}{}

\newcommand{\shiftedText}[2]{{\hspace{#1}#2}}
\newcommand{\calcHShift}[1]{\settowidth{\txtWidth}{#1}\setlength{\txtHShift}{-0.5\txtWidth}}

\newcommand{\TextCenter}[3]{{\HalfLength{#2}{T0}%
\HalfLength{#3}{T1}\addtocounter{T1}{-3}\calcHShift{#1}%
\put(\value{T0},\value{T1}){\shiftedText{\txtHShift}{#1}}}}

\newcommand{\BlockA}[2]{{\YoungScale\bep(\Length{#1},\Length{#2}){\Add{A1}{#1}{1}\Add{A2}{#2}{1}}%
\multiput(0,0)(10,0){\value{A1}}{\line(0,1){\Length{#2}}}\multiput(0,0)(0,10){\value{A2}}{\line(1,0){\Length{#1}}}%
\setcounter{YoungHeight}{\Length{#2}}\setcounter{YoungWidth}{\Length{#1}}\eep}}

\newcommand{\BlockB}[4]{{\YoungScale\Add{B3}{\Length{#2}}{\Length{#4}}%
\bep(\Length{#1},\value{B3})\put(0,\Length{#4}){\BlockA{#1}{#2}}%
\put(0,0){\BlockA{#3}{#4}}\setcounter{YoungHeight}{\value{B3}}\setcounter{YoungWidth}{\Length{#1}}\eep}}

\newcommand{\RectT}[3]{\bep(\Length{#1},\Length{#2})\put(0,0){\line(1,0){\Length{#1}}}\put(0,0){\line(0,1){\Length{#2}}}%
\put(\Length{#1},\Length{#2}){\line(-1,0){\Length{#1}}}\put(\Length{#1},\Length{#2}){\line(0,-1){\Length{#2}}}#3{#1}{#2}\eep}

\newcommand{\RectBRowUp}[4]{{\bep(\Length{#1},20)\put(0,0){\RectT{#2}{1}{\TextCenter{#4}}}%
\put(0,10){\RectT{#1}{1}{\TextCenter{#3}}}\eep}}

\newcommand{\YoungB}{\BlockA{2}{1}}

\newcommand{\YoungBB}{\BlockA{2}{2}}

\newcommand{\Youngfourrow}{\BlockB{4}{1}{0}{0}}

\usepackage{youngtab}
\usepackage{empheq}
\usepackage{sidecap}
\usepackage{ifpdf}
\usepackage{verbatim}
\usepackage{color}
\usepackage{relsize}
\usepackage{tikz-cd}
\usepackage{titling}
\usepackage{hyperref,setspace}
\usepackage{soul}
\usepackage{cite}
\usepackage{graphicx}
\usepackage{lipsum}
\usepackage{multicol}
\usepackage{yfonts}
\usepackage{stmaryrd}
\usepackage{mathrsfs}

\newcommand{\Ucal}{\mathcal{U}}

\newcommand{\Ecal}{\mathcal{E}}
\tikzcdset{arrow style=math font}
\newcommand{\hatphi}{\hat{\phi}}
\definecolor{darkgreen}{RGB}{9, 120, 9}
\definecolor{darkred}{RGB}{120, 9, 9}

\newcommand{\tK}{\mathtt{K}}

\newcommand{\PS}{\mathbb{PS}}

\newcommand{\RR}{\mathbb{R}}

\newcommand{\Hcal}{\mathcal{H}}
\newcommand{\Bcal}{\mathcal{B}}
\newcommand{\Fcal}{\mathcal{F}}
\newcommand{\Ocal}{\mathcal{O}}
\newcommand{\Ncal}{\mathcal{N}}

\newcommand{\PTc}{\mathcal{PT}}

\newcommand{\SCF}{\text{SCF}}
\newcommand{\PT}{\mathbb{PT}}

\newcommand{\PPb}{\overline{\mathbb{P}}}
\newcommand{\CC}{\mathbb{C}}

\newcommand{\projectivespace}{\mathbb{CP}}

\newcommand{\Mcal}{\mathcal{M}}

\newcommand{\Jcal}{\mathcal{J}}

\newcommand{\Acal}{\mathcal{A}}

\newcommand{\pl}{\partial}
\newcommand{\pvec}{\boldsymbol{p}}

\newcommand{\AB}{\mathbb{A}}

\newcommand{\Tr}{\text{Tr}}

\newcommand{\Ccal}{\mathcal{C}}
\newcommand{\Dcal}{\mathcal{D}}
\newcommand{\Ical}{\mathcal{I}}
\newcommand{\CP}{\mathbb{CP}}

\newcommand{\ta}{\mathtt{a}}
\newcommand{\tb}{\mathtt{b}}
\newcommand{\tc}{\mathtt{c}}
\newcommand{\td}{\mathtt{d}}
\newcommand{\tx}{\mathtt{x}}
\newcommand{\Xcal}{\mathcal{X}}
\newcommand{\BG}{\text{BG}}
\newcommand{\sub}{\text{sub}}
\newcommand{\lead}{\text{lead}}
\newcommand{\TA}{\mathtt{A}}
\newcommand{\TB}{\mathtt{B}}

\def\Acurl{\mathscr{A}}

\def\Ccurl{\mathscr{C}}

\def\Fcurl{\mathscr{F}}

\def\TY{\mathtt{Y}}

\def\msu{\mathfrak{su}}
\def\mso{\mathfrak{so}}
\def\hs{\mathfrak{hs}}
\def\msp{\mathfrak{sp}}

\def\mhs{\mathfrak{hs}}
\def\ths{\mathfrak{ths}}

 \def\one{\mbox{1 \kern-.59em {\rm l}}}

\def\R{{\mathbb R}} \def\C{{\mathbb C}} 
\newcommand{\diag}{\text{diag}}

 \newcommand{\Kcurl}{\mathscr{K}}

\def\a{\alpha}  \def\b{\beta}
\def\g{\gamma} 
 \def\d{\delta} 
\def\k{\kappa}
\def\l{\lambda}  
\def\ty{\mathtt{y}}

  \def\cC{{\cal C}} \def\cM{{\cal M}}

\begin{document}

\begin{flushright}
 UWThPh-2022-6 \\
\end{flushright}
\hfill
\vskip 0.01\textheight
\begin{center}
{\Large\bfseries 
A Twistorial Description of the IKKT-Matrix Model}\\

\vskip 0.03\textheight

Harold C. Steinacker\,\footnote{Email: harold.steinacker@univie.ac.at}${}^{\symrook}$ and Tung Tran\,\footnote{Email: vuongtung.tran@umons.ac.be}${}^{\symknight}$

\vskip 0.04\textheight
{${}^{\symrook}$ Department of Physics, University of Vienna, \\
Boltzmanngasse 5, A-1090 Vienna, Austria}
\vskip 0.03\textheight
{${}^{\symknight}$ Service de Physique de l’Univers, Champs et Gravitation,\\
Université de Mons, 20 place du Parc, 7000 Mons, Belgium}
\end{center}

\begin{abstract} 

We consider the fuzzy 4-sphere $S_N^4$ as a background in the IKKT matrix model,
and explore the relation between $S_N^4$ and fuzzy twistor space in the semi-classical limit.
A novel description for the IKKT-matrix model in terms of spinorial indices is given, which is reminiscent of $\Ncal=4$ super-symmetric Yang-Mills (SYM) in $4d$. On fuzzy twistor space, the interactions of the IKKT model are of gravitational type.  The higher-spin (HS) gauge theory emerging in this limit from the IKKT model, denoted as HS-IKKT, on fuzzy twistor space is shown to be a higher-spin extension of $\Ncal=4$ 
SYM, with vertices that have more than two derivatives. We obtain its (Euclidean) spacetime action using the Penrose transform. Although this is a gravitational theory, it shares many  features with the higher-spin extensions of 
Yang-Mills in $4d$ flat space obtained in \cite{Krasnov:2021nsq,Tran:2021ukl}. The tree-level amplitudes of the HS-IKKT  are studied in the semi-classical flat limit. 
The self-dual gauge sector of the IKKT model is  obtained by dropping some parts of the cubic- and the quartic interactions, which is shown 
to reduce to a $\Bcal\Fcal$-type action on commutative deformed projective twistor space.

\end{abstract}
\newpage

\tableofcontents
\section{Introduction}

Various No-go theorems in flat space \cite{Coleman:1967ad,Weinberg:1964ew} and AdS space \cite{Maldacena:2011jn} have been the main 
arguments  obstructing the
construction of viable massless interacting higher-spin theories using field theory approaches.\footnote{See \cite{Bekaert:2010hw} for a review.} Building toy models of higher-spin theories which can avoid No-go theorems usually requires  to give up at least one of the important features of field theory, notably unitarity and locality.
A few examples of higher-spin theories with local interactions are (quasi-)topological theories \cite{Blencowe:1988gj,Bergshoeff:1989ns,Campoleoni:2010zq,Henneaux:2010xg,Pope:1989vj,Fradkin:1989xt,Grigoriev:2019xmp,Grigoriev:2020lzu,Ponomarev:2016lrm,Ponomarev:2017nrr,Krasnov:2021nsq}, or higher-spin extensions of Weyl gravity \cite{Segal:2002gd,Tseytlin:2002gz,Bekaert:2010ky}. In any case, the (holographic) S-matrix turns out to be trivial or simple, which indicates that possible interactions are severely constrained by higher-spin symmetry (an infinite-dimensional symmetry), and are forced to cancel each other out in the physical amplitudes.

Nevertheless, various attempts
during recent years towards a construction of interacting higher-spin theories using Fronsdal fields as the main objects \cite{Bekaert:2010hp,Fotopoulos:2010ay,Boulanger:2015ova,Roiban:2017iqg,Bekaert:2015tva,Sleight:2017pcz} taught us some lessons on how to build up toy models for higher-spin theories:
\begin{itemize}
    \item First of all, if we willingly forgo covariance, then the light-front approach \cite{Bengtsson:1983pd,Bengtsson:1986kh,Fradkin:1991iy,Metsaev:1991mt,Metsaev:1991nb,Ponomarev:2016lrm,Ponomarev:2017nrr} is the very first approach that provides positive results on perturbatively local interacting higher-spin theories with propagating degrees of freedom. The chiral higher-spin theories \cite{Metsaev:1991mt,Metsaev:1991nb,Ponomarev:2016lrm,Metsaev:2018xip,Skvortsov:2018uru} are the first theories that can avoid No-go theorems in both flat and AdS spaces. The flat space chiral theories were shown to be integrable in \cite{Ponomarev:2017nrr} and  proven to be UV finite at one-loop in \cite{Skvortsov:2018jea,Skvortsov:2020wtf,Skvortsov:2020gpn}. We expect that the chiral theories are one-loop exact.
    \item  Secondly, one can start with an auxiliary space where non-locality is under control, and find a map to spacetime with the requirement that the interacting vertices in spacetime should not be too non-local. In particular, twistor space provides such a framework to construct (covariant) theories of interacting higher-spin fields in spacetime. For instance, by deforming the complex structure on twistor space, one can obtain conformal higher-spin gravity in $AdS_4$ \cite{Haehnel:2016mlb,Adamo:2016ple}. The higher-spin extensions of (self-dual) Yang-Mills (HS-(SD)YM) \cite{Krasnov:2021nsq,Tran:2021ukl} and self-dual gravity (HS-SDGRA) \cite{Krasnov:2021nsq} were obtained recently, using also some methods with deep roots in twistor theory. The main advantage of constructing higher-spin theories using twistor theory is that we can carefully maintain the covariance.
    
\end{itemize}

On the other hand,
the IKKT-matrix model \cite{Ishibashi:1996xs} -- which can be viewed as an alternative and constructive description of type IIB superstring theory --
was recently shown to induce a higher-spin gauge theory on a fuzzy 4-sphere/hyperboloid $S_N^4/H_N^4$  in the large-$N$ (semi-classical) limit \cite{Sperling:2017dts,Sperling:2017gmy,Sperling:2018xrm,Sperling:2019xar}. We shall refer to this higher-spin gauge theory as HS-IKKT for short. In four dimensions, the HS-IKKT contains $\Ncal=4$ super-symmetric Yang-Mills (SYM) as a subsector. Therefore, it can be thought of as a higher-spin extensions of $\Ncal=4$ SYM on a fuzzy manifold. However, the interactions of the HS-IKKT are of gravitational type, since its vertices can contain more than two-derivatives. Fuzzy $S_N^4$ or $H^4_N$ can be understood as quantized $S^2_N$-bundle over the base manifold, which is either $S^4$ or $H^4$. Since $S^2$ is isomorphic to $\CP^1$, the total space is nothing but a fuzzy twistor space $\CP^3_N$. 
Being related to both type IIB string theory and twistor theory, the IKKT-matrix model offers a remarkable opportunity to construct a complete higher-spin theory in spacetime with interactions reaching up to the quartic. In particular, previous analysis \cite{Steinacker:2019awe} showed that there is no ghost (no physical
mode with negative norm) in the HS-IKKT, albeit on a slightly different background. Therefore, it is reasonable to expect that the higher-spin gauge theory that emerges from the IKKT-matrix model is a rare example of a local higher-spin theory that may avoid the No-go theorems in some way.

 Up to now, the HS-IKKT-matrix model on the fuzzy 4-sphere has been mostly studied using the representation of $\mso(5)$ algebra,
 where $5$ is the dimensions of the ambient space of the $4$-dimensional spacetime manifold. 
The rotational symmetry is recovered through the extra structure of covariant background geometries which induces a higher-spin theory, see \cite{Steinacker:2019fcb} for a review. One noteworthy feature of the HS-IKKT model is that the tower of highes spin modes is truncated, due to the non-commutativity of the coordinates. However, one can recover the usual spectrum of higher-spin theories in the large $N$ limit where matrices are effectively commutative. The space of functions is then taking value in a higher-spin algebra associated to $\mso(5)$. The cases of the fuzzy 4-sphere and 4-hyperboloid were studied in \cite{Sperling:2017gmy,Sperling:2018xrm}. Instead of using $\mso(5)$ to study the HS-IKKT on $S^4_N$, it may be more suggestive to consider $\msp(4)\simeq \mso(5)$ as an alternative realization\footnote{We will be somewhat cavalier on the real structure in this paper. Most considerations will be restricted to the Euclidean case, but might be extended by some sort of analytic continuation. $\msp(4)$ is understood as the appropriate real sector of  $\msp(4)_\C$.}.
Noting that $Sp(4)\subset SU(4)$, and therefore, we can refer to $Sp(4)$-vectors as twistors, which we will denote as $Z^{\Acal}$. Since $Sp(4)\supset SU(2)_L\times SU(2)_R$, we can write the space of functions on a fuzzy sphere as polynomials in terms of spinors. 
This will significantly simplify the analysis of higher-spin modes arising from the IKKT model.

In this work, we will study the HS-IKKT on a fuzzy 4-sphere using the spinorial representation $SU(2)_L\times SU(2)_R$ of $Sp(4)$. All the higher-spin modes described by spinors are said to live in \textit{balanced weight representation} (BWR) on fuzzy twistor space. Upon integrating out all fiber coordinates, which are the auxiliary spinors on  fuzzy twistor space, we end up with a spacetime description of the HS-IKKT model. These spacetime higher-spin fields live in the \textit{maximally unbalanced representation} (MUR) of the Lorentz group \cite{Krasnov:2021nsq} --- a representation inspired by twistor theory \cite{Hitchin:1980hp,Eastwood:1981jy,Woodhouse:1985id}. It has a crucial property of allowing us to control spins and derivatives in the interactions almost independently.
Written in terms of spinorial indices, the (HS)-IKKT model can be further decomposed into a self-dual sector and a non-self-dual one. We exhibit the similarity between the (self-dual) IKKT and (self-dual) $\Ncal=4$ SYM in $4d$. Moreover, the higher-spin extensions of (self-dual) Yang-Mills HS-(SD)YM obtained in \cite{Krasnov:2021nsq,Tran:2021ukl} can be understood as the deformed gauge sectors of the the (self-dual) HS-IKKT in the semi-classical and flat (SCF) limit, after integrating out fibre coordinates. We also show that the action of the self-dual Yang-Mills sector of the (HS)-IKKT can be rewritten as a deformed $\Bcal\Fcal$ action on commutative twistor space. As a consequence, it is natural to conjecture that the self-dual $\Ncal=4$ HS-IKKT is a deformed Chern-Simons theory on super twistor space $\CP^{3|4}$, along  the lines of \cite{Witten:2003nn,Boels:2006ir}. From a geometrical perspective, the self-dual HS-IKKT should be integrable. Finally, the action of the HS-IKKT model written in terms of spinorial indices shows that it has higher-derivative vertices, where the interactions at the lowest order are of gravitational (two-derivative) type due to the Poisson brackets. It is plausible that even though the HS-IKKT has higher-derivative interactions at the quartic, it should be a local interacting higher-spin theory.

The paper is organized as follows. In section \ref{sec:2}, we will give a brief review of the fuzzy 4-sphere and explain how spacetime emerges from it. We also discuss an alternative realization of  $S^4_N$ using the representation theory of $\msp(4)$, and briefly discuss the space of functions on $S^4_N$ in terms of $\msp(4)$. In section \ref{sec:3}, we describe how the incidence relations of twistor theory can be understood in terms of  a  projection of fuzzy twistor space $\CP^3_N$ to $S^4_N$  via the Hopf fibration followed by a  stereographic projection, in the large $N$ limit. We also briefly study the complex structure of $\CP^3_N$. Next, we describe the spinorial representations for higher-spin valued functions, fermionic modes and vector modes on $S_N^4$. The spinorial effective vielbein, metric and torsion are also discussed. Section \ref{sec:4} is dedicated to rewriting the IKKT-matrix model in terms of spinorial indices. As a result, one can write the IKKT model as a self-dual sector plus a non-self-dual one. We then study the twistorial higher-spin theory induced from the IKKT model and perform the Penrose transform to obtain the spacetime action for the HS-IKKT model. In the SCF limit, we study the simplest example of the 3-point scattering amplitude for the Yang-Mills sector of the HS-IKKT in section \ref{sec:5}.
Next, in section \ref{sec:6}, the self-dual gauge sector of the HS-IKKT in spacetime is shown to be a deformed $\Bcal\Fcal$ theory on commutative twistor space. Finally we conclude in section \ref{sec:7}. Various technicalities are collected in the Appendix.

\paragraph{Conventions.} Let us briefly introduce our convention of indices used in this paper. First of all, we will use the Latin letters $a,b$ as $SO(5)$-indices and the capital letters $A,B$ for $SO(6)$. The Greek indices $\mu,\nu$ are used as spacetime indices, while $\alpha,\beta$ (and their primes) are spinorial indices. We will use the curly Latin indices $\Acal,\Bcal$ as $\msp(4)$- and twistor indices. The background vielbein contain spacetime indices $\mu,\nu$ and typewriter type font letters $\ta,\tb$ stand for tangent space indices. We note that $\ta,\tb=1,2,3,4$ in our paper. Next, indices that are symmetrized are denoted by the same Greek letters, e.g. $A_{\alpha}B_{\alpha}$ denotes $\frac{1}{2}(A_{\alpha_1}B_{\alpha_2}+A_{\alpha_2}B_{\alpha_1})$. Fully symmetric rank-$s$ tensor will be denoted by $T_{\alpha(s)}=T_{\alpha_1...\alpha_s}$. Lastly, we denote the Poisson bracket as $\{,\}$ and the anti-commutator as $\{,\}_+$. We will write non-commutative coordinates as capital letters, e.g. $Y,X,P,Q$ while we will denote them as lower-case letters, e.g. $y,x,p,q$, in the semi-classical limit. Coordinates with vectorial indices, for example $Y^a$, have dimension of length ,while coordinates with spinorial indices are dimensionless, e.g. $Y^{\alpha\alpha'}$. We denote the dimensionless coordinates in the semi-classical limit by lower-case typewriter font letters $\ty,\tx$.

 \section{Preliminaries}\label{sec:2}

\subsection{The IKKT model, matrix backgrounds and emergent gauge theory}

To set up the stage, we briefly recall how the IKKT matrix model leads to a gauge theory on emergent space(time) backgrounds.
 The $SO(10)$-invariant action of the Euclidean IKKT model reads
\begin{align}\label{eq:IKKTaction}
    S=\Tr\Big([Y^{\boldsymbol{I}},Y^{\boldsymbol{J}}][Y_{\boldsymbol{I}},Y_{\boldsymbol{J}}]+\bar{\Psi}^{\Acal}\gamma^{\boldsymbol{I}}_{\Acal\Bcal}[Y_{\boldsymbol{I}},\Psi^{\Bcal}]\Big)\,,\qquad {\boldsymbol{I}}=1,...,10\,.
\end{align}
Here the  $Y^{\boldsymbol{I}}$ are $N\times N$ hermitian matrices, and
 $\Psi^{\Bcal}$ are matrix-valued spinors\footnote{Strictly speaking the model should be considered in Minkowski signature, where the fermions are Majorana-Weyl spinors of $SO(9,1)$. Then the $S^4_N$ background should be replaced by $H^4_n$ \cite{Sperling:2018xrm}. Since we focus on the bosonic sector,
 there is no obstacle going to the Euclidean case.}.
For our purpose,
the most important feature of IKKT-type matrix models is that they define a gauge theory on suitable matrix backgrounds. Such a background is defined by a set of 10 ``almost-commutative'' matrices $\bar Y^{\boldsymbol{I}}$, and typically defines a noncommutative or quantized space(time)  as follows \cite{Steinacker:2020nva,Ishiki:2015saa,Schneiderbauer:2016wub,Berenstein:2012ts}: 
One can define optimally localized quasi-coherent states $|y\rangle \in{\cal H}$, which are 
approximate common eigenstates of the $Y^a$, localized at some point in target space
\begin{align}
 y^{\boldsymbol{I}} = \langle y|Y^{\boldsymbol{I}}|y\rangle \quad \in  \R^{9,1} \ .
 \label{coherent-expect}
\end{align}
These $y^{\boldsymbol{I}}$ sweep out some variety $\cM$ in target space, and
\begin{align}
 Y^{\boldsymbol{I}} \sim y^{\boldsymbol{I}} :\quad \cM \hookrightarrow \R^{9,1}
 \label{Y-embed}
\end{align}
is interpreted as quantized embedding of some ``brane'' $\cM$ in target space $\R^{9,1}$. More generally,
one can then associate classical functions to the matrices via
\begin{align}
\begin{split}
{\rm Mat}(\Hcal) &\sim \cC(\cM) \\
\Phi &\sim \langle y|\Phi|y\rangle = \phi(y) \ ,
\end{split}
\end{align}
and the matrix algebra 
${\rm Mat}(\Hcal)$ generated by the  $Y^{\boldsymbol{I}}$ is interpreted as quantized algebra of functions on $\cM$.
The non-commutativity 
\begin{align}\label{eq:noncom}
    [Y^{\boldsymbol{I}},Y^{\boldsymbol{J}}] =: i\theta^{\boldsymbol{IJ}}
\end{align}
amounts to a quantized Poisson structure on $\cM$.
In this way, a fuzzy notion of geometry is extracted from  nearly-commuting matrix configurations in the matrix model. 
A priori, such a Poisson structure breaks Lorentz invariance. 
This is mitigated on covariant quantum spaces such as $S^4_N$ which carry a collection of such the Poisson structures
\begin{align}\label{eq:noncom}
    [Y^a,Y^b]= i\theta^{ab}= ir^2 M^{ab}\,,\qquad a=1,...,5\,,
\end{align}
which form an $S^2$ bundle; in the case of fuzzy $S^4_N$ under consideration, the
$M^{ab}$ are the generators of $\mso(5)$, and $r>0$ is a natural length scale. We will be related to twistor space in section \ref{sec:3}.

Adding fluctuations $\bar Y^a + A^a$ to the background, the action defines a non-commutative Yang-Mills-type gauge theory on $\cM$ \cite{Aoki:1999vr}, with the gauge transformations $U^{-1}(\bar Y^a + A^a)U$.
On the $S^4_N$ background, this was elaborated in \cite{Sperling:2017gmy}, leading to a tower of 4 tangential (off-shell) higher-spin modes. We will reconsider this in the following using a spinorial approach, which considerably simplifies the analysis  in the flat limit.


\subsection{Fuzzy 4-sphere \texorpdfstring{$S^4_N$}{SN} and \texorpdfstring{$\mso(5)$}{so5} representations}
We recall the definition of a fuzzy 4-sphere $S_N^4$, using $\mso(5)$ representation theory \cite{Grosse:1996mz,Castelino:1997rv}. A $4$-dimensional sphere of radius $R$ living in a 5-dimensional flat ambient space $\RR^5$ obeys to the following constraint
\begin{align}\label{eq:4sphereambient}
    Y_{a}Y^{a}= R^2\,,\qquad a=1,...,5\,.
\end{align}
Here, $Y^a$ are the $N\times N$ Hermitian matrices and are the coordinates of the ambient space endowed with the metric $\eta^{ab}=\diag(+,+,+,+,+)$. By requiring $Y^a$ to transform as vectors under $SO(5)$ equipped with the generators $M_{ab}$, we have the following algebra
\begin{subequations}\label{eq:so(5)algebra}
\begin{align}
    [M_{ab},M_{cd}]&=i(M_{ad}\delta_{bc}-M_{ac}\delta_{bd}-M_{bd}\delta_{ac}+M_{bc}\delta_{ad})\,,\\
    [M_{ab},Y_c]&=i(Y_a\delta_{bc}-Y_b\delta_{ac})\,,\\
    [Y_a,Y_b]&=ir^2 M_{ab}\,.
\end{align}
\end{subequations}
The relations \eqref{eq:4sphereambient} and \eqref{eq:so(5)algebra} define a $S_N^4$. The above algebra of $\mso(5)$ can be embedded into an $\mso(6)$ algebra 
\begin{align}
    [J_{AB},J_{CD}]=i(J_{AD}\delta_{BC}-J_{AC}\delta_{BD}-J_{BD}\delta_{AC}+J_{BC}\delta_{AD})\,,\qquad A=(a,6)\,,
\end{align}
by the following identifications
\begin{align}
    M_{ab}=J_{ab}\,,\qquad Y_a = r J_{a6} \,,
\end{align}
where the ``6" is an additional direction. 
To obtain a 4-sphere, we must choose the 
highest weight irreducible representation  of $\mso(6)\cong \msu(4)$, say 
$\Xi=(N,0,0)$, denoted by $\Hcal_N$ henceforth. Then the following relations hold
\begin{align}
\label{eq:highestweightS4N}
    Y_aY^a=R_N^2= \frac{r^2}{4}N(N+4)\,,\qquad \epsilon_{abcde}M^{ab}M^{cd}&=\frac{4}{r}(N+2)Y_e\, .
\end{align}
which provide the basis for the interpretation as fuzzy 4-sphere.


\subsection{\texorpdfstring{$ \mso(5)\simeq \msp(4)$ as subalgebra of $\msu(4)$ }{dictionary}}\label{sec:2.3}
There is another way to describe  $S^4_N$ in term of $\msp(4)$ instead of $\mso(5)$ as in \cite{Sperling:2017dts,Sperling:2017gmy}, which is more natural from the spinorial point of view. We note that our realization is slightly different with the previous literature \cite{Claus:1999xr,Fernando:2009fq,Govil:2013uta}. Consider first the $\mso(5)\simeq \msp(4)$ gamma matrices $\g_a, \ a=1,...,5$, which satisfy the Clifford algebra 
\begin{align}
 (\{ \gamma_a,\gamma_b\}_+)^{\Acal}_{\ \Bcal} = 2\d_{ab}\delta^{\Acal}_{\ \Bcal}\,,\qquad \Acal,\Bcal=0,1,2,3\,.
\end{align}
One useful realization for the $\gamma$ matrices is the chiral representation. Explicitly,
\begin{align}
 (\gamma_{m})^{\Acal}_{\ \Bcal}= -i
\begin{pmatrix}
 0 & (\sigma_m)^{\alpha}_{\ \beta'} \\ -(\sigma_m)^{\alpha'}_{\ \beta} & 0
\end{pmatrix}\,,
\qquad
(\gamma_4)_{\ \Bcal}^{\Acal} =
\begin{pmatrix}
 0 & \one_2 \\ \one_2 & 0
\end{pmatrix}, 
\qquad
(\gamma_{5})^{\Acal}_{\ \Bcal}=
\begin{pmatrix}
 \one_2 & 0 \\ 0 &-\one_2
\end{pmatrix} \ .
\label{gamma-so5-explicit}
\end{align}
There is a unique $\mso(5)$-invariant tensor in $(4)\otimes (4)$, given by
\begin{align}
    C_{\Acal\Bcal}=-C_{\Bcal\Acal}=\begin{pmatrix}
     \epsilon_{\alpha\beta} &0\\
     0& \epsilon_{\alpha'\beta'}
    \end{pmatrix}\, ,
    \label{C-matrix}
\end{align}
which we can use to 
 raise and lower $\Acal,\Bcal$ indices as
\begin{align}
    U_{\Acal}C^{\Bcal\Acal}=U^{\Bcal}\,,\qquad U^{\Acal}C_{\Acal\Bcal}=U_{\Bcal}\,.
\end{align}
Then the $\gamma$-matrices 
are anti-symmetric and traceless, i.e.
\begin{align}
    \gamma_a^{\Acal\Bcal}= - \gamma_a^{\Bcal\Acal}\,,\qquad \gamma_a^{\Acal\Bcal}C_{\Acal\Bcal}=0\, .
\end{align}
Our convention for the $\epsilon$ tensor is  $\epsilon^{\alpha\beta}=\epsilon_{\alpha\beta}$ is that $\epsilon^{01}=-\epsilon^{10}=1$. 
Spinorial indices
are raised and lowered as follows
\begin{align}
    u^{\alpha}=u_{\beta}\epsilon^{\alpha\beta}\,,\qquad u_{\alpha}=u^{\beta}\epsilon_{\beta\alpha}\,.
\end{align}
Then the generators
\begin{align}
 \Sigma_{ab}^{\Acal\Bcal} &=-\Sigma_{ba}^{\Acal\Bcal}=\Sigma^{\Bcal\Acal}_{ab}= \frac{i}{4}[\gamma_a,\gamma_b]^{\Acal\Bcal}\, 
\end{align}
provide the spinorial representation of $\mso(5)\simeq \msp(4)$. We, then, consider the following identifications that map $\mso(6)$ generators to $\msu(4)$ ones
\begin{align}
    Y^{\Acal\Bcal}=-Y^{\Bcal\Acal}=r^{-1}Y^a\gamma_a^{\Acal\Bcal}\,,\qquad L^{\Acal\Bcal}= L^{\Bcal\Acal}=\frac{1}{2} M^{ab}\Sigma^{\Acal\Bcal}_{ab}\, 
     \,.
\end{align}
Note that the $Y^{\Acal\Bcal}$ satisfy the hermiticity relations
\begin{align}
 (Y^{\Acal\Bcal})^\dagger 
  &= - (C^{-1} Y C)_{\Bcal\Acal}\,,
  \label{gamma-AS-real}
\end{align}
where $\dagger$ denotes the hermitian conjugation of the matrices. Roughly speaking, we have changed the symmetries of the generators $Y$ and $M$ as 
\begin{align}
    Y^a\sim {\tiny\Yvcentermath1 \yng(1)} \ \mapsto \ Y^{\Acal\Bcal}\sim {\tiny\Yvcentermath1 \yng(1,1)} \,,\qquad M^{ab}\sim {\tiny\Yvcentermath1 \yng(1,1)}\ \mapsto \ L^{\Acal\Bcal}\sim {\tiny\Yvcentermath1 \yng(2)}\,.
\end{align}
The $\msu(4)$ algebra reads
\cite{Sharapov:2019pdu}
\begin{subequations}
\begin{align}
    [L^{\Acal\Bcal},L^{\Ccal\Dcal}]&=i(L^{\Acal\Ccal}C^{\Bcal\Dcal}+L^{\Acal\Dcal}C^{\Bcal\Ccal}+L^{\Bcal\Dcal}C^{\Acal\Ccal}+L^{\Bcal\Ccal}C^{\Acal\Dcal})\,,\\
    [L^{\Acal\Bcal},Y^{\Ccal\Dcal}]&=i(Y^{\Acal\Ccal}C^{\Bcal\Dcal}+Y^{\Bcal\Ccal}C^{\Acal\Dcal}-Y^{\Acal\Dcal}C^{\Bcal\Ccal}-Y^{\Bcal\Dcal}C^{\Acal\Ccal})\,,\\
    [Y^{\Acal\Bcal},Y^{\Ccal\Dcal}]&=i(L^{\Acal\Ccal}C^{\Bcal\Dcal}-L^{\Acal\Dcal}C^{\Bcal\Ccal}-L^{\Bcal\Ccal}C^{\Acal\Dcal}+L^{\Bcal\Dcal}C^{\Acal\Ccal})\,.
\end{align}
\end{subequations}
Here, we recognize the $L^{\Acal\Bcal}$ as
$\msp(4)$ generators, and $Y^{\Acal\Bcal}$ as ``vectors'' that transform  under $\msp(4)$. 

\subsection{Higher-spin modes on \texorpdfstring{$S_N^4$}{SN4}}
\label{sec:hs-modes}
The space of functions $\Ccurl$ consists of higher-spin modules which are polynomials in $L^{\Acal\Bcal}$ and $Y^{\Acal\Bcal}$. Using \eqref{eq:sp(4)relations}, we can write $\Ccurl$ as
\begin{align}
    \begin{split}
    \Ccurl&=\sum_{k,m}  f_{\Acal(k)\Bcal(2m)|\Ccal(k)}Y^{\Acal\Ccal}...Y^{\Acal\Ccal}L^{\Bcal\Bcal}...L^{\Bcal\Bcal}=\bigoplus_{k,m}\ \parbox{75pt}{{\bep(70,30)\unitlength=0.4mm%
\put(0,3){\RectBRowUp{7}{4}{$k+2m$}{$k$}}%
\eep}}\quad  \,.
    \end{split}
\end{align}
Note that the spectrum looks similar to $\Ccurl(\mso(5))$ in \cite{Sperling:2017gmy} in terms of Young diagrams. Consider a subspace $\ths\subset \Ccurl$ with the following higher-spin modules
\begin{align}
    \ths=\sum_m \psi_{\Bcal(2m)}L^{\Bcal\Bcal}...L^{\Bcal\Bcal}=\bigoplus_m\ \parbox{50pt}{{\bep(\Length{6},6)\unitlength=0.4mm\put(0,-1.5){\RectT{5}{1}{\TextCenter{$2m$}}}\eep}}\,\quad  ,
\end{align}
where the coefficients $\psi_{\Bcal(2m)}$ are totally symmetric and traceless. Then, the above subspace $\ths$ defines a truncated higher-spin algebra. Compare with the case of $\mso(5)$, we have the following dictionary
\begin{align}
    \ths(\mso(5))=\bigoplus_m\ \parbox{60pt}{{\bep(70,30)\unitlength=0.4mm%
\put(0,3){\RectBRowUp{5}{5}{$m-1$}{$m-1$}}%
\eep}}\qquad \longleftrightarrow \qquad \ths(\msp(4))=\bigoplus_m\ \parbox{60pt}{{\bep(\Length{6},6)\unitlength=0.4mm\put(0,-1.5){\RectT{5}{1}{\TextCenter{$2m$}}}\eep}}\,.
\end{align}

\paragraph{Relation with $\mhs(\msp(4))$.} We note that, at large $N$ limit, the above truncated higher-spin algebra is identical with the usual higher-spin algebra defined by the quotient of the universal enveloping algebra of $\msp(4)$ by the two-sided Joseph ideal \cite{Eastwood:2002su,Joung:2014qya,Didenko:2014dwa} generated by 
\begin{subequations}
\begin{align}
   C_2&=-\frac{1}{2}W_{\Acal\Bcal}W^{\Acal\Bcal}\,,\\
   \  \parbox{10pt}{{\bep(10,20)\unitlength=0.3mm%
\put(0,1){\RectBRowUp{1}{1}{}{}}%
\eep}}&= \{W_{[\Acal\Mcal},W_{\Bcal]}^{\  \Mcal}\}_+-\frac{5}{4}C_{\Acal\Bcal}\,,\\
    \parbox{18pt}{{\bep(10,20)\unitlength=0.3mm%
    \put(0,1){\YoungBB}%
    \eep}}&=\{W_{[\Acal}^{\ \ [\Bcal},W_{\Ccal]}^{\ \Dcal]}\}_+-\frac{1}{4}\{W_{[\Acal\Mcal},W^{\ \Mcal}_{ \Ccal]}\}_+C^{\Bcal\Dcal}-\frac{1}{4}\{W_{\Mcal}^{\ [\Bcal},W^{\Mcal\Dcal]}\}_+C_{\Acal\Ccal}-\text{trace}\,,
\end{align}
\end{subequations}
in the units of the cosmological constant that is 1. Here, $W$ are generators of $\msp(4)$ algebra in 4-dimensional target space that obey
\begin{align}\label{AdS4algebra}
    [W^{\Acal\Bcal},W^{\Ccal\Dcal}]=W^{\Acal\Dcal}C^{\Bcal\Ccal}+W^{\Acal\Ccal}C^{\Bcal\Dcal}+W^{\Bcal\Dcal}C^{\Acal\Ccal}+W^{\Dcal\Ccal}C^{\Acal\Dcal}\,.
\end{align}
The higher-spin algebra generated by the $\msp(4)$ generators $W^{\Acal\Bcal}$ is defined as
\begin{align}
    \mhs(\msp(4))&=\frac{\Ucal(\msp(4))}{\langle \Ical\rangle}=\bigoplus_m\ \parbox{50pt}{{\bep(\Length{6},6)\unitlength=0.4mm\put(0,-1.5){\RectT{5}{1}{\TextCenter{$2m$}}}\eep}}\,\quad,
\end{align}
where the universal enveloping algebra $\Ucal(\msp(4))$ reads
\begin{align}
    \begin{split}
    \Ucal(\msp(4))&=\bullet\oplus\parbox{20pt}{\YoungB}\oplus\big(\parbox{20pt}{\YoungB}\otimes \parbox{20pt}{\YoungB}\big)_S\oplus ...\\
    &=\bullet\oplus \parbox{20pt}{\YoungB}\oplus\Big(\bullet\oplus\  \parbox{40pt}{\Youngfourrow} \oplus\parbox{10pt}{{\bep(20,20)\unitlength=0.35mm%
\put(0,0){\RectBRowUp{1}{1}{}{}}%
\eep}}\oplus\parbox{20pt}{\YoungBB}\ \Big)\oplus ...\,.
    \end{split}
\end{align}
Here, $()_S$ denotes the symmetrized tensor product of $W_{\Acal\Bcal}$,\footnote{Any anti-symmetrization between the generators $W$ will reduce to lower orders due to the algebra \eqref{AdS4algebra}.}  and the first bullet $\bullet$ is the singlet of $\Ucal(\msp(4))$ while the second is the quadratic Casimir operator $C_2$. The two-sided Joseph ideal is defined as
\begin{align}
    \langle \Ical\rangle =\Ucal(\msp(4))\otimes \Big((C_2-\lambda)\, \oplus\  \parbox{15pt}{{\bep(10,20)\unitlength=0.4mm%
\put(0,-){\RectBRowUp{1}{1}{}{}}%
\eep}}\oplus \  \parbox{26pt}{{\bep(10,20)\unitlength=0.4mm%
\put(0,-){\YoungBB}%
\eep}}\Big)\otimes \Ucal(\msp(4))\,,\qquad \lambda=-\frac{5}{2}\,.
\end{align}
The main difference between $\ths(\msp(4))$ and $\mhs(\msp(4))$ is that $\ths(\msp(4))$ is defined on the ambient space $\RR^5$ while $\hs(\msp(4))$ is defined on spacetime $S^4$. Moreover, the latter is infinite-dimensional while the former is not. However, these two higher-spin algebras coincide in the large $N$ limit.

We note that the above realization of $\ths(\msp(4))$ allows us to make a connection with fuzzy twistor space $\CP^3_N$. Roughly speaking, $\CP^3\subset \CC^4$ is spanned by $\msp(4)$ or $\msu(4)$ vectors $Z^{\Acal}$ and their dual vectors $\hat{Z}^{\Acal}$. These are known as twistors. Then, functions on $\CP^3_N$ are represented by ``balanced" polynomials of $Z^{\Acal},\hat{Z}^{\Acal}$ with cutoff at $N$. 
In particular, the space of functions on $\CP_N^3$ reads (cf. \eqref{eq:Fockspace})
\begin{align}\label{eq:spectrumtruncated}
    \begin{split}
    \CP^3_N &=End(\Hcal_N)=(0,0,N)_{\msu(4)}\otimes (N,0,0)_{\msu(4)} =\sum_{n=1}^{N}(n,0,n)_{\msu(4)}\,
     \\
     &=\sum_{n=0}^N f_{\Acal(n)\Bcal(n)}Z^{\Acal}...Z^{\Acal}\hat{Z}^{\Bcal}...\hat{Z}^{\Bcal}\,.
     \end{split}
\end{align}
It is clear from \eqref{eq:spectrumtruncated} that the spectrum of higher-spin modes on $\CP^3_N$ is bounded from above, which is an appealing feature of the IKKT-matrix model compared to usual higher-spin theories (see further discussion in \cite{Steinacker:2015dra}). 

\subsection{Spacetime geometry in the semi-classical limit} 
From the relation \eqref{eq:highestweightS4N}, we see that  the natural length scale $r$ scales as $r^2\sim \frac{R^2}{4N^2}$. In the semi-classical (large $N$) limit, the coordinates can be considered as commutative. In this limit, we replace capital letters to the normal ones, namely $Y^a\mapsto y^a$, and replace the commutator \eqref{eq:noncom} with the Poisson bracket $i\{\,,\}$
\begin{align}
    \{y^a,y^b\}= r^2 m^{ab}\,.
    \label{Poisson-S4N}
\end{align}
The Poisson bracket is the only structure which exhibits the non-commutativity of the geometry in the semi-classical limit,
while the functions are considered as commutative. This is in contrast to the fuzzy or noncommutative case, where the functions do not commute and the higher spin modes discussed in section \ref{sec:hs-modes} are truncated. In the following, we will use $y_{\mu}$ with $\mu=1,2,3,4$ to represent spacetime coordinates and write \eqref{eq:4sphereambient} as
\begin{align}
    y_{\mu}y^{\mu}+y_5^2=R^2\,.
\end{align}
To describe a 4-dimensional sphere in the target space, we can choose the following stereographic parametrization
\begin{align}
\label{eq:Yparametrization}
    y^{\mu}=\frac{2R^2x^{\mu}}{(R^2+x^2)}\,,\qquad y^5=\frac{R(R^2-x^2)}{(R^2+x^2)}\,,\qquad x^2=x_{\mu}x^{\mu}\,.
\end{align}
The background metric is therefore
\begin{align}\label{eq:Lorentzmetric}
    ds^2=\Big(\frac{\pl y^a}{\pl x^{\mu}}\frac{\pl y^b}{\pl x^{\nu}}\eta_{ab}\Big)dx^{\mu}dx^{\nu}:=g_{\mu\nu}dx^{\mu}dx^{\nu}=\frac{4R^4dx_{\mu}dx^{\mu}}{(R^2+x^2)^2} \,.
\end{align}
The vielbein reads
\begin{align}
    \bar{e}^{\ta}_{\mu}=\frac{2R^2}{(R^2+x^2)}\d^{\ta}_{\mu} \,,\qquad \ta=1,2,3,4\,.
\end{align}
Note that the typewriter type font indices $\ta,\tb$ are different with the indices $a,b$ on ambient space. They are used to describe tangent space of the $S^4$ in the semi-classical limit. The above is also known as the \textit{affine} patch of the $S^4$. We note that the metric is conformally flat, which 
is a nice feature of the stereographic projection.
Clearly the $SO(4)$ which stabilizes $y^5$ acts linearly on the $x^\mu$, leaving $dx^\mu dx_\mu$ invariant, while $SO(5)$  acts as part of the conformal group of the flat 4-dimensional metric $dx^\mu dx_\mu$.
\paragraph{Flat limit.} From \eqref{eq:Lorentzmetric}, it is obvious that at the limit where $R\rightarrow \infty$, we obtain the flat metric as
\begin{align}
    ds^2=dx_{\mu}dx^{\mu}\,,
\end{align}
where we have rescaled $x^{\mu}$ with a factor of $1/2$. 
In this limit, the $SO(5)$ isometry of $S^4$ reduces to the $ISO(4)$ isometry of flat $\RR^4$. We will often take the semi-classical (large $N$) limit first before considering the flat limit. 
The combination of both limits, 
 i.e. the semi-classical and flat limit, will be denoted  as SCF limit. The SCF limit will be useful to study scattering amplitudes of the HS-IKKT in spacetime in section \ref{sec:6}.
\paragraph{Higher-rank tensors.} 
In the ambient space formulation,
a tangential, traceless and divergence-free symmetric rank-$s$ tensor on $S^4_N$ is characterized by \cite{Sperling:2017dts,Sperling:2017gmy}
\begin{subequations}
\begin{align}
    y_{b}T^{ba(s-1)}&=0\,,\\
    T_{cda(s-2)}\eta^{cd}&=0\,,\\
    \eth^bT_{ba(s-1)}&=0\,,
\end{align}
\end{subequations}
where derivations $\eth^a$ are defined through \cite{Sperling:2018xrm}
\begin{align}
    r^2m^{ab}\eth_b\bullet:=\{y^a,\bullet\}\,.
\end{align}
To obtain a rank-$s$ covariant tensor $T_{\mu_1...\mu_s}$ on $S^4$ from an $SO(5)$-tensor $T_{a_1...a_s}$, we can use the following pullback
\begin{align}
    \pi^*:T_{a_1...a_s}\mapsto T_{\mu_1...\mu_s}=\frac{\pl y^{a_1}}{\pl x^{\mu_1}}...\frac{\pl y^{a_s}}{\pl x^{\mu_s}}\ T_{a_1...a_s}\,.
\end{align}

\section{Fuzzy twistor space and spinors}\label{sec:3}
In this section, we first give a brief review on commutative twistor space. Then, we describe what is a quantized (or fuzzy) twistor space using the Hopf fibration and spinors. Next, we define the complex structures of the fuzzy twistor space in the semi-classical limit where we have $S^4$ as a classical base manifold and the $S^2\simeq \CP^1$ as the fibers at each point on the manifold $S^4$. We also study spinorial effective vielbein, metric and torsion in the semi-classical limit as a preparation for the next section.
\subsection{Commutative twistor space} Consider homogeneous coordinates on the usual complex projective space $\projectivespace^3$
\begin{align}
    Z^{\Acal} = (Z^1,Z^2,Z^3,Z^4)=(\lambda^{\alpha},\mu^{\alpha'})\in \CC^4\,,\qquad Z^i\neq 0\ \ (i=1,2,3,4)\,.
    \label{complex-Z}
\end{align}
Here $\lambda^{\alpha},\mu^{\alpha'}$ are Weyl spinors of opposite chirality.
We note that the twistor $Z^{\Acal}$ transforms in the fundamental representation of $\msu(4)$, which will be useful to define quantized or fuzzy twistor space in the following. Let us consider the complex conjugation of the twistor $Z^{\Acal}$ denoted as $\bar{Z}_{\Acal}$. Then,
\begin{align}
\bar Z_{\Acal}&= ({\bar\lambda}_{\alpha},
{\bar\mu}_{\alpha'})\,,
\end{align}
and  it transforms in the anti-fundamental representation of $\msu(4)$. By 
restricting ourselves to $\msp(4)\subset \msu(4)$, we 
can use the anti-symmetric matrix $C^{\Acal\Bcal}$
to define the dual twistor $\hat{Z}^{\Acal}$ of $Z^{\Acal}$ as
\begin{align}
    \hat{Z}^{\Acal}=\bar{Z}_{\Bcal}C^{\Acal\Bcal}\,.
\end{align}
In terms of spinors, the dual twistor $\hat Z^{\Acal}$ reads
\begin{align}
    \hat{Z}^{\Acal}&=(\hat{\lambda}^{\alpha},\hat{\mu}^{\alpha'})\,.
\end{align}
Here, 
\begin{align}
    \hat{\lambda}^{\alpha}=\bar{\lambda}_{\beta}\epsilon^{\alpha\beta}\,,\qquad \qquad \hat{\mu}^{\alpha'}=\bar{\mu}_{\beta'}\epsilon^{\alpha'\beta'}\,.
\end{align}
The above is also known as the quaternionic conjugation in twistor literature.
Note that the inner product between the twistor $Z$ and its complex conjugate (or dual) twistor is $SU(4)$-invariant
\begin{align}\label{eq:numberoperator}
\begin{split}
N:= \bar{Z}_{\Acal} Z^{\Acal} &=-\langle\bar{\l} \l \rangle -[ \bar{\mu}\mu]\,,\\
&=-\langle \hat{\lambda}\lambda\rangle-[\hat{\mu}\mu]=-\hat{Z}^{\Acal}Z_{\Acal}\,,
\end{split}
\end{align}
and defines an $S^7 \subset \C^4$. The angle and square brackets are defined as
\begin{align}
    \langle u v\rangle = u^{\alpha}v_{\alpha}\,,\qquad \qquad [uv]=u^{\alpha'}v_{\alpha'}\,.
\end{align}
Here $N$ is a dimensionless number, which will be quantized in the fuzzy case. We define twistor space $\PT$ as the open subset of $\projectivespace^3$ where\footnote{See, e.g. \cite{Adamo:2017qyl,Krasnov:2020lku}, for a nice review on twistor theory.}
\begin{align}
    \PT=\{Z^{\Acal}\in \projectivespace^3|\lambda^{\alpha}\neq 0 \ \text{and} \ N\neq0\}\,,
\end{align}
with the projective line $\{\lambda^{\alpha}= 0\}$  removed.\footnote{From the point of view of the Hopf map \eqref{Hopf-map} this amounts to removing the South pole, which is mapped to infinity by the stereographic projection.} The radius of $S^4$ is 
obtained using the Fierz identity as
\begin{align}
    y_a y^a=R^2 
    = \Big(\frac{r N}{2}\Big)^2 \, \qquad \Rightarrow \qquad R=\frac{rN}{2}\,.
\end{align}
Beside the $SU(4)$-invariant product \eqref{eq:numberoperator}, we also have an inner product that is $Sp(4)$-invariant\footnote{This relation  corresponds to \eqref{eq:highestweightS4N} from the $S^4_N$ point of view.} by considering another twistor $Z_2\neq Z_1$
 \begin{align}
 Z_1^\Acal Z_2^\Bcal C_{\Acal\Bcal} = -[ \mu_1 \mu_2] - \langle \l_1 \l_2\rangle\,.
\end{align}
Now we can understand
the correspondence between twistor space and spacetime 
through the following incident relation:
\begin{align}
\label{eq:incident}
    \mu^{\alpha'}= \tx^{\alpha\alpha'}\lambda_{\alpha}\,.
\end{align}
The inverse of the above reads
\begin{align}
\label{eq:inverseincident}
    \tx^{\alpha\alpha'}= \frac{\hat{\lambda}^{\alpha}\mu^{\alpha'}-\lambda^{\alpha}\hat{\mu}^{\alpha'}}{\langle \hat{\lambda}\lambda\rangle}\,.
\end{align}
This can be understood in terms of the Hopf map
\begin{align}\label{Hopf-map}
    \begin{split}
    \projectivespace^1\xhookrightarrow{}\projectivespace^3 \simeq S^7/_{U(1)}&\rightarrow S^4\,,  \\
     Z^{\Acal} &\mapsto  y^{a} :=\frac r2{\bar Z}_{\Acal} (\gamma^{a})^{\Acal}_{\ \Bcal} Z^{\Bcal}=-\frac{r}{2}\hat{Z}^{\Acal}(\gamma^a)_{\Acal\Bcal}Z^{\Bcal}\,,
   \end{split}
\end{align}
cf. \eqref{fuzzy-S4-osc} and \eqref{fuzzy-Hopf}. We can make the Hopf map more explicit by using the chiral basis of the $\gamma$-matrices in \eqref{gamma-so5-explicit}. They become manifestly anti-symmetric if the first index is lowered with the $\msp(4)$-invariant matrix $C_{\Acal\Bcal}$. Explicitly,
 \begin{align}
  (\gamma_{m})_{\Acal \Bcal}= 
\begin{pmatrix}
 0 & (\tilde\sigma_m)_{\alpha\beta'} \\ 
 -(\tilde\sigma_m)_{\beta'\alpha} & 0
\end{pmatrix}\,,
\quad
 (\gamma_{4})_{\Acal \Bcal}= 
\begin{pmatrix}
 0 & -\epsilon_{\a\a'} \\ 
 \epsilon_{\a\a'} & 0
\end{pmatrix}\,,
\quad
(\gamma_{5})_{\Acal \Bcal}=
\begin{pmatrix}
 -\epsilon_{\a\b} & 0 \\ 0 &\epsilon_{\a'\b'}
\end{pmatrix}
\label{gamma-so5-explicit-AS}
 \end{align}
where $\tilde\sigma^m_{\alpha\alpha'}=-i(\sigma_m)^{\bullet}_{\ \alpha'}\epsilon_{\bullet\alpha}$. This allows us to define a new basis of Pauli's matrices as
 \begin{align}\label{eq:sigmabasis}
 \hat{\sigma}^{\mu}_{\alpha\alpha'}=(\tilde\sigma^m_{\alpha\alpha'},\epsilon_{\alpha\alpha'})=(i\sigma_3,\one_2,-i\sigma_1,i\sigma_2) \ .
 \end{align}
 Comparing  with (\ref{eq:inverseincident}), we recognize
\begin{align}\label{eq:yinspinor}  
    \begin{split}
y^{\mu} =   -\frac r2 \hat Z^{\Acal} (\gamma^{\mu})_{\Acal\Bcal} Z^{\Bcal}
 = \frac r2 \langle \lambda\hat{\lambda}\rangle (\hat\sigma^{\mu})_{\alpha\alpha'}\tx^{\alpha\alpha'}\,,\qquad \qquad  \mu=1,...,4\,,
\end{split}
\end{align}
and, 
\begin{align}
    \begin{split}
 y^5 = - \frac r2 \hat{Z}^{\Acal}(\gamma^5)_{\Acal\Bcal} Z^{\Bcal}
 &= \frac r2([ \hat{\mu}\mu] - \langle\hat{\l} \l\rangle)= -(R+r \langle\hat{\l} \l\rangle)\geq -R\,.
  \label{y5-spinors-explicit}
    \end{split}
\end{align}
using \eqref{eq:inverseincident} and \eqref{eq:numberoperator}. The relation \eqref{y5-spinors-explicit} implies
\begin{align}\label{eq:spinorproduct}
\langle \lambda \hat{\lambda}\rangle =\frac N2\Big(1+\frac{y_5}{R}\Big) = \frac{NR^2}{R^2+x^2}\,,
\end{align}
where we have used the stereographic parametrization \eqref{eq:Yparametrization}. From \eqref{eq:numberoperator}, we can deduce that
\begin{align}\label{eq:muproduct}
    [\mu\hat{\mu}]=\frac{Nx^2}{R^2+x^2}\,.
\end{align}
Note that we can further rewrite \eqref{eq:yinspinor} as
\begin{align}\label{eq:ymu}
    y_{\mu} =  \frac{Nr}{4}\big(1+\frac{y_5}{R}\big) (\hat\sigma_{\mu})_{\alpha\alpha'}\tx^{\alpha\alpha'}
    =\frac{NrR^2}{2(R^2+x^2)}(\hat\sigma_{\mu})_{\alpha\alpha'}\tx^{\alpha\alpha'}=\frac{2R^2x_{\mu}}{(R^2+x^2)}\,
\end{align}
which is in agreement with \eqref{eq:Yparametrization} where we have defined
\begin{align}
    x_{\mu}=
     \frac{rN}4(\hat\sigma_{\mu})_{\alpha\alpha'}\tx^{\alpha\alpha'}
    =
    \frac R2(\hat\sigma_{\mu})_{\alpha\alpha'}\tx^{\alpha\alpha'} \ .
\end{align}%
Therefore, the
incidence relation (\ref{eq:incident}) of the  twistor construction corresponds precisely to the Hopf map followed by a stereographic projection. By denoting $x^2=x_{\mu}x^{\mu}$, and $\tx^2=\tx_{\alpha\alpha'}\tx^{\alpha\alpha'}$, we see that
\begin{align}\label{eq:xtranslation}
    x^2=\frac{R^2}{4}\tx^2\,.
\end{align}
In other words, by moving from Lorentz indices to spinorial indices, the coordinates $x^{\mu}$ is automatically rescaled with a factor of $\frac{R}{2}$. Phrased differently, $x^{\mu}$ has dimension of length while $\tx^{\alpha\alpha'}$ is dimensionless.
In the following, we will let $\lambda,\hat{\lambda}$ be the coordinates of $\CP^1$. Then, the equation \eqref{eq:spinorproduct} allows us to parametrize the spinors $\lambda,\hat{\lambda}$ in a projective way as 
\begin{align}\label{eq:spinorparametrize}
    \lambda_{\alpha}\mapsto\frac{R}{\sqrt{R^2+x^2}}\lambda_{\alpha}:=\frac{R}{\sqrt{R^2+x^2}}\binom{we^{i\theta}}{-1}\,,\qquad \hat{\lambda}_{\alpha}\mapsto\frac{R}{\sqrt{R^2+x^2}}\hat{\lambda}_{\alpha}:=\frac{R}{\sqrt{R^2+x^2}}\binom{1}{we^{-i\theta}}\,,
\end{align}
where $(1+w^2)=N$ and $w\in \RR^*$, $\theta\in [0,2\pi]$. On the other hand, from \eqref{eq:muproduct}, we can parametrize the spinors $\mu,\hat{\mu}$ as
\begin{align}\label{eq:muparametrization}
    \mu^{\alpha'}\mapsto \frac{|x|}{\sqrt{R^2+x^2}}\mu^{\alpha'}:=\frac{|x|}{\sqrt{R^2+x^2}}\binom{a}{b}\,,\qquad \hat{\mu}^{\alpha'}\mapsto \frac{|x|}{\sqrt{R^2+x^2}}\hat{\mu}^{\alpha'}:=\frac{|x|}{\sqrt{R^2+x^2}}\binom{-\bar{b}}{\bar{a}}\,,
\end{align}
where $|a|^2+|b|^2=N$ for $a,b\in \CC^*$. As we will see, it will be convenient to perform the Penrose transform with the above quaternionic parametrization. Henceforth, all spinors are weightless (without conformal factor) unless otherwise stated. 

\subsection{Quantized twistor space} 
\paragraph{Poisson structure and functions on $\CP^3$.} To understand the quantization of twistor space, we must first describe the space of functions on $\CP^3$. It acquires a Poisson structure from the $\msu(4)$-invariant canonical brackets
\begin{align}\label{eq:fuzzytwistor}
    \{Z^{\Acal},\bar{Z}_{\Bcal}\} = -i\delta^{\Acal}_{\ \Bcal}\,,\qquad \{Z^{\Acal},\hat{Z}^{\Bcal}\}=-iC^{\Acal\Bcal}\,.
\end{align}
Then  the "number" generator
\begin{align}
\hat \Ncal := \bar{Z}_{\Acal} Z^{\Acal}
 =-[ \hat{\mu}\mu] -\langle\hat{\l} \l  \rangle\,
 = - \hat{Z}^{\Acal} Z_{\Acal}
\label{N-hat-class}
\end{align}
defines a gradation via 
\begin{align}\label{eq:gradeoftwistors}
    \{\hat\Ncal,\hat{Z}_{\Acal}\} = +i\hat{Z}_{\Acal}\,, \qquad 
     \{\hat\Ncal,Z^{\Acal}\} =- iZ^{\Acal}\,.
\end{align}
Then functions on $\CP^3$ are spanned by polynomials in $\hat{Z}, Z$ that have total grade zero:
\begin{align}\label{eq:blancedcondition}
    \Ccurl := \Big\{P(\hat Z,Z)| \ \{\hat \Ncal,P(\hat{Z},Z)\} = 0\Big\}\,.
\end{align}
It is worth to emphasize that only polynomials that are balanced in $Z,\hat{Z}$ will respect the definition of $\CP^3_N$. This property will be "forwarded" to the quantization space. Due to our definition of the twistor $Z^{\Acal}=(\lambda^{\alpha},\mu^{\alpha'})$, \eqref{eq:gradeoftwistors} implies that $(\lambda,\mu)$ have grade plus one while $(\hat{\lambda},\hat{\mu})$ have grade minus one in the quantized case where we have $i\{,\}\mapsto [,]$.

\paragraph{Quantized twistor space.} In the non-commutative case, we can define the following relations that describes quantized (or fuzzy) twistor space \cite{Penrose:1972ia,Claus:1999xr,Hannabuss:2001xj,Heckman:2011qt}
\begin{align}\label{eq:fuzzytwistor}
    [Z^{\Acal},\bar{Z}_{\Bcal}]=\delta^{\Acal}_{\ \Bcal}\,,\qquad \quad [Z^{\Acal},\hat{Z}^{\Bcal}]=  C^{\Acal\Bcal}\,
\end{align}
which arise from the above Poisson structure.
The second relation above follows directly from the definition of the dual twistor $\hat{Z}^{\Acal}=\bar{Z}_{\Bcal}C^{\Acal\Bcal}$, and $C^{\Acal\Bcal}$ is given explicitly in \eqref{C-matrix}. Henceforth, we will use explicitly $Z^{\Acal}$ and its dual $\hat{Z}^{\Acal}$ to describe quantized twistor space. Let us impose the following $\msu(4)$-invariant constraint
\begin{align}
   \hat \Ncal = \bar{Z}_{\Acal} Z^{\Acal}
   = -\hat{Z}^{\Acal} Z_{\Acal} =  N
   \label{grade-constraint-1}
\end{align}
which holds on the $N$-particle Fock space 
\begin{align}\label{eq:Fockspace}
  \Hcal_N  =(0,0,N)_{\msu(4)}\,
  = (0,0,1)^{\otimes_{\text{sym}} N}.
\end{align}
We can understand $\Hcal_N$ as the space generated by $N$ creation operators $\hat{Z}^{\Acal}_m$ for $m=1,...,N$, i.e. $\Hcal_N=\hat{Z}_1...\hat{Z}_N|0\rangle$\,. 
Then the algebra
\begin{align}
    \Ccurl := End(\Hcal_N) \ 
\end{align}
is recognized as quantized space of functions on $\CP^3$, which is known as fuzzy $\CP^3_N$ \cite{Sperling:2017dts,Sperling:2017gmy,Balachandran:2001dd,Medina:2012cs}. We once again emphasize that $ \Ccurl$
comprises only balanced polynomials in $\hat Z,Z$. 
This reflects the classical definition $\CP^3 \cong S^7/_{U(1)}$ (\ref{Hopf-map}). Because any polynomial with more than $N$ annihilation operators will vanish identically upon normal ordering, $\Ccurl$ is a finite-dimensional space. 
For this reason, the fuzzy 
 $\CP^3_N$ is nothing but quantized twistor space, 
 or more precisely it is a quantization of compactified twistor space. An uncompactified version of the fuzzy twistor space can be defined similarly in terms of $\CP^{2,1}$, see for example \cite{Sperling:2018xrm}.

The fuzzy twistor space can be understood in terms of the following non-commutative version of the Hopf fibration
\begin{align}\label{fuzzy-Hopf}
    \begin{split}
    \projectivespace^1_N\xhookrightarrow{}\projectivespace^3_N \simeq S^7/_{U(1)}&\rightarrow S^4\,,  \\
     Z^{\Acal} &\mapsto Y^{a}=\frac{r}{2} \hat{Z} \gamma^a Z\,.
     \end{split}
\end{align}
Consider a point $x_0\in S^4$ as a reference point. Then the fiber over $x_0$ is determined by \cite{Steinacker:2015dra}
\begin{align}
    \hat{Z}\gamma^5Z=1\,.
\end{align}
The above defines $S^3$, which can be reduced to $S^2$ by quotienting out $U(1)$. Hence, the fuzzy $S^4_N$ can be understood as  projection of fuzzy twistor space $\projectivespace^3_N$. We note that in the non-commutative case, the incident relation \eqref{eq:incident} does not have a well-defined inverse. Moreover, since there is not a geometry in the usual sense, differential forms and complex structures are not defined a priori. We will therefore restrict ourselves mostly to the  
semi-classical or
large $N$ regime, where the non-commutative structure reduces to classical Poisson geometry.

The above construction can also be described in terms of two spinorial creation- and anihilation operators,
noting that \eqref{eq:fuzzytwistor} implies the commutation relations 
\begin{align}\label{eq:fuzzyfiber}
    [\lambda^{\alpha},\hat{\lambda}^{\beta}] &=\epsilon^{\alpha\beta}, \qquad 
    [\mu^{\alpha'},\hat{\mu}^{\beta'}] = \epsilon^{\alpha'\beta'} \,,
\end{align}
which are quantizations of the  Poisson structure
\begin{align}
   \{\lambda^{\alpha},\hat{\lambda}^{\beta}\} = -i
    \epsilon^{\alpha\beta}\,, \qquad 
    \{\mu^{\alpha'},\hat{\mu}^{\beta'}\} 
    = -i \epsilon^{\alpha'\beta'} \, .
    \label{spinor-Poisson}
\end{align}
Hence, the fuzziness of twistor space is encoded by the non-commutativity of the spinors $\lambda$ and $\mu$.\footnote{The spinors $\lambda,\mu$ are also known as doubletons \cite{Gunaydin:1998jc,Gunaydin:1998sw,Sezgin:2001zs}.} 
The space of functions can also be described in terms of these spinors
\begin{align}
    \Ccurl=\sum\limits_{n}\varpi^{\alpha(n)\beta(n),\alpha'(n)\beta'(n)}\lambda_{\alpha}...\lambda_{\alpha}\mu_{\alpha'}...\mu_{\alpha'}\hat{\lambda}_{\beta}...\hat{\lambda}_{\beta}\hat{\mu}_{\beta'}...\hat{\mu}_{\beta'}\,.
\end{align}
Here, any normal ordering can be chosen, and we have spelled out the balance condition explicitly. For convenience, we note the following useful relations
\begin{align}
    \{\langle \hat{\lambda}\lambda\rangle, \l^\a\} = 
      +i \l^\a \,,  \qquad \qquad 
 \{\langle \hat{\lambda}\lambda\rangle, \hat\l^\a\}
 =-i \hat\l_\a \,.
\end{align}
We also find
\begin{align}
 \{[\hat{\mu}\mu], \mu^{\a'}\} & =  i \mu^\a \ , \qquad \qquad 
 \{[\hat{\mu}\mu], \hat\mu^{\a'}\}  =  -i \hat\mu^\a \,.
\end{align}
As a consequence, 
\begin{align}
    \{\hat \Ncal,\tx^{\a\a'}\} = 0\,,
\end{align}
which means $\tx^{\alpha\alpha'}$ has grade zero, as it must.

\paragraph{Balanced weight representations (BWR) for higher-spin modes.} Using the incident relation \eqref{eq:incident} in the semi-classical limit, we can cast any function $\omega(\lambda,\mu;\hat{\lambda},\hat{\mu})$ into $\omega(\tx,\lambda,\hat{\lambda})$. In particular, the space of functions on twistor space comprises of polynomials purely in terms of the spinors $\lambda,\hat{\lambda}$,
\begin{align}\label{balancedC}
    \Ccurl=\sum_{n}f^{\alpha(n)\beta(n)}(\tx)\lambda_{\alpha}...\lambda_{\alpha}\hat{\lambda}_{\beta}...\hat{\lambda}_{\beta}\,.
\end{align}
In the semi-classical limit, the coefficient $f^{\alpha(n)\beta(n)}(\tx)$ becomes a tensorial field in spacetime, which for irreducible modes is totally symmetric in all $2n$ indices\footnote{Note that any anti-symmetric combination of $\a$ and $\b$ reduces to $\langle \l\hat\l\rangle$, which would have lower spin due to \eqref{eq:spinorproduct}.}; this provides the link to the maximally unbalanced (MUR) representation in section \ref{sec:unbalanced}. Besides the space of functions $\Ccurl$, we also need to have the notion of the space of higher-spin valued vector modes, which we will denote $\Acurl$. Using the same argument about balanced weight representation (BWR), our space of vector-modes reads
\begin{align}\label{balancedA}
    \Acurl=\sum_{m=n}A^{\alpha(m)\beta(n)\gamma,\gamma'}(\tx)\lambda_{\alpha}...\lambda_{\alpha}\hat{\lambda}_{\beta}...\hat{\lambda}_{\beta}\,.
\end{align}
Here, $\gamma,\gamma'$ are two independent indices which can be converted into the usual Lorentz index $\mu$ by Pauli's matrices. In spacetime, the coefficient $A^{\alpha(m),\alpha'}$ is a tensorial field that is symmetric in the first group of unprimed indices and represent generalized gauge potentials. We also need the spaces of higher-spin valued (fermionic) spinor modes,
\begin{align}\label{balancedF}
    \Fcurl=\begin{cases}\sum_{m=n}\chi^{\alpha(m)\beta(n)\gamma}(\tx)\lambda_{\alpha}...\lambda_{\alpha}\hat{\lambda}_{\beta}...\hat{\lambda}_{\beta}\,,\\
    \sum_{m=n}\widetilde{\chi}^{\alpha(m)\beta(n),\gamma'}(\tx)\lambda_{\alpha}...\lambda_{\alpha}\hat{\lambda}_{\beta}...\hat{\lambda}_{\beta}\,,
    \end{cases}
\end{align}
To this end, let us make the following remark. Due to the condition of balancing the weight in twistors (or spinors) oscillators, we have the BWR on fuzzy twistor space. However, after integrating out all fibre coordinates $(\lambda,\hat{\lambda})$, we end up with the MUR in spacetime. It is astonishing how the Penrose transform can help us move from one to another representation (see section \ref{sec:4}).
\subsection{Complex structures of the fuzzy twistor space in the semi-classical limit} 
At large $N$, where coordinates are effectively commutative, we can define a symplectic form $\Omega$ on fuzzy twistor space as \cite{Penrose:1967wn,Penrose:1972ia}
\begin{align}
    \Omega=d\hat{Z}^\Acal\wedge dZ_{\Acal}=(1+\tx^2)\Big[D\hat{\lambda}^{\alpha}\wedge D\lambda_{\alpha}+\hat{\lambda}_{\alpha}\frac{d\tx^{\alpha\alpha'}\wedge d\tx^{\beta}_{\ \alpha'}}{(1+\tx^2)^2}\lambda_{\beta}\Big]\,,
\end{align}
where we have used the incident relation \eqref{eq:incident} and $\tx^2:= \tx^{\alpha\alpha'}\tx_{\alpha\alpha'}$. This is nothing but the Kirillov-Kostant symplectic form on $\CP^3$, which is underlying fuzzy  $\CP^3_N$. Here,
\begin{align}
    D\hat{\lambda}^{\alpha}=d\hat{\lambda}^{\alpha}+\frac{d\tx_{\beta\beta'}\tx^{\alpha\beta'}}{(1+\tx^2)}\hat{\lambda}^{\beta}\,,\qquad \qquad 
    D\lambda_{\alpha}=d\lambda_{\alpha}+\frac{\tx_{\alpha\beta'}d\tx^{\beta\beta'}}{(1+\tx^2)}\lambda_{\beta}\,.
\end{align}
We note that for $K=\hat{Z}_{\Acal}Z^{\Acal}$,
\begin{align}\label{eq:complexstructure}
    \Omega=\bar{\pl}\pl K\,\,,\qquad \pl=dZ^\Acal\frac{\pl}{\pl Z^\Acal}\,,\qquad \bar{\pl}=d\hat{Z}^\Acal\frac{\pl}{\pl \hat{Z}^\Acal}\,.
\end{align}
Hence, in the semi-classical limit, we have the following fibration $$\projectivespace^1_N\xhookrightarrow{}\projectivespace^3_N \rightarrow S^4\,,$$
Here, $S^4$ is the base space and $\CP^1_N$ are the fibers. Moreover, we will take $\bar{\pl}$ as our definition of integrable complex structure on $\CP^3_N$ since $\bar{\pl}^2=0$.

\subsection{Background  kinetic term in semi-classical limit}\label{sec:3.4}
For the $S^4$ background, the kinetic term of the matrix model arises from  the Poisson brackets with the background configuration $Y^a\equiv y^a$, which acts on the fields $\phi\in\Ccurl$ as
\begin{align}
    \{y^a,\phi\} = r^2 m^{ab}\eth_b \phi\,,\qquad a=1,2,3,4,5\,.
\end{align}
In terms of components, the above reduces to 
\begin{subequations}
\begin{align}\label{eq:PoissonbracketsLorentzindices}
     \{y_\mu,\phi\} &=  \Big\{\frac{2R^2}{R^2+x^2}x_{\mu},\phi\Big\}\,,\qquad \mu=1,2,3,4\,,\\
     \{y_5,\phi\} &=
    \Big\{\frac{R(R^2-x^2)}{R^2 + x^2},\phi\Big\}\,,
\end{align}
\end{subequations}
where $x_{\mu}\equiv \frac{R}{2} \sigma_{\mu}^{\alpha\alpha'}\tx_{\alpha\alpha'}$ and we have used \eqref{eq:Yparametrization}. As discussed above, the twistor correspondence is equivalent to a Hopf map followed by a rescaling of $\frac{2R^2}{(R^2+x^2)}$. We note that in the flat limit,
\begin{align}
    \{y_{\mu},\phi\}\mapsto 2\{x_{\mu},\phi\}\,,\qquad \{y_5,\phi\}\mapsto 0\,.
\end{align}
\paragraph{Spinorial effective vielbein and derivativation.} To evaluate the Poisson bracket $\{y^{\mu},\phi\}$ in terms of spinorial indices, namely  $\{\ty^{\alpha\alpha'},\phi(\tx|\lambda,\hat{\lambda})\}$, we can use \eqref{eq:yinspinor} to see that
\begin{align}\label{eq:yinspinors}
    \ty^{\alpha\alpha'}=-(\hat{\lambda}^{\alpha}\mu^{\alpha'}-\lambda^{\alpha}\hat{\mu}^{\alpha'})\,,\qquad y_{\mu}=\frac{r}{2}(\sigma_{\mu})_{\alpha\alpha'}\ty^{\alpha\alpha'}\,.
\end{align}
Then, we first evaluate
\begin{align}\label{eq:Poissononcoordinates1}
    \begin{split}
    \{\ty^{\alpha\alpha'},\ty^{\beta\beta'}\}&= 2i(\lambda^{(\alpha}\hat{\lambda}^{\beta)}\epsilon^{\alpha'\beta'}+\mu^{(\alpha'}\hat{\mu}^{\beta')}\epsilon^{\alpha\beta})\,,
     \end{split}
\end{align}
where $A^{(\alpha}B^{\beta)}=\frac{1}{2}(A^{\alpha}B^{\beta}+A^{\beta}B^{\alpha})$ as explained in the convention. Next, we find
\begin{subequations}\label{eq:Poissononcoordinates2}
\begin{align}
     \{\ty^{\alpha\alpha'},\lambda^{\beta}\}
    &=+i\epsilon^{\alpha\beta}\mu^{\alpha'}\,,\\
    \{\ty^{\alpha\alpha'},\hat{\lambda}^{\beta}\}&=-i\epsilon^{\alpha\beta}\hat{\mu}^{\alpha'}\,.
\end{align}
\end{subequations}
Now, we make a crucial observation that follows directly from \eqref{eq:yinspinor}. Instead of working with $\hs$-valued functions on $S^4$ where the fibers depend on coordinates $\tx^{\alpha\alpha'}\in S^4$, we can also work with $\hs$-valued functions $\varphi(\ty)$ on the ambient space $\RR^4\subset \RR^5$ and make a conformal transformation that maps $\varphi(\ty)$ to $\phi(\tx)$ later. The reason is that the fibers will be the same everywhere on $\RR^4$. Hence, $\CP^3=\CP^1\times \RR^4$ as in the flat limit. It is an important difference to
the $S^4_N$ point of view in \cite{Sperling:2017dts,Sperling:2017gmy}, where the internal fiber depends on $\tx\in S^4$. The total space is now an (infinite-dimensional) bundle 
$\Kcurl=\Ccurl(\CP^1)\times \RR^4$. Since it is now a Cartesian product, our bundle is a trivial one. It, then, makes sense to consider $\varphi(\ty)$
as a section of the bundle of $\hs$-valued functions  over $\R^4$. Using the above information, we can compute $\{\ty^{\alpha\alpha'}, \varphi\}$ where $\varphi(\ty|\lambda,\hat{\lambda})\in \Ccurl$ as
\begin{align}\label{nabla-def-1}
    \begin{split}
    \{\ty^{\alpha\alpha'},\varphi\}:&=\Big(\{\ty^{\alpha\alpha'},\ty^{\beta\beta'}\}\frac{\pl}{\pl \ty^{\beta\beta'}}+\{\ty^{\alpha\alpha'},\lambda^{\beta}\}\frac{\pl}{\pl \lambda^{\beta}}+\{\ty^{\alpha\alpha'},\hat{\lambda}^{\beta}\}\frac{\pl}{\pl \hat{\lambda}^{\beta}}\Big)\varphi\,\\
    &=\Ecal^{\alpha\alpha'|\beta\beta'}\pl_{\beta\beta'}\varphi+\Ecal^{\alpha\alpha'|\beta}\frac{\pl}{\pl \lambda^{\beta}}\varphi+\hat{\Ecal}^{\alpha\alpha'|\beta}\frac{\pl}{\pl \hat{\lambda}^{\beta}}\varphi\,.
    \end{split}
\end{align}
The Poisson bracket for the fifth coordinate reads
\begin{align}\label{eq:5thderivation}
    \begin{split}
    \{\ty^5,\varphi\} &:=\Big(\{\ty^5,\ty^{\beta\beta'}\}\frac{\pl}{\pl \ty^{\beta\beta'}}+\{\ty^5,\lambda^{\alpha}\}\frac{\pl}{\pl \lambda^{\alpha}}+\{\ty^5,\hat{\lambda}^{\alpha}\}\frac{\pl}{\pl \hat{\lambda}^{\alpha}}\Big)\varphi\,\\
    &=\Ecal^{5|\beta\beta'}\pl_{\beta\beta'}\varphi+\Ecal^{5|\beta}\frac{\pl}{\pl \lambda^{\beta}}\varphi+\hat{\Ecal}^{5|\beta}\frac{\pl}{\pl \hat{\lambda}^{\beta}}\varphi\,,
    \end{split}
\end{align}
where 
\begin{subequations}
\begin{align}
    \{\ty^5,\ty^{\alpha\alpha'}\}&=-i(\hat{\lambda}^{\alpha}\mu^{\alpha}+\lambda^{\alpha}\hat{\mu}^{\alpha'})\,,\\
    \{\ty^5,\lambda^{\alpha}\}&=+i\lambda^{\alpha}\,,\\
    \{\ty^5,\hat{\lambda}^{\alpha}\}&=-i\hat{\lambda}^{\alpha}\,.
\end{align}
\end{subequations}
We will refer $\Ecal$ 
to as the effective spinorial vielbeins. Explicitly,
\begin{subequations}\label{eq:effectivevielbein}
\begin{align}
    \Ecal^{\alpha\alpha'|\beta\beta'}:&=\{\ty^{\alpha\alpha'},\ty^{\beta\beta'}\}=2i(\lambda^{(\alpha}\hat{\lambda}^{\beta)}\epsilon^{\alpha'\beta'}+\mu^{(\alpha'}\hat{\mu}^{\beta')}\epsilon^{\alpha\beta})\,\label{eq:1steffectivevielbein},\\
    \Ecal^{5|\alpha\alpha'}:&=\{\ty^5,\ty^{\alpha\alpha'}\}=-i(\hat{\lambda}^{\alpha}\mu^{\alpha'}+\lambda^{\alpha}\hat{\mu}^{\alpha'})\,.
\end{align}
\end{subequations}
We note that in the flat limit where $R\rightarrow \infty$, the effective vielbein $\Ecal^{\alpha\alpha'|\beta\beta'}$ coincides with the tensor $J^{\alpha\alpha'|\beta\beta'}$ in \cite{Woodhouse:1985id} up to a conformal rescaling. It is an important fact that can help us obtain the action of the HS-IKKT in flat space, see section \ref{sec:4}.

\paragraph{On higher-spin valued derivation.} The above suggests a natural way to define a frame and a derivation $\eth$ for the $\hs$-valued functions $\varphi$ on $\RR^4$ as 
\begin{align}
    \{\ty_{\alpha\alpha'},\varphi\}
    := \Ecal_{\alpha\alpha'}^{\quad  \beta\beta'}\eth_{\b\b'} \varphi
    =  \Ecal_{\alpha\alpha'}^{\quad  \beta\beta'}\Big( \pl_{\b \b'} + \Sigma_{\b\b'}  \Big)\varphi\,,
     \label{hs-cov-deriv-def}
\end{align}
where 
\begin{align}\label{connection-hs-def}
\Ecal^{\alpha\alpha'\beta\beta'}\Sigma_{\b\b'}\varphi := \Big(\{\ty^{\alpha\alpha'},\lambda^{\beta}\}\frac{\pl}{\pl \lambda^{\beta}}+\{\ty^{\alpha\alpha'},\hat{\lambda}^{\beta}\}\frac{\pl}{\pl \hat{\lambda}^{\beta}}\Big)\varphi\,.
\end{align}
The above Poisson brackets will be the same for $A^{\alpha\alpha'}(\ty|\lambda,\hat{\lambda})\in \Acurl$ and $\{\chi^{\alpha}(\ty|\lambda,\hat{\lambda}),\widetilde{\chi}^{\alpha'}(\ty|\lambda,\hat{\lambda})\}\in \Fcurl$ (cf. \eqref{balancedA}, \eqref{balancedF}). Here, it is clear that the $\Sigma$ operator acts only on the fibre and can be thought of as "spin" operator. By contracting with $\Ecal^{\kappa\kappa'}_{\ \ \ \alpha\alpha'}$, we obtain the following expression
\begin{align}
    \begin{split}
    g^{\kappa\kappa'\beta\beta'}\Sigma_{\beta\beta'}\varphi&=\Ecal^{\kappa\kappa'}_{\ \ \ \alpha\alpha'}\Big(\Ecal^{\alpha\alpha'|\bullet}\frac{\pl}{\pl \lambda^{\bullet}}+\hat{\Ecal}^{\alpha\alpha'|\bullet}\frac{\pl}{\pl \hat{\lambda}^{\bullet}}\Big)\varphi\\
    &=-i\Big(\lambda^{\kappa}\hat{\lambda}_{\alpha}+\lambda_{\alpha}\hat{\lambda}^{\kappa})\Big(\mu^{\kappa'}\frac{\pl}{\pl \lambda_{\alpha}}-\hat{\mu}^{\kappa'}\frac{\pl}{\pl \hat{\lambda}_{\alpha}}\Big)\varphi+i[\hat{\mu}\mu]\Big(\mu^{\kappa'}\frac{\pl}{\pl \lambda_{\kappa}}+\hat{\mu}^{\kappa'}\frac{\pl}{\pl \hat{\lambda}_{\kappa}}\Big)\varphi\,.
    \end{split}
\end{align}
Here, $g^{\alpha\alpha'\beta\beta'}$ is the effective metric in the tangential direction that will be defined in \eqref{eq:tensoreffectivemetric1}. Due to the parametrization \eqref{eq:muparametrization}, we see that the rhs. vanishes in the flat limit. Hence, to a good approximation, the contribution of $\Sigma$ can be neglected when the radius $R$ is large enough. This observation will be useful when we study the scattering amplitudes of the HS-IKKT model in the semi-classical and flat limit. 
\paragraph{The effective metric.} Combining \eqref{nabla-def-1} and \eqref{eq:5thderivation},  the effective metric is obtained as follow
\begin{align}\label{eq:preeffectivemetric}
    \begin{split}
    \{\ty^{\zeta\zeta'},\vartheta\}\{\ty_{\zeta\zeta'},\vartheta\}+\{y^5,\vartheta\}\{y^5,\vartheta\}&=\Ecal^{\zeta\zeta'|\alpha\alpha'}\pl_{\alpha\alpha'}\vartheta\, \Ecal_{\zeta\zeta'|\beta\beta'}\pl^{\beta\beta'}\vartheta+\Ecal^{5|\alpha\alpha'}\pl_{\alpha\alpha'}\vartheta\Ecal_{5|\beta\beta'}\pl^{\beta\beta'}\vartheta\\
    &=:g^{\alpha\alpha'\beta\beta'}+\varrho^{\alpha\alpha'\beta\beta'}\\
    &=:G^{\alpha\alpha'\beta\beta'}\pl_{\alpha\alpha'}\vartheta \,\pl_{\beta\beta'}\vartheta\,,
    \end{split}
\end{align}
where $\vartheta(\ty)$ is some scalar field. Explicitly, the effective metric is  (see the derivation in Appendix \ref{app:C})
\begin{align}\label{fullmetric}
    G^{\alpha\alpha'\beta\beta'}(\ty)&=N^2\epsilon^{\alpha\beta}\epsilon^{\alpha'\beta'}-\ty^{\alpha\alpha'}\ty^{\beta\beta'}\,.
\end{align}
It is remarkable that the total effective metric depends only on the coordinates $\ty^{\alpha\alpha'}$ of $\RR^4$  and not on the ``internal" spinors, which is in
consistent with (2.33) in \cite{Steinacker:2016vgf}. In terms of $\tx^{\alpha\alpha'}$, the effective metric reads
\begin{align}\label{eq:effectivemetricx}
    G^{\alpha\alpha'\beta\beta'}(\tx)=\langle \hat\l\l\rangle^2\Big(\frac{N^2}{\langle \hat \l \l\rangle^2}\epsilon^{\alpha\beta}\epsilon^{\alpha'\beta'}-\tx^{\alpha\alpha'}\tx^{\beta\beta'}\Big)\,.
\end{align}
It is clear from \eqref{eq:spinorproduct} and \eqref{eq:effectivemetricx} that $\langle \hat \l \l\rangle$ plays the role of the conformal factor.
\paragraph{Torsion.} On the fuzzy $S^4$ background, the torsion related to the Witzenb\"ock connection can be computed as \cite{Steinacker:2020xph}
\begin{align}\label{eq:torsion}
    \begin{split}
    \{\{\ty^{\beta\beta'},\ty^{\zeta\zeta'}\},\ty^{\bullet\bullet'}\}\frac{\pl \tx^{\alpha\alpha'}}{\pl \ty^{\bullet\bullet'}}&=\frac{1}{\langle \hat \l \l \rangle}\Big(\epsilon^{\alpha\zeta}\epsilon^{\beta'\zeta'}\ty^{\beta\alpha'}+\epsilon^{\alpha'\beta'}\epsilon^{\beta\zeta}\ty^{\alpha\zeta'}+\epsilon^{\beta\zeta}\epsilon^{\alpha'\zeta'}\ty^{\alpha\beta'}+\epsilon^{\alpha\beta}\epsilon^{\beta'\zeta'}\ty^{\zeta\alpha'}\Big)\\
    &=\frac{2}{\langle \hat \l \l \rangle}\big[\epsilon^{\beta'\zeta'}\epsilon^{\alpha(\beta}\ty^{\zeta)\alpha'}+\epsilon^{\beta\zeta}\epsilon^{\alpha'(\beta'}\ty^{\alpha\zeta')}\big]\,.
    \end{split}
\end{align}
We would like to emphasize that the Weitzenb\"ock connection only makes sense for 4 dimensional frames $\Ecal^{\alpha\alpha'\beta\beta'}$. However, as shown above, the $5th$ direction is crucial for obtaining the effective metric. Hence, it would be interesting to develop further the definition of torsion and covariant derivative in \cite{Steinacker:2020xph} for the decomposition of fuzzy twistor space considered in this paper. We will return to this question in a future work.






\section{A twistorial description of the (HS)-IKKT}\label{sec:4} 
In this section, we first rewrite the IKKT-matrix model using $\msp(4)$-indices which can be reduced further down to spinorial indices. By dropping some terms in the interactions with the requirement that the action is still gauge invariant and has the same degrees of freedom, we obtain the self-dual sector of the IKKT-matrix model. In the semi-classical limit, we perform the Penrose transform to obtain an effective spacetime action for the (self-dual) IKKT model. As discussed in the previous sections, all functions, fermionic spinor modes and vector modes of the IKKT model takes values in $\Ccurl,\Fcurl,\Acurl$, respectively, which can be thought of as quantized functions on twistor space or as $\hs$-valued functions on $\R^4$ in the SCF limit. Therefore, the spacetime action for the IKKT model is a higher-spin theory that exhibits many twistorial features. We will also briefly discuss about the maximally unbalanced representation (MUR) in $4d$. It is interesting to note that the Penrose transform will move fields from BWR on $\CP^3_N$ to MUR in spacetime.



\subsection{Spinorial representation for the action of the IKKT model}
\label{sec:action-spinor}
We have learnt that the $S^4_N$ generators $Y_{a}$ can be mapped to $Y^{\Acal\Bcal}=P^{\Acal\Bcal}+Q^{\Acal\Bcal}$. Due to the realization of the $\mso(5)$ $\gamma$-matrices, we can decompose 
\begin{align}\label{eq:YABdecomposition}
    Y^{\Acal\Bcal}=P^{\Acal\Bcal}+Q^{\Acal\Bcal}=\begin{pmatrix}
     0 & P^{\alpha\beta'}\\
     -P^{\beta'\alpha}&0
    \end{pmatrix}+\begin{pmatrix}
     Q^{\alpha\beta} & 0\\
     0& Q^{\alpha'\beta'}
    \end{pmatrix}\,,
\end{align}
where $P$ are off-diagonal, and $Q$ represents the fifth direction and is diagonal. In particular,
\begin{subequations}
\begin{align}
    P^{\alpha\alpha'}&=P^{\mu}\hat{\sigma}_{\mu}^{\alpha\alpha'}\,,\qquad \qquad  \hat\sigma_{\mu}^{\alpha\alpha'}=(i\sigma_3,\one{_2},-i\sigma_1,i\sigma_2)\\
    Q^{\alpha\beta} &= - Y_5\epsilon^{\alpha\beta} , \qquad \qquad \ \,
     Q^{\alpha'\beta'} = + Y_5\epsilon^{\alpha'\beta'} \,.
\end{align}
\end{subequations}
The remaining 5 coordinates of $SO(10)$ in \eqref{eq:IKKTaction} that $SO(5)$ does not act on will be denoted as $\widetilde{Y}_i$ for $i=6,7,8,9,10$. They will be treated as scalar fields. The $\gamma$-matrices that associate to these extra coordinates will form another $\mso(5)\simeq \msp(4)$ algebra. Hence, we can write them as 
\begin{align}\label{eq:extrascalars}
    \widetilde{Y}^i\gamma_i^{\Ical \Jcal}\mapsto \phi^{\Ical \Jcal}\,,\qquad \qquad  \Ical,\Jcal=1,2,3,4\,.
\end{align}
Here, the $\Ical,\Jcal$ indices are understood as the indices that associates to the internal symmetry group $SU(4)$ of $\Ncal=4$ SYM. In principle, we would like to have 6 scalar fields transform in the adjoint of $SU(4)$. However, since the external $SO(5)$ acts on one of the scalar fields, it breaks the internal group $SU(4)$ explicitly. Therefore, we cannot treat $Q^{\Acal\Bcal}$ and $\phi^{\Ical\Jcal}$ on the same footing at this stage. Note that the internal group $SU(4)$, can be recovered in the flat limit, see e.g. \cite{Steinacker:2010rh}. Hence, in the flat limit the IKKT is said to be higher-spin extensions of $\Ncal=4$ SYM. After the substitution of \eqref{eq:YABdecomposition} and \eqref{eq:extrascalars}, the IKKT model becomes
\begin{align}
    \begin{split}
    S&=\Tr\Big([P^{\Acal\Ccal},P^{\Bcal\Dcal}][P_{\Acal\Ccal},P_{\Bcal\Dcal}]+2[P^{\Acal\Ccal},Q^{\Bcal\Dcal}][P_{\Acal\Ccal},Q_{\Bcal\Dcal}]+2[P^{\Acal\Ccal},\phi^{\Ical\Jcal}][P_{\Acal\Ccal},\phi_{\Ical\Jcal}]\Big)\\
    &+\Tr\Big(\bar{\Psi}^{\Acal}[P_{\Acal\Bcal},\Psi^{\Bcal}]+\bar{\Psi}^{\Acal}[Q_{\Acal\Bcal},\Psi^{\Bcal}]+\bar{\Psi}^{\Ical}[\phi_{\Ical\Jcal},\Psi^{\Jcal}]\Big)\\
    &+\Tr\Big([Q^{\Acal\Ccal},Q^{\Bcal\Dcal}][Q_{\Acal\Ccal},Q_{\Bcal\Dcal}]+2[Q^{\Acal\Ccal},\phi^{\Ical\Jcal}][Q_{\Acal\Ccal},\phi_{\Ical\Jcal}]+[\phi^{\Ical\Jcal},\phi^{\Mcal\Ncal}][\phi_{\Ical\Jcal},\phi_{\Mcal\Ncal}]\Big)\,.
    \end{split}
\end{align}
Using \eqref{eq:YABdecomposition}, we can further write the IKKT-model in terms of spinorial indices as
\begin{align}
    \begin{split}
    S&=\Tr\Big(4[P^{\alpha\alpha'},P^{\kappa\kappa'}][P_{\alpha\alpha'},P_{\kappa\kappa'}]+8[P^{\alpha\alpha'},Q^{\beta\beta'}][P_{\alpha\alpha'},Q_{\beta\beta'}]+16[P^{\alpha\alpha'},\phi^{IJ}][P_{\alpha\alpha'},\phi_{IJ}]\\
    &+32\bar{\chi}^{\alpha}[P_{\alpha\beta'},\widetilde{\chi}^{\beta'}]+16\widetilde{\bar{\chi}}^{\alpha'}[Q_{\alpha'\beta'},\widetilde{\chi}^{\beta'}]+16\bar{\chi}^{\alpha}[Q_{\alpha\beta},\chi^{\beta}]+16\bar{\chi}^I[\phi_{IJ},\chi^J]+16\widetilde{\bar{\chi}}^{I'}[\phi_{I'J'},\widetilde{\chi}^{J'}]\\
    &+4[Q^{\alpha\beta},Q^{\gamma\delta}][Q_{\alpha\beta},Q_{\gamma\delta}]+8[Q^{\alpha\beta},\phi^{IJ}][Q_{\alpha\beta},\phi_{IJ}]+16[\phi^{IJ},\phi^{MN}][\phi_{IJ},\phi_{MN}]\Big)\,.
    \end{split}
\end{align}
where $\Psi^{\Acal}=(\chi^{\alpha},\widetilde{\chi}^{\alpha'})$ and $\Psi^{\Ical}=(\chi^{I},\widetilde{\chi}^{I'})$ for $I,I'=1,2$. From the discussion in the subsection \ref{sec:2.3}, we note that $\phi^{\Ical\Jcal}=-\phi^{\Jcal\Ical}$ can be written in terms of $2\times 2$ block matrices $\phi^{IJ}$ and $\phi^{I'J'}=-\phi^{IJ}$.\footnote{These block matrices can be diagonal or off-diagonal inside the $4\times 4$ $\phi^{\Ical\Jcal}$ matrices but the detail does not effect the computation below.} Note that we have rescaled $\Psi\rightarrow 4\Psi$ for later convenience. Let us explicitly compute the Yang-Mills part of the above action. First of all,
\begin{align}
  [P^{\alpha\alpha'},P^{\kappa\kappa'}]
   = \epsilon^{\a\k}[P^{\g(\alpha'},P_\g^{\ \kappa')}]
    + \epsilon^{\a'\k'}[P^{(\a\g'},P^{\kappa)}_{\  \g'}]
     = \epsilon^{\a\k}F^{\alpha'\k'} + \epsilon^{\a'\k'}F^{\alpha\k}\,.
\end{align}
where our definition of the field strength $F$ is
\begin{align}
    F^{\alpha\k}=[P^{(\alpha}_{\ \ \gamma'},P^{\k)\gamma'}]\,,
\end{align}
which is symmetric in $\a$ and $\k$.
Since $P^{\alpha\alpha'}=-P^{\alpha'\alpha}$, we obtain
\begin{align}
    [P^{\alpha\alpha'},P^{\kappa\kappa'}][P_{\alpha\alpha'},P_{\kappa\kappa'}]=4F_{\alpha\kappa}F^{\alpha\kappa}\,.
\end{align}
Now, we can consider the following fluctuation
\begin{align}
    \binom{P^{\alpha\alpha'}}{Q^{\alpha\beta}}=\binom{\TY^{\alpha\alpha'}}{\TY_5\epsilon^{\alpha\beta}}+\binom{A^{\alpha\alpha'}}{\hat\phi\epsilon^{\alpha\beta}}\,,
\end{align}
where $\TY$ describes the background and $(A,\hatphi)$ stand for fluctuations. Note that $\hatphi$ is the scalar field that $SO(5)\sim Sp(4)$ acts on. The above action can be simplified further to
\begin{align}\label{eq:intermediate1}
    \begin{split}
    S=&\Tr\Big(\frac{1}{2}F_{\alpha\alpha}F^{\alpha\alpha}+\frac{1}{2}[P^{\alpha\alpha'},\hatphi][P_{\alpha\alpha'},\hatphi]+\frac{1}{2}[P^{\alpha\alpha'},\phi^{IJ}][P_{\alpha\alpha'},\phi_{IJ}]+\bar{\chi}^{\alpha}[P_{\alpha\beta'},\widetilde{\chi}^{\beta'}]\\
    &+\frac{1}{2}[\TY_5,P^{\alpha\alpha'}][\TY_5,P_{\alpha\alpha'}]+\frac 12 [\TY_5,\hat{\phi}][\TY_5,\hatphi]+\frac{1}{4}[\TY_5,\phi^{IJ}][\TY_5,\phi_{IJ}]-\frac{1}{2}\widetilde{\bar{\chi}}_{\alpha'}[\TY_5,\widetilde{\chi}^{\alpha'}]\\
    &+\frac{1}{2}\bar{\chi}_{\alpha}[\TY_5,\chi^{\alpha}]-\frac{1}{2}\widetilde{\bar{\chi}}_{\alpha'}[\hatphi,\widetilde{\chi}^{\alpha'}]+\frac{1}{2}\bar{\chi}_{\alpha}[\hatphi,\chi^{\alpha}]+\frac 12 \bar{\chi}^I[\phi_{IJ},\chi^J]+\frac 12 \widetilde{\bar{\chi}}^{I'}[\phi_{I'J'},\widetilde{\chi}^{J'}]\\
    &+\frac{1}{2}[\hatphi,\hatphi][\hatphi,\hatphi]+\frac{1}{2}[\hatphi,\phi^{IJ}][\hatphi,\phi_{IJ}]+\frac{1}{2}[\phi^{IJ},\phi^{MN}][\phi_{IJ},\phi_{MN}]\Big)\,.
    \end{split}
\end{align}
We note that the commutator $[\hatphi,\hatphi]$ is non-trivial since $\hatphi$ takes value in $\hs$. In terms of the fluctuations,
the field strength takes the explicit form
\begin{align}
    \begin{split}
    F^{\alpha\k} &= [\TY^{(\alpha}_{\ \ \gamma'},\TY^{\k)\gamma'}]
   + [\TY^{(\alpha}_{\ \ \gamma'},A^{\k)\gamma'}]
    +  [A^{(\alpha}_{\ \ \gamma'},\TY^{\k)\gamma'}]
     +  [A^{(\alpha}_{\ \ \gamma'},A^{\k)\gamma'}]\\
     &=[\TY^{(\alpha}_{\ \ \gamma'},\TY^{\k)\gamma'}]
   + 2[\TY^{(\alpha}_{\ \ \gamma'},A^{\k)\gamma'}]
    +[A^{(\alpha}_{\ \ \gamma'},A^{\k)\gamma'}]
     \end{split}
\end{align}
where lifting and lowering a pair of spinorial indices comes with a minus sign. Then
\begin{align}
    \begin{split}
    \Tr(F_{\alpha\kappa}F^{\alpha\kappa})&=4\Tr [\TY^{(\alpha}_{\ \gamma'},A^{\kappa)\gamma'}][\TY_{(\alpha\zeta'},A_{\kappa)}^{\ \,\zeta'}]+2[\TY^{(\alpha}_{\ \gamma'},\TY^{\kappa)\gamma'}][A_{(\alpha\zeta'},A_{\kappa)}^{\ \, \zeta'}]\,\\
    &+4\Tr[\TY^{(\alpha}_{\ \gamma'},A^{\kappa)\gamma'}][A_{(\alpha\zeta'},A_{\kappa)}^{\ \,\zeta'}]+\Tr[A^{(\alpha}_{\ \gamma'},A^{\kappa)\gamma'}][A_{(\alpha\zeta'},A_{\kappa)}^{\ \,\zeta'}]\,.
    \end{split}
\end{align}
Due to symmetrization, 
the above includes terms of the form
\begin{align}
4\Tr [\TY^{(\alpha}_{\ \gamma'},A^{\kappa)\gamma'}][\TY_{(\alpha\zeta'},A_{\kappa)}^{\ \,\zeta'}] 
=  2[\TY^{\alpha}_{\ \ \gamma'},A^{\k\gamma'}] [\TY_{\a\d'},A_{\k}^{\ \ \d'}] 
 + 2 \Tr [\TY^{\alpha}_{\ \ \gamma'},A^{\k\gamma'}] [\TY_{\k\d'},A_{\alpha}^{\  \d'}] 
\end{align}
etc. which is in complete analogy to the field strength in non-commutative gauge theory, cf. \cite{Sperling:2017gmy,Blaschke:2011qu}. 
To  shorten the expressions, we shall use the convention $F^{\alpha\alpha}=F^{(\alpha\kappa)}$ to express the symmetrization over unprimed spinorial indices. Then, the Yang-Mills part of the IKKT action reads
\begin{align}
    \begin{split}
    \frac{1}{2}\Tr\Big(F_{\alpha\alpha}F^{\alpha\alpha}\Big)=\bar{S
    }^F_{\BG}&+\Tr\Big(2[\TY_{\alpha\kappa'},A_{\alpha}^{\ \kappa'}][\TY^{\alpha}_{\ \zeta'},A^{\alpha\zeta'}]+[\TY_{\alpha\kappa'},\TY_{\alpha}^{\ \kappa'}][A^{\alpha}_{\ \zeta'},A^{\alpha\zeta'}]\\
    &\qquad +2[\TY_{\alpha\kappa'},A_{\alpha}^{\ \kappa'}][A^{\alpha}_{\ \zeta'},A^{\alpha\zeta'}]+\frac 12[A_{\alpha\kappa'},A_{\alpha}^{\ \kappa'}][A^{\alpha}_{\ \zeta'},A^{\alpha\zeta'}]\Big)\,,
    \end{split}
\end{align}
where $\bar{S}^F_{\BG}$ consists of terms that are 0th order or 1st order in fields. It can be considered as the background action for the Yang-Mills part of the IKKT-matrix model. Notice that in the present spinorial formalism,
there is no explicit "gauge-fixing" term of the form $[\TY_{\alpha\alpha'},A^{\alpha\alpha'}]^2$ which appears in \cite{Sperling:2017dts,Sperling:2017gmy}. The term $[\TY_{\alpha\kappa'},\TY_{\alpha}^{\ \kappa'}][A^{\alpha}_{\ \zeta'},A^{\alpha\zeta'}]$ looks non-standard but is familiar in non-commutative gauge theory and matrix models. We can avoid to deal with this troublesome term as follow.
\paragraph{First-order formulation.} By introducing an auxiliary field $\Bcal_{\alpha\alpha}$, we can absorb the term $[\TY^{\alpha}_{\ \kappa'},\TY^{\alpha\kappa'}][A_{\alpha\tau'},A_{\alpha}^{\ \tau'}]$ into the background and write the Yang-Mills part as
\begin{align}
    S^{\text{YM}}=\Tr\Big(\Bcal_{\alpha\alpha}F^{\alpha\alpha}-\frac{1}{2}\Bcal_{\alpha\alpha}\Bcal^{\alpha\alpha}\Big)\,.
\end{align}
The above action is invariant under 
\begin{align}\label{eq:gaugeBF}
    U^{-1}F^{\alpha\alpha}U\,,\qquad \qquad U^{-1}\Bcal_{\alpha\alpha}U\,,
\end{align}
where $U=e^{i\xi}$ for $\xi$ is some $\hs$-valued gauge parameter. Next, we consider a fluctuation of the $\Bcal$ field as 
\begin{align}
    \Bcal_{\alpha\alpha}=\bar{B}_{\alpha\alpha}+B_{\alpha\alpha}\,,
\end{align}
where $\bar{B}$ can be understood as the background of the $\Bcal$ field. Then, in terms of components, the Yang-Mills action in first order can be written as
\begin{align}\label{eq:actionIKKT}
    \begin{split}
    S^{\text{YM}}=S_{\text{BG}}^{\text{YM}}[A,\bar{B},B,\TY]+\Tr\Big(2B_{\alpha\alpha}[\TY^{\alpha}_{\ \alpha'},A^{\alpha\alpha'}]+B_{\alpha\alpha}[A^{\alpha}_{\ \alpha'},A^{\alpha\alpha'}]-\frac{1}{2}B_{\alpha\alpha}B^{\alpha\alpha}\Big)\,,
    \end{split}
\end{align}
where the background Yang-Mills action $S_{\text{BG}}^{\text{YM}}[A,\bar{B},B,\TY]$ is 0th order or 1st order in fluctuation modes. Observe that there is no gauge fixing term or the troublesome term $[\TY,\TY][A,A]$ since they are absorbed by the background. 
Therefore, the total action $S=S_{\BG}+S_2+S_3+S_4$ consists of: (1) the quadratic action
\small
\begin{align}
    \begin{split}
    S_2=&\Tr\Big(2B_{\alpha\alpha}[\TY^{\alpha}_{\ \gamma'},A^{\alpha\gamma'}]-\frac{1}{2}B_{\alpha\alpha}B^{\alpha\alpha}+\bar{\chi}^{\alpha}[\TY_{\alpha\beta},\widetilde{\chi}^{\beta' }]+\frac{1}{2}[\TY^{\alpha\alpha'},\hat{\phi}][\TY_{\alpha\alpha'},\hatphi]+\frac{1}{2}[\TY^{\alpha\alpha'},\phi^{IJ}][\TY_{\alpha\alpha'},\phi_{IJ}]\\
    &+\frac{1}{2}[\TY_5,A^{\alpha\alpha'}][\TY_5,A_{\alpha\alpha'}]+\frac 12 [\TY_5,\hatphi][\TY_5,\hatphi]+\frac{1}{2}[\TY_5,\phi^{IJ}][\TY_5,\phi_{IJ}]-\frac{1}{2}\widetilde{\bar{\chi}}_{\alpha'}[\TY_5,\widetilde{\chi}^{\alpha'}]+\frac{1}{2}\bar{\chi}_{\alpha}[\TY_5,\chi^{\alpha}]\Big)\,,
    \end{split}
\end{align}
\normalsize
(2) the cubic action
\begin{align}
    \begin{split}
    S_3&=\Tr\Big(B_{\alpha\alpha}[A^{\alpha}_{\ \gamma'},A^{\alpha\gamma'}]+[\TY^{\alpha\alpha'},\hatphi][A_{\alpha\alpha'},\hatphi]+[\TY^{\alpha\alpha'},\phi^{IJ}][A_{\alpha\alpha'},\phi_{IJ}]\\
    &+\overline{\chi}^{\alpha}[A_{\alpha\beta'},\widetilde{\chi}^{\beta' }]-\frac{1}{2}\widetilde{\bar{\chi}}_{\alpha'}[\hatphi,\widetilde{\chi}^{\alpha' }] +\frac{1}{2}\bar{\chi}_{\alpha}[\hatphi,\chi^{\alpha }]+\frac 12 \bar{\chi}^I[\phi_{IJ},\chi^J]+\frac 12 \widetilde{\bar{\chi}}^{I'}[\phi_{I'J'},\widetilde{\chi}^{J'}]\Big)\,,
    \end{split}
\end{align}
and (3) the quartic action
\begin{align}
    \begin{split}
    S_4=&\Tr\Big(\frac{1}{2}[A^{\alpha\alpha'},\hatphi][A_{\alpha\alpha'},\hatphi]+\frac{1}{2}[A^{\alpha\alpha'},\phi^{IJ}][A_{\alpha\alpha'},\phi_{IJ}]\\
    &+\frac{1}{2}[\hatphi,\hatphi][\hatphi,\hatphi]+ \frac{1}{2}[\hatphi,\phi^{IJ}][\hatphi,\phi_{IJ}]+\frac{1}{2}[\phi^{IJ},\phi^{MN}][\phi_{IJ},\phi_{MN}]\Big)\,.
    \end{split}
\end{align}


\paragraph{The self-dual sector.} Similar to the story of $\Ncal=4$ SYM in \cite{Chalmers:1996rq}, we can also obtain the self-dual sector of the IKKT-matrix model by dropping some of the terms in the action \eqref{eq:actionIKKT} with the requirement that the reduced action is still gauge-invariant and has the same degrees of freedom as before. The self-dual sector of the IKKT-model reads
\begin{align}\label{eq:quantizedactionSD}
    \begin{split}
    S_{SD}=&\Tr\Big(B_{\alpha\alpha}F^{\alpha\alpha}+\frac{1}{2}[P^{\alpha\alpha'},\hatphi][P_{\alpha\alpha'},\hatphi]+\frac{1}{2}[P^{\alpha\alpha'},\phi^{IJ}][P_{\alpha\alpha'},\phi_{IJ}]+\bar{\chi}^{\alpha}[P_{\alpha\beta'},\widetilde{\chi}^{\beta'}]\\
    &+\frac{1}{2}[\TY_5,A^{\alpha\alpha'}][\TY_5,A_{\alpha\alpha'}]+\frac 12 [\TY_5,\hat{\phi}][\TY_5,\hatphi]+\frac{1}{2}[\TY_5,\phi^{IJ}][\TY_5,\phi_{IJ}]\\
    &-\frac{1}{2}\widetilde{\bar{\chi}}_{\alpha'}[\TY_5,\widetilde{\chi}^{\alpha'}]-\frac{1}{2}\widetilde{\bar{\chi}}_{\alpha'}[\hatphi,\widetilde{\chi}^{\alpha'}]+\frac{1}{2}\widetilde{\bar{\chi}}^{I'}[\phi_{I'J'},\widetilde{\chi}^{J'}]\Big)\,.
    \end{split}
\end{align}
We note that unlike the case of $4d$ self-dual $\Ncal=4$ SYM, the action for self-dual IKKT on the $S^4$ has extra contributions from $[\TY_5,\cdot]$. These contributions, however, vanish in the semi-classical and flat (SCF) limit.\footnote{See also the work of \cite{Heckman:2011qt} from a different perspective.}

\paragraph{The SCF limit.} From \eqref{eq:Yparametrization}, we see that at large $R$, $\ty_5$ scales as $R$. Hence, $\{R,\cdot\}=0$ as expected. To make sure the contribution from $\{x^2,\cdot\}$ does not contribute in the SCF limit, let us compute it explicitly,
\begin{align}\label{eq:checkingy5bracket1}
    \begin{split}
    \{\frac{Rx^2}{R^2+x^2},\bullet\}&=R\tx_{\alpha\alpha'}\{\tx^{\alpha\alpha'},\ty^{\kappa\kappa'}\}\frac{\pl}{\pl \ty^{\kappa\kappa'}
    }\bullet=-\frac{R}{\langle \hat \l\l\rangle}\tx_{\alpha\alpha'}\Ecal^{\alpha\alpha'\kappa\kappa'}\frac{\pl}{\pl \ty^{\kappa\kappa'}}\bullet\sim \Ocal (r)
    \end{split}
\end{align}
where we have use \eqref{eq:spinorproduct}. Since it scales as $\Ocal(r)$, all of the Poisson brackets involving $\ty_5$ in the SCF limit can be neglected, i.e. $\{\ty_5,\cdot\}\sim 0$. Hence, the \eqref{eq:quantizedactionSD} action reduces to
\begin{align}\label{eq:SDIKKT4dreduction}
    \begin{split}
    S_{SD}\simeq &\int\,\Big(B_{\alpha\alpha}F^{\alpha\alpha}+\frac{i}{2}\{P^{\alpha\alpha'},\hatphi\}\{P_{\alpha\alpha'},\hatphi\}+\frac{i}{2}\{P^{\alpha\alpha'},\phi^{IJ}\}\{P_{\alpha\alpha'},\phi_{IJ}\}\\
    &\qquad +\bar{\chi}^{\alpha}\{P_{\alpha\beta'},\widetilde{\chi}^{\beta'}\}-\frac{1}{2}\widetilde{\bar{\chi}}_{\alpha'}\{\hatphi,\widetilde{\chi}^{\alpha'}\}+\frac{1}{2}\widetilde{\bar{\chi}}^{I'}\{\phi_{I'J'},\widetilde{\chi}^{J'}\}\Big)\,,
    \end{split}
\end{align}
 which is reminiscent of self-dual $\Ncal=4$ SYM in $4d$ \cite{Chalmers:1996rq,Witten:2003nn,Heckman:2011qt}. However, the interactions are gravitational due to the Poisson brackets.


\subsection{The HS-IKKT on twistor space in the semi-classical limit} 
As mentioned, at large $N$ where functions are effectively commutative, one can replace the commutators by the Poisson brackets, i.e. $ [,]\mapsto i\{,\}\,$. We also replace 
\begin{align}\label{eq:yandx}
    \TY^{\alpha\alpha'}\mapsto \ty^{\alpha\alpha'}=-\langle \hat\l\l\rangle\tx^{\alpha\alpha'}\,.
\end{align}
by using eq. \eqref{eq:yinspinors}. Since there is an emergence of geometry in this limit, see discussion in sections \ref{sec:2} and \ref{sec:3}, we can consider fields with smooth enough distribution. Hence, we can insert an integral as to average out these distribution. Therefore, the action for the IKKT model on the infinite dimensional bundle $\Kcurl=\Ccurl(\CP^1)\times \RR^4$ becomes
\begin{align}\label{eq:IKKTN=4poissonsemi}
    \begin{split}
    S=&\int_{\Kcurl}\,\Big[ 2B_{\alpha\alpha}\{\ty^{\alpha}_{\ \alpha'},A^{\alpha\alpha'}\}+B_{\alpha\alpha}\{A^{\alpha}_{\ \alpha'},A^{\alpha\alpha'}\}+\frac{i}{2}B_{\alpha\alpha}B^{\alpha\alpha}\\
    &+\frac{i}{2}\{P^{\alpha\alpha'},\hatphi\}\{P_{\alpha\alpha'},\hatphi\}+\frac{i}{2}\{P^{\alpha\alpha'},\phi^{IJ}\}\{P_{\alpha\alpha'},\phi_{IJ}\}+\frac{i}{2}\{\ty_5,A^{\alpha\alpha'}\}\{\ty_5,A_{\alpha\alpha'}\}\\
    &+\frac i2 \{\ty_5,\hat{\phi}\}\{\ty_5,\hatphi\}+\frac{i}{2}\{\ty_5,\phi^{IJ}\}\{\ty_5,\phi_{IJ}\}+\bar{\chi}^{\alpha}\{\ty_{\alpha\beta'},\widetilde{\chi}^{\beta'}\}+\bar{\chi}^{\alpha}\{A_{\alpha\beta'},\widetilde{\chi}^{\beta'}\}\\
    &-\frac{1}{2}\widetilde{\bar{\chi}}_{\alpha'}\{\ty_5,\widetilde{\chi}^{\alpha'}\}+\frac{1}{2}\bar{\chi}_{\alpha}\{\ty_5,\chi^{\alpha}\}-\frac{1}{2}\widetilde{\bar{\chi}}_{\alpha'}\{\hatphi,\widetilde{\chi}^{\alpha'}\}+\frac{1}{2}\bar{\chi}_{\alpha}\{\hatphi,\chi^{\alpha}\}+\frac 12 \bar{\chi}^I\{\phi_{IJ},\chi^J\}\\
    &+\frac 12 \widetilde{\bar{\chi}}^{I'}\{\phi_{I'J'},\widetilde{\chi}^{J'}\}+\frac{i}{2}\{\hatphi,\hatphi\}\{\hatphi,\hatphi\}+\frac{i}{2}\{\hatphi,\phi^{IJ}\}\{\hatphi,\phi_{IJ}\}+\frac{i}{2}\{\phi^{IJ},\phi^{MN}\}\{\phi_{IJ},\phi_{MN}\}\Big]\,.
    \end{split}
\end{align}
As discussed above, since fields are $\hs$-valued, all of the Poisson brackets are non-trivial. Next, we want to have a measure that is $SU(4)$-invariant and have complex form degree 3 since the total space is a $\CP^3_N$. A nature candidate for the measure is \cite{Witten:2003nn}
\begin{align}\label{eq:DZmeasure}
    D^3Z=\epsilon_{\Acal\Bcal\Ccal\Dcal}Z^{\Acal}dZ^{\Bcal}dZ^{\Ccal}dZ^{\Dcal}=\frac{R^4\langle \lambda d\lambda\rangle \wedge [d \mu \wedge d \mu]}{(R^2+x^2)^2}=\frac{R^4\lambda_{\alpha}\lambda_{\beta}\langle \lambda d \lambda \rangle \wedge d\tx^{\alpha\alpha'} \wedge d\tx^{\beta}_{\ \alpha'}}{(R^2+x^2)^2}\,.
\end{align}
The above define a holomorphic measure on our fuzzy twistor space. We also have an anti-holomorphic measure which is $D^3\bar{Z}$ that is also invariant under $SU(4)$ and has form degree $(0,3)$. In terms of spinors, we just need to replace $\lambda$ by $\hat{\lambda}$ and $\mu$ for $\hat{\mu}$ to describe $D^3\bar{Z}$. Notice that we have a conformal factor of $\frac{R^4}{(R^2+x^2)^{2}}$ from the parametrization \eqref{eq:spinorparametrize} and \eqref{eq:muparametrization}, which addresses the fact that our target space is a 4-sphere. It is easy to see that we can have a smooth flat limit when $R\rightarrow \infty$. The total measure for the above integral reads
\begin{align}\label{eq:Upsilonmeasure}
    \Delta:=D^3Z\wedge D^3\bar{Z}\,.
\end{align}
Note that the measure $\Delta$ is unique and is a $(3,3)$-form. The above action \eqref{eq:IKKTN=4poissonsemi} reveals an interesting feature of the fuzzy twistor construction. Specifically, the spacetime action of the IKKT model
is already recognized in \eqref{eq:IKKTN=4poissonsemi} without the need of referring to the twistor cohomology. Therefore, we can have the same measure for both the self-dual sector and the non-self-dual one. This is different from the usual twistor construction of non-self-dual theories, see e.g. \cite{Mason:2005zm}. 

By varying the action \eqref{eq:IKKTN=4poissonsemi} with respect to $B$, we obtain the free equations of motion for $A^{\alpha\alpha'}$ as
\begin{align}
    \{\ty^{\alpha}_{\ \alpha'},A^{\alpha\alpha'}\}=0\,.
\end{align}
The free equation of motion for the $B$ field reads
\begin{align}
    2\{\ty^{\alpha}_{\ \alpha'},B_{\alpha\alpha}\}+i\Box_5A_{\alpha\alpha'}=0\,,
\end{align}
where $\Box_5\,\bullet=\{\ty_5,\{\ty_5,\bullet\}\}$. Let us also list the free equations of motion for the scalar fields and the fermions. The free EOMs for the scalar fields reads
\begin{subequations}
\begin{align}
    (\Box+\Box_5)\hatphi&=0\,,\\
    (\Box+\Box_5)\phi^{IJ}&=0\,,
\end{align}
\end{subequations}
where $\Box\,\bullet=\{\ty^{\alpha\alpha'},\{\ty_{\alpha\alpha'},\bullet \}\}$. The free EOMs for the fermions are
\begin{subequations}
\begin{align}
    \{\ty_{\alpha\alpha'},\widetilde{\chi}^{\alpha'}\}+\frac{1}{2}\{\ty_5,\chi^{\alpha}\}=0\,,\\
    \{\ty_{\alpha\alpha'},\bar{\chi}^{\alpha}\}-\frac{1}{2}\{\ty_5,\widetilde{\bar{\chi}}_{\alpha'}\}=0\,.
\end{align}
\end{subequations}
It may seem strange that a (gravitational) Yang-Mills theory is described in terms of first-order equations of motion. However, there is no contradiction since the  missing ``momentum'' degrees of freedom of the gauge field $A$ are encoded in the $B$ modes.
It is also interesting to note that even though our target space is a 4-sphere, the free equations of motion for higher-spin fields are also coupled to the transversal modes on the $5th$ direction.

\paragraph{The semi-classical and flat (SCF) limit of IKKT-matrix model.} The contributions from the $5th$ direction will vanish in the flat limit as discussed above. In the SCF limit, the free equations of motion read
\begin{subequations}
\begin{align}
    \{\ty^{\alpha}_{\ \alpha'},A^{\alpha\alpha'}\}&=0\,, &&
    &\{\ty^{\alpha}_{\ \alpha'},B_{\alpha\alpha}\}&=0
    \label{eq:freeAnadB}\,,\\
    \Box \,\hatphi&=0\,, && &\Box\, \phi^{IJ}&=0\,,\\
    \{\ty_{\alpha\alpha'},\widetilde{\chi}^{\alpha'}\}&=0\,, && &\{\ty_{\alpha\alpha'},\bar{\chi}^{\alpha}\}&=0\,.
\end{align}
\end{subequations}
 The equation of $A^{\alpha\alpha'}$ is invariant under 
\begin{align}
    \delta A^{\alpha\alpha'}=\{\ty^{\alpha\alpha'},\xi\}\,,
\end{align}
where $\xi$ is some $\hs$-valued section on $\Kcurl$. Let us prove the above statement by computing explicitly
\begin{align}\label{eq:gaugeinvariantxi}
    \{\ty^{\alpha}_{\ \alpha'},\{\ty^{\alpha\alpha'},\xi(\tx)\}\}\sim \frac{1}{2}\{\{\ty^{\alpha}_{\ \alpha'},\ty^{\alpha\alpha'}\},\tx^{\kappa\kappa'}\}\frac{\pl}{\pl \tx^{\kappa\kappa'}}\xi
    =T^{\alpha\ \, \alpha\alpha'\kappa\kappa'}_{\ \alpha'}\frac{\pl}{\pl \tx^{\kappa\kappa'}}\xi=\frac{4}{\langle \hat \l \l\rangle}\epsilon^{\kappa \alpha}\ty^{\alpha\kappa'}\frac{\pl}{\pl \tx^{\kappa\kappa'}}\xi\,.
\end{align}
Here, the torsion $T$ is defined in \eqref{eq:torsion} and 
we have used Jacobi identity. The torsion vanishes in the flat limit since it scales at $1/R$. We note that the contribution from the spin operator $\Sigma$ (cf. subsection \ref{sec:3.4}) can be neglected in the same limit. Moreover, the external $SO(5)$ group is degenerated in this limit and no longer acts on $\hatphi$. Together, $\hatphi$ and $\phi^{IJ}$ will become the usual six adjoint scalars $\phi^{\Ical\Jcal}=-\phi^{\Jcal\Ical}$ of the internal symmetry group $SU(4)$. Hence, the action for the IKKT-matrix model in this limit becomes
\begin{align}\label{eq:actionIKKTinSCFlimit}
    \begin{split}
    S_{\SCF}=\int_{\Kcurl}\,&\Big[ B_{\alpha\alpha}F^{\alpha\alpha}+\frac{i}{2}B_{\alpha\alpha}B^{\alpha\alpha}+i\{P^{\alpha\alpha'},\phi^{\Ical\Jcal}\}\{P_{\alpha\alpha'},\phi_{\Ical\Jcal}\}+2\bar{\chi}^{\alpha}\{P_{\alpha\beta'},\widetilde{\chi}^{\beta'}\}\\
    &+ \bar{\chi}^{\Ical}\{\phi_{\Ical\Jcal},\chi^{\Jcal}\}+ \widetilde{\bar{\chi}}^{\Ical}\{\phi_{\Ical\Jcal},\widetilde{\chi}^{\Jcal}\}+\frac{i}{2}\{\phi^{\Ical\Jcal},\phi^{\Mcal\Ncal}\}\{\phi_{\Ical\Jcal},\phi_{\Mcal\Ncal}\}\Big]\,,
    \end{split}
\end{align}
where we have rescale $\phi\rightarrow 2\phi$ and $\chi\rightarrow \sqrt{2}$ for convenience. 

\subsection{$\hs$-valued eigenmodes of the first-order equations} \label{sec:analysisontwistorspace}


In this section, we will provide the explicit solutions of the first-order equations of motion \eqref{eq:freeAnadB} as
 functions 
 on twistor space, or equivalently as $\hs$-valued  functions  (or 1--forms) on $\R^{3,1}$. This is analogous to the tower of $\hs$ solutions found in \cite{Sperling:2018xrm} translated to the spinor formalism, which simplifies in the flat limit as some extra terms in the equations of motion disappear as discussed above. Recall that the gauge fields $A$ can be expanded in  terms of
\begin{align}
A^{\alpha,\alpha'}(\tx)&= \sum_s A^{\kappa(s)\tau(s)|\alpha,\alpha'}(\tx)\lambda_{\kappa}...\lambda_{\kappa}\hat{\lambda}_{\tau}...\hat{\lambda}_{\tau}\,.
\label{planewave-gaugemodes-C}
\end{align}
We can again assume that $A^{\kappa(s)\tau(s)|\alpha,\alpha'}(\tx)$ is totally symmetric in $\kappa(s)\tau(s)$, but the $(\a,\a')$ are independent indices. Therefore, there are 4 independent (off-shell) $A$ modes, corresponding precisely to the 4 tangential modes  identified in \cite{Sperling:2018xrm}. The present spinorial formalism allows a more transparent organization of these modes in terms of the following two modes (in the maximally unbalanced representation):
\begin{subequations}
\begin{align}
A_{(1)}^{\alpha\alpha'} \ &= \  
A^{\kappa(2s)\alpha,\alpha'}\lambda_{\kappa}^s\hat{\lambda}^s_{\kappa}\,, \\
A_{(2)}^{\alpha\alpha'} \ &= \ \epsilon^{\a\k} \TA^{\kappa(2s-1),\alpha'}\lambda_{\kappa}^s\hat{\lambda}^s_{\kappa}=\lambda^{\alpha}\TA^{\kappa(2s-1),\alpha'}\lambda_{\kappa}^{s-1}\hat{\l}_{\kappa}^s+\hat{\l}^{\alpha}\TA^{\kappa(2s-1),\alpha'}\lambda_{\kappa}^s\hat{\l}^{s-1}_{\kappa} \,.
\end{align}
\label{A-ansatz-1-2}
\end{subequations}
Here $A^{\kappa(2s)\alpha,\alpha'}$ is totally symmetric in $\kappa(2s)\alpha$.
We shall sometime use the notation $\simeq$ to denote the SCF limit, and 
\begin{align}
    \lambda^s_{\alpha}=\LaTeXunderbrace{\lambda_{\alpha}...\lambda_{\alpha}}_{s\ \text{times}}\,,    
\end{align}
to shorten our expressions. To check completeness, we note that $A_{(1)}^{\alpha\alpha'}$ provides $2(2s+2)$ components, and 
$A_{(2)}^{\alpha\alpha'}$ provides other $2(2s)$ ones. Taken  together, this provides $4(2s+1)$ components, which is the correct number of components in \eqref{planewave-gaugemodes-C} and is consistent with \cite{Sperling:2017gmy}. Similarly, we also have 2 eigenmodes of the $B_{\alpha\alpha}$, they are
\begin{subequations}
\begin{align}
    B_{(1)}^{\alpha\bullet}&=B^{\k(2s)\alpha\bullet}\lambda_{\kappa}^s\hat\l^s_{\kappa}\,,\label{B-mode-1}\\
    B_{(2)}^{\a \bullet}&=\epsilon^{\bullet \k}\TB^{\kappa(2s-1)\alpha}\lambda_{\kappa}^s\hat\l^s_{\kappa}\,. \label{B-mode-2}
\end{align}
\end{subequations}
Here, $\bullet$ is the index of $B$ that contract with the index of the coordinate $\ty^{\bullet}_{\ \alpha'}$. Now, let us discuss about the symmetry of the free action. Besides the usual gauge transformation $\{\ty^{\alpha}_{\ \alpha'},\xi\}$ where $\xi$ is some section on $\Kcurl$, the tensorial fields $A^{\kappa(2s)|\alpha,\alpha'}$ also have an algebraic gauge symmetry. The gauge transformation for $\delta A^{\kappa(2s)|\alpha,\alpha'}$ reads
\begin{align}
    \delta_{\xi,\vartheta}A^{\kappa(2s)|\alpha,\alpha'}=\{\ty^{\alpha\alpha'},\xi^{\kappa(2s)}\}+\epsilon^{\kappa\alpha}\vartheta^{\kappa(2s-1),\alpha'}\,.
\end{align}
The algebraic symmetry $\vartheta$ ensures that the (unwanted) second eigenmode $A_{(2)}$ does not propagate. Indeed, as a simple exercise, one can check that 
\begin{align}
    \begin{split}
    \delta_{\vartheta}S_2&=\int \Delta\, (\lambda^{\zeta})^n(\hat{\lambda}^{\zeta})^n \epsilon^{\alpha\beta}\,B_{\zeta(2n)\alpha|\bullet}\{\ty^{\bullet}_{\ \alpha'},\vartheta^{\beta(2m-1),\alpha'}\}\,(\lambda_\beta)^m(\hat{\l}_{\beta})^m\\
    &\sim \int \Delta\, B^{\nu}_{\ \nu\alpha(2n)}\{\ty^{\alpha}_{\ \alpha'},\vartheta^{\alpha(2n-1),\alpha'}\}=0\,,
    \end{split}
\end{align}
since $B$ is traceless when we consider it as element in the MUR. Here, we have used \eqref{eq:magic} to derive the above expression. The second mode of $B$ plays the role of a Lagrangian multiplier and provides us the usual generalized Lorenz gauge condition of the form
\begin{align}
    \int \Delta \,\TB_{\alpha(2n-1)}\{\ty_{\alpha\alpha'},A^{\alpha(2n),\alpha'}\}\,.
\end{align}
Therefore, only the first eigenmodes of $A^{\alpha\alpha'}$ and $B^{\alpha\alpha}$ propagate.

\paragraph{Equations of motion.} Now, we consider the equations of motion. 
Since non-trivial solutions exist only in Minkowski
signature, the following considerations are somewhat formal, assuming a suitable analytic continuation. For the $A_{(1)}^{\alpha\alpha'}$ modes, we have
\begin{align}
    \begin{split}
    \{\ty^{\a}_{\ \a'},A_{(1)}^{\alpha\alpha'}\} 
    &=  \{\ty^{\a}_{\ \a'}, A^{\kappa(2s)\alpha,\alpha'}(\tx)\lambda_{\kappa}...\lambda_{\kappa}\hat{\lambda}_{\k}...\hat{\lambda}_{\k} \} \\ 
    &=  \{\ty^{\a}_{\ \a'}, A^{\kappa(2s)\a,\alpha'}(\tx) \} \lambda_{\kappa}...\lambda_{\kappa}\hat{\lambda}_{\k}...\hat{\lambda}_{\k}  \\
    &=-\frac{1}{\langle \hat\l\l\rangle} \Ecal_{\ \a'}^{\a \  |\beta\beta'}\frac{\pl}{\pl \tx^{\beta\beta'}}A^{\kappa(2s)\alpha,\alpha'}(\tx)
    \lambda_{\kappa}...\lambda_{\kappa}\hat{\lambda}_{\k}...\hat{\lambda}_{\k} \,,
    \end{split}
    \label{eom-A1-mode}
\end{align}
where we have used \eqref{eq:yandx} and 
\begin{align}
    \begin{split}
  A^{\kappa(2s)\alpha,\alpha'} \{\ty^{\a}_{\ \a'}, \lambda_{\kappa}...\lambda_{\kappa}\hat{\lambda}_{\k}...\hat{\lambda}_{\k} \} &=isA^{\kappa(2s-1)\alpha(2),\alpha'}\big(\mu_{\alpha'}\lambda_{\kappa}^{s-1}\hat{\l}_{\kappa}^s-\hat{\mu}_{\alpha'}\lambda_{\k}^s\hat{\l}^{s-1}_{\kappa}\big)\\
  &=is\,\tx_{\kappa\alpha'} A^{\kappa(2s-1)\alpha(2),\alpha'}\lambda_{\k}^{s-1}\hat{\l}^{s-1}_{\k}\simeq 0\,.
  \end{split}
\end{align}
This leads to the first-order  equation of motion 
\begin{align}
    \Ecal_{\ \a'}^{\a \  |\beta\beta'}\frac{\pl A^{\kappa(2s)\a,\alpha'}}{\pl \tx^{\beta\beta'}} 
    \simeq
    \lambda^{(\a}\hat{\lambda}^{\beta)}\frac{\pl A^{\kappa(2s)\a,\alpha'}}{\pl \tx^{\beta\alpha'}}
    =0
\end{align}
in the flat limit, cf. \eqref{eq:1steffectivevielbein}. Note that the fields $A^{\kappa(2s)\alpha,\alpha'}$ does not depend on the fibre coordinates. 
The above equation admits the following plane-wave solution:
\begin{align}\label{eq:positivehelicityansatz}
    \begin{split}
    A^{\alpha,\alpha'}_{(1)} &=A^{\kappa(2s)\alpha,\alpha'} 
    \lambda_{\kappa}^s\hat{\lambda}^s_{\kappa}\,,\\
    A^{\alpha(2s+1),\alpha'}&=\frac{\zeta^{\alpha}...\zeta^{\alpha}\tilde{\upsilon}^{\alpha'}}{\langle \zeta \upsilon\rangle^{2s+1}}\,e^{i\upsilon^{\alpha}\tx_{\alpha\alpha'}\tilde{\upsilon}^{\alpha'}}\,
    \end{split}
\end{align}
in terms of $2s+1$ auxiliary spinors $\zeta$, which generates the monomials of degree $2s+1$.
It is easy to check that this ansatz indeed satisfies \eqref{eom-A1-mode}:
\begin{align}
       \begin{split}
    \{\ty^{\a}_{\ \a'},A_{(1)}^{\alpha\alpha'}\} 
  &\simeq 
  \lambda^{(\a}\hat{\lambda}^{\beta)}
  \frac{\pl}{\pl \tx^{\beta\a'}}A^{\kappa(2s)\alpha,\alpha'}
    \lambda_{\kappa}...\lambda_{\kappa}\hat{\lambda}_{\k}...\hat{\lambda}_{\k} \\
    &= i \lambda^{(\a}\hat{\lambda}^{\beta)}
 \zeta^{\k}...\zeta^{\k}
    \lambda_{\kappa}...\lambda_{\kappa}\hat{\lambda}_{\k}...\hat{\lambda}_{\k}
    \tilde{\upsilon}^{\alpha'}
    \tilde{\upsilon}_{\alpha'}\upsilon_{\b}
 e^{i\upsilon^{\alpha}\tx_{\alpha\alpha'}\tilde{\upsilon}^{\alpha'}} = 0\,
    \end{split}
    \label{}
\end{align}
in the SCF limit. For later use, we note that $A_{(1)}$ also satisfies the gauge-fixing condition 
\begin{align}
     \frac{\pl}{\pl \tx^{\a\a'}}A_{(1)}^{\a,\alpha'} = 0\,.
     \label{A1-solutions-gaugefix}
\end{align}
This solution encodes $2(2s+2)$ components, and therefore 
provides all components of the first eigenmode $A_{(1)}^{\alpha\alpha'}$. We will show that after the Penrose transform, the above ansatz \eqref{eq:positivehelicityansatz} is indeed the plane-wave solution for positive helicity (gauge) fields in spacetime, see subsection \ref{sec:Penrose}. 
Similarly, we find the following solution for the $B_{(1)}^{\alpha\alpha}$ mode as
\begin{align}\label{eq:negativehelicityansatz}
    B^{\alpha(2s)}=\upsilon^{\alpha}...\upsilon^{\alpha}e^{i\upsilon^{\kappa}\tx_{\kappa\kappa'}\tilde{\upsilon}^{\kappa'}}\,.
\end{align}
Therefore we have obtained the most general solutions of the first-order equations of motion for the $A$ and $B$ fields.

Finally, to count the propagating degrees of freedom for the $A$ and $B$ fields, we can follow the instruction in \cite{Kaparulin:2012px}. There are in total $2s+1$ equations in $\{\ty^{\alpha}_{\ \alpha'},A^{\alpha(2s-1),\alpha'}\}$ and there are $2s-1$ number of components in the gauge symmetry generator $\xi$. Hence, we have 
\begin{align}
    \frac{2s+1-(2s-1)}{2}=1
\end{align}
propagating degree of freedom for the $A$ field. For the $B$ field, there are in total $4s$ equations in $\{\ty^{\gamma}_{\ \alpha'},B_{\gamma\alpha(2s-1)}\}$ and there are $2s-1$ fuzzy Bianchi identities\footnote{We call them Bianchi identities in the sense of \eqref{eq:gaugeinvariantxi}.}
\begin{align}
    \{\ty^{\gamma\alpha'},\{\ty^{\gamma}_{ \ \alpha'},B_{\gamma\gamma\alpha(2s-2)}\}\}\simeq 0\,.
\end{align}
These are 2nd order identities. Therefore, we have 
\begin{align}
    \frac{4s-2(2s-1)}{2}=1
\end{align}
propagating degree of freedom for the $B$ field. Together, $A$ and $B$ (which correspond to positive/negative helicity fields) describe massless higher-spin fields that have two dof as they must. The factor of $1/2$ accounts for the fact that we are counting the degrees of freedom on phase space, where the fields $(A,B)$ play the roles of coordinates. It may seem strange that the $B$-field carries one degree of freedom, since it is an auxiliary field. This can be explained as follows. In the original Yang-Mills term $F_{\alpha\alpha}F^{\alpha\alpha}$, the $A$ field should carry 2 dof. However, after the introducing of the $B$ field,  extra relations and gauge symmetries arise. As a consequence the $B$ field takes away 1 dof from the $A$ field, to describe the negative helicity mode. 


To compare these modes with the results in \cite{Sperling:2018xrm}, we recall that the present spinorial approach is geared towards the on-shell modes. A systematic account of off-shell modes in a Euclidean second-order formulation was given in \cite{Sperling:2017gmy,Sperling:2018xrm} without using the spinor formalism, which leads to  4 towers of off-shell $\hs$ modes. This is consistent with the above 
counting of $4(2s+1)$ modes in \eqref{A-ansatz-1-2}. The discussion of on-shell modes strictly speaking makes sense only in Minkowski signature, which is done here formally by assuming a suitable analytic extension of the spinors. A different way to carry over the higher-spin structures to Minkowski signature within the IKKT model was discussed in \cite{Sperling:2019xar,Steinacker:2019awe}, which is not equivalent to the present approach.
A spinorial re-formulation of that approach would 
certainly be very useful, and is postponed to future work.

\subsection{Maximally unbalanced representation in spacetime} \label{sec:unbalanced} 

Before performing the Penrose transform to obtain the effective spacetime action for the IKKT-matrix model in the semi-classical limit, we would like to discuss briefly about the maximally unbalanced representation (MUR). In $4d$ spacetime, a field that lives in an irreducible finite dimensional representation can be characterized by two numbers $(m,n)$. These two numbers represent for the number of unprimed and primed indices in a tensorial field $T^{\alpha(m),\alpha'(n)}$, respectively. Here, the tensorial fields $T^{\alpha(m),\alpha'(n)}$ are totally symmetric in each group of indices.
Then, instead of dealing with general objects in $S(m,n)$, we would like to locate ourselves at one ``corner" of $S(m,n)$, say $S(m,0)$ and $S(m,1)$. Together, $S(m,0)$ and $S(m,1)$ define MUR \cite{Krasnov:2021nsq}.

The fundamental objects in $S(m,0)$ are the higher-spin generalization of Maxwell- and Weyl- tensors, denoted as $B^{\alpha(m)}$. On the other hand, the fundamental objects in $S(m,1)$ are the higher-spin generalizations of the gauge potential $A^{\alpha,\alpha'}$ which we will denote as $A^{\alpha(m),\alpha'}$. The free equations of motion for $B$ and $A$ are known long ago from twistor theory \cite{Penrose:1965am,Hughston:1979tq,Eastwood:1981jy,Woodhouse:1985id}. We will simply quote them here
\begin{subequations}
\begin{align}
    \nabla_{\beta\alpha'}B^{\beta\alpha(m-1)}&=0\,,\\
    \nabla^{\alpha}_{\ \alpha'}A^{\alpha(m-1),\alpha'}&=0\,,\qquad \qquad  \delta A^{\alpha(m-1),\alpha'}=\nabla^{\alpha\alpha'}\xi^{\alpha(m-2)}\,.
\end{align}
\end{subequations}
The gauge invariance of the equations above requires half of the Weyl tensor to vanish. In this case, it is the component $B_{\alpha(4)}$ of the Weyl tensor. There are two remarkable features of the MUR:
\begin{itemize}
    \item It allows us to treat spins and derivatives in the interactions almost independently. This is not possible in the approaches using Fronsdal fields as the main objects to construct toy models for higher-spin theories. There, the number of derivatives rise with spins and there is no standard 2-derivatives interaction that is important to describe gravitational interactions of higher-spin fields \cite{Aragone:1979hx,Berends:1984rq,Fradkin:1987ks,Boulanger:2006gr,Boulanger:2008tg,Zinoviev:2008ck,Manvelyan:2010je,Conde:2016izb}.
    \item It is closely related to twistor cohomology. Hence, one can employ techniques used in twistor theory to construct spacetime action for higher-spin theories, see e.g. \cite{Tran:2021ukl,Haehnel:2016mlb,Adamo:2016ple}. However, the twistor construction using MUR is mainly designed for chiral theories, see e.g. \cite{Mason:2005zm,Mason:2007ct,Haehnel:2016mlb,Adamo:2016ple}.
\end{itemize}

\subsection{Spacetime actions for the (self-dual) HS-IKKT model}\label{sec:Penrose}

\paragraph{HS-IKKT on twistor space.} As mentioned, in the large-$N$ regime, there is an emergence of higher-spin modes from the IKKT-matrix model. From the analysis in section \ref{sec:3} and \ref{sec:analysisontwistorspace}, the higher-spin extensions of the fields in \eqref{eq:IKKTN=4poissonsemi} reads
\begin{subequations}\label{eq:fieldcontents}
\begin{align}
   B_{\alpha\alpha}&\mapsto \sum_sB_{\alpha\alpha\kappa(s)\tau(s)}\lambda^{\kappa}...\lambda^{\kappa}\hat{\lambda}^{\tau}...\hat{\lambda}^{\tau}\,, && &A^{\alpha,\alpha'}&\mapsto \sum_s A^{\kappa(s)\tau(s)\alpha,\alpha'}\lambda_{\kappa}...\lambda_{\kappa}\hat{\lambda}_{\tau}...\hat{\lambda}_{\tau}\,,\\
    \bar{\chi}^{\alpha}&\mapsto \sum_m\bar{\chi}^{\alpha\kappa(m)\tau(m)}\lambda_{\kappa}...\lambda_{\kappa}\hat{\lambda}_{\tau}...\hat{\lambda}_{\tau}\,, &&
    &\tilde{\chi}^{\alpha'}&\mapsto \sum_m\tilde{\chi}^{\kappa(m)\tau(m)\alpha'}\lambda_{\kappa}...\lambda_{\kappa}\hat{\lambda}_{\tau}...\hat{\lambda}_{\tau}\,,\\
    \hatphi&\mapsto \sum_n\hatphi^{\kappa(n)\tau(n)}\lambda_{\kappa}...\lambda_{\kappa}\hat{\lambda}_{\tau}...\hat{\lambda}_{\tau}\,, && &\phi_{IJ}&\mapsto \sum_n\phi_{IJ}^{\kappa(n)\tau(n)}\lambda_{\kappa}...\lambda_{\kappa}\hat{\lambda}_{\tau}...\hat{\lambda}_{\tau}\,,
\end{align}
\end{subequations}
for $s,m,n\geq 0$. Note that at $s=0$ we have $B_{\alpha\alpha}$. Since fields on fuzzy twistor space live in the BWR, their higher-spin extensions result in a remarkable simple action of the HS-IKKT matrix model where there is an equal in the numbers of $\lambda$ and $\hat{\lambda}$. Note that due to the balance of weights between $\lambda$ and $\hat{\lambda}$, the higher-rank tensors will increase with \textit{integers} in spins. Therefore, the higher-spin version of the IKKT model schematically reads
\begin{align}\label{eq:schematicalform}
    S=\sum_{s}\int_{\Kcurl} \Delta\,\frac{\lambda_{\alpha}...\lambda_{\alpha}\hat{\lambda}^{\beta}...\hat{\lambda}^{\beta}}{\langle \hat{\lambda}\lambda\rangle^s}(\textit{higher-rank tensors})_{\beta(s)}^{\quad \ \alpha(s)}\,.
\end{align}
The integral over the fibres in \eqref{eq:schematicalform} can be done with the help of the Penrose transform. Here, we insert the conformal factors $\langle \hat \l\l\rangle$ so that the integrals over $\CP^1$ are weightless in fibre coordinates. It can be understood as some suitable field re-definitions. 

One of the intriguing features of the HS-IKKT is that the Poisson brackets provide more structures that mix left- and right-handed spinors (cf. subsection \ref{sec:3.4}) through the spin operator $\Sigma$. For this reason, the spacetime action resulting from the Penrose transform of the HS-IKKT model is not standard compared to usual field theories even in the SCF limit where the contribution from $\Sigma$ is approximately negligible. However,  the "asymptotic states" for the HS-IKKT model in SCF limit are identical with the usual ones of free higher-spin theory in \cite{Krasnov:2021nsq}. Another intriguing feature of the HS-IKKT model is that the cubic and quartic vertices in the IKKT model are naturally higher-spin extensible since we just need to contract all unprimed indices in all possible way. We will elaborate this fact below.
\paragraph{The Penrose transform and spacetime actions.} Using the incident relations \eqref{eq:incident}, we can rewrite the measure $D^3Z \wedge D^3\bar{Z}$ as
\begin{align}\label{eq:measureconformal}
    D^3Z\wedge D^3\bar{Z}=\frac{R^8d^4\tx}{(R^2+x^2)^4}\frac{\langle \lambda d\lambda\rangle \wedge \langle \hat{\lambda} d\hat{\lambda}\rangle }{\langle \hat{\lambda}\lambda\rangle^2}=\frac{16d^4\tx}{(1+\tx^2)^4}\frac{\langle \l d\l\rangle\wedge \langle \hat \l d \hat \l \rangle}{\langle \hat \l \l \rangle^2}\,,
\end{align}
which is just another way of writing the symplectic volume form on $\CP^3$. Note that we have used \eqref{eq:xtranslation} and rescaled $\tx \rightarrow 2\tx$ to obtain the second equality. In the following, we will denote 
\begin{align}
    dX=\frac{16d^4\tx}{(1+\tx^2)^4}\,,\qquad \qquad \tK = \frac{\langle \lambda d\lambda\rangle \wedge \langle \hat{\lambda}d\hat{\lambda}\rangle }{\langle \hat \l\l\rangle^2}\,,
\end{align}
for convenience. Next, we will integrate out fibre coordinates using the following integral \cite{Woodhouse:1985id,Boels:2006ir,Jiang:2008xw}\footnote{This averaging operator was previously used in the semi-classical limit of the fuzzy geometry $S^4_N$ and $H^4_N$ with different notation denoted as $[\cdot]_0$, see \cite{Sperling:2018xrm} for more information.}
\begin{align}\label{eq:bridge}
    \int_{\projectivespace^1}\tK\, \frac{\hat{\lambda}^{\alpha}...\hat{\lambda}^{\alpha}\,\lambda_{\beta}...\lambda_{\beta}}{\langle \hat{\lambda}\lambda\rangle^{m}}\,\delta_{m,n
    }=-\frac{2\pi i}{(m+1)}\epsilon_{\beta}^{\ \alpha}...\epsilon_{\beta}^{\ \alpha}\,.
\end{align}
Loosely speaking, the above integral simply tells us that all fibre coordinates will transform into the Van der Waerden symbols $\epsilon$ in spacetime. It is the "vehicle" that allows us to move from BWR on twistor space to MUR on spacetime. The spacetime action for the self-dual HS-IKKT then reads
\begin{align}\label{eq:HSIKKTN=4noSUSY}
    \begin{split}
    S_{SD}=\int dX\llangle[\Big]\Big(B_{\alpha\alpha}F^{\alpha\alpha}&+\bar{\chi}^{\alpha}\{P_{\alpha\beta'},\widetilde{\chi}^{\beta' }\}+\frac{i}{2}\{P^{\alpha\alpha'},\hatphi\} \{P_{\alpha\alpha'},\hatphi\}-\frac{1}{2} \widetilde{\bar{\chi}}_{\alpha'}\{\hatphi,\widetilde{\chi}^{\alpha' }\}\\
    &\qquad \quad +\frac{i}{2}\{P^{\alpha\alpha'},\phi^{IJ}\} \{P_{\alpha\alpha'},\phi_{IJ}\}+\frac 12 \widetilde{\bar{\chi}}^{I'}\{\phi_{I'J'},\widetilde{\chi}^{J'}\} \Big)\rrangle[\Big]\,.
    \end{split}
\end{align}

Here, the doubled angle bracket $\llangle ... \rrangle$ means all possible contractions between unprimed indices from higher-spin extensions of the fields in \eqref{eq:fieldcontents}. The action for the non-self-dual part reads
\begin{align}\label{eq:nonselfdualHSIKKT}
    \begin{split}
    S_{n.SD}=\int dX \llangle[\Big]\Big(&\frac{i}{2}B_{\alpha\alpha}B^{\alpha\alpha}+\frac{1}{2}\bar{\chi}_{\alpha}\{\hatphi,\chi^{\alpha }\}+\frac 12 \bar{\chi}^I\{\phi_{IJ},\chi^J\}+\frac{i}{2}\{\hatphi,\hatphi\}\{\hatphi,\hatphi\}\\
    &+\frac{i}{2}\{\hatphi,\phi^{IJ}\}\{\hatphi,\phi_{IJ}\}+\frac{i}{2}\{\phi^{IJ},\phi^{MN}\}\{\phi_{IJ},\phi_{MN}\}\Big)\rrangle[\Big]\,.
    \end{split}
\end{align}
After integrating out the higher-spin fields $B$, we obtain spacetime action for the HS-IKKT model 
on $S^4$ as
\begin{align}\label{eq:N=4HSSYM}
    \begin{split}
    S&=\int dX \llangle[\Big]\Big(\frac{i}{2}F_{\alpha\alpha}F^{\alpha\alpha}+\bar{\chi}^{\alpha}\{P_{\alpha\alpha'}\widetilde{\chi}^{\alpha' }\}+\frac{i}{2}\{P^{\alpha\alpha'},\hatphi\}\{ P_{\alpha\alpha'},\hatphi\}-\frac{1}{2} \widetilde{\bar{\chi}}_{\alpha'}\{\hatphi,\widetilde{\chi}^{\alpha' }\}\\
    &\qquad \quad +\frac{i}{2}\{P^{\alpha\alpha'},\phi^{IJ}\} \{P_{\alpha\alpha'},\phi_{IJ}\}+\frac 12 \widetilde{\bar{\chi}}^{I'}\{\phi_{I'J'},\widetilde{\chi}^{J'}\}+\frac{1}{2}\bar{\chi}_{\alpha}\{\hatphi,\chi^{\alpha }\}+\frac 12 \bar{\chi}^I\{\phi_{IJ},\chi^J\}\\
    &\qquad \quad +\frac{i}{2}\{\hatphi,\hatphi\}\{\hatphi,\hatphi\}+\frac{i}{2}\{\hatphi,\phi^{IJ}\}\{\hatphi,\phi_{IJ}\}+\frac{i}{2}\{\phi^{IJ},\phi^{MN}\}\{\phi_{IJ},\phi_{MN}\} \Big)\rrangle[\Big]\,.
    \end{split}
\end{align}
It is noteworthy to mention that the action of the HS-IKKT in spacetime has minimally two derivative in the interactions. Hence, the HS-IKKT is a gravitational higher-spin theory as pointed out previously in \cite{Sperling:2018xrm,Steinacker:2019awe,Steinacker:2017vqw}.

\paragraph{The flat limit of HS-IKKT.} As discussed above, together with 5 other scalar fields associated to the remaining extra dimensions, we have in total 6 scalar fields that transform in the adjoint representation of $SU(4)_R$ --- the $R$-internal symmetry group. Moreover, $\chi,\tilde{\chi}$ also carry internal indices and transform in the fundamental representation of $SU(4)_R$. The spacetime effective action of the HS-IKKT in the SCF limit reads
\begin{align}\label{eq:IKKTN=4SCFspacetime}
    \begin{split}
    S=\int dX\,\llangle[\Big]&\Big[ B_{\alpha\alpha}F^{\alpha\alpha}+\frac{i}{2}B_{\alpha\alpha}B^{\alpha\alpha}+i\{P^{\alpha\alpha'},\phi^{\Ical\Jcal}\}\{P_{\alpha\alpha'},\phi_{\Ical\Jcal}\}+2\bar{\chi}^{\alpha}\{P_{\alpha\beta'},\widetilde{\chi}^{\beta'}\}\\
    &\qquad + \bar{\chi}^{\Ical}\{\phi_{\Ical\Jcal},\chi^{\Jcal}\}+ \widetilde{\bar{\chi}}^{\Ical}\{\phi_{\Ical\Jcal},\widetilde{\chi}^{\Jcal}\}+\frac{i}{2}\{\phi^{\Ical\Jcal},\phi^{\Mcal\Ncal}\}\{\phi_{\Ical\Jcal},\phi_{\Mcal\Ncal}\}\Big]\rrangle[\Big]\,,
    \end{split}
\end{align}
for $\Ical,\Jcal,\Mcal,\Ncal=1,2,3,4$. This is reminiscent of the action for $\Ncal=4$ SYM in $4d$ \cite{Chalmers:1996rq,Witten:2003nn,Boels:2006ir}. Therefore, we can view the (self-dual) HS-IKKT as higher-spin extension of $\Ncal=4$ (self-dual) SYM in the SCF limit but with gravitational interactions.

\paragraph{Yang-Mills sector.} In this work, we will consider only the Yang-Mills sector of the HS-IKKT model in the SCF limit while saving other sectors for future work. Namely, we will consider only the following action
\begin{align}
    S=\int_{\Kcurl} dX\,\tK  \,\llangle[\Big]\Big(2B_{\alpha\alpha}\{\ty^{\alpha}_{\ \alpha'},A^{\alpha\alpha'}\}+B_{\alpha\alpha}\{A^{\alpha}_{\ \alpha'},A^{\alpha\alpha'}\}+\frac{i}{2}B_{\alpha\alpha}B^{\alpha\alpha}\Big)\rrangle[\Big]\,.
\end{align}
We recall that the contribution from the spin operator $\Sigma$ can be neglected in the SCF limit, hence all fibre dependence drops out from the Poisson brackets in the kinetic term. From \eqref{eq:freeAnadB}, the kinetic action reads
\begin{align}
    S_2=\sum_{m,n}\int 2\lambda^{\beta}...\lambda^{\beta}\hat{\lambda}^{\gamma}...\hat{\lambda}^{\gamma}\lambda_{\zeta}...\lambda_{\zeta}\hat{\lambda}_{\tau}...\hat{\lambda}_{\tau}\,\big(B_{\alpha\alpha\beta(m)\gamma(m)}\Ecal^{\alpha\ \bullet\bullet'}_{\ \alpha'}\pl_{\bullet\bullet'}A^{\zeta(n)\tau(n)\alpha,\alpha'}\big)\,,
\end{align}
where we have shown in the subsection \ref{sec:analysisontwistorspace} that $A$ and $B$ are totally symmetric in their unprimed indices. Using \eqref{eq:effectivevielbein}, \eqref{A1-solutions-gaugefix} and \eqref{eq:bridge}, we end up with the following free spacetime action
\begin{align}
    S_2=2\int d^4\tx\,B_{\alpha(2s)}\pl^{\alpha}_{\ \alpha'}A^{\alpha(2s-1),\alpha'}\,.
    \label{free-action-index-BF}
\end{align}
Note that we have discarded all terms that look like the generalized Lorenz gauge condition, 
and we have used the fact that $\mu\sim \Ocal(R^{-\frac{1}{2}})$ cf. \eqref{eq:muparametrization}. 
The free action \eqref{free-action-index-BF} is precisely the one obtained in \cite{Krasnov:2021nsq}. Moreover, it can be shown that the solutions for the free equations of motion for the spacetime fields $A$ and $B$ are exactly the ones in \eqref{eq:positivehelicityansatz} and \eqref{eq:negativehelicityansatz}. It is  remarkable that even though the free equations of motion for the higher-spin gauge potentials $A$ and Lagrangian multiplier fields $B$ in the HS-IKKT model on $\Kcurl$ are more complicated compared to those of \cite{Krasnov:2021nsq}, the spacetime EOMs are exactly the same with the standard EOMs of free higher-spin fields in flat space after the Penrose transform. We also stress once again that the solutions \eqref{eq:positivehelicityansatz} and \eqref{eq:negativehelicityansatz} belong to MUR.\footnote{The free spacetime action for $\hs$-valued scalar fields $\phi^{\Ical\Jcal}$ and fermionic fields $\chi,\widetilde{\chi}$ can also be obtained similarly.} Next, the term $B^2$ on twistor space reduces to\footnote{The coefficients resulting from the Penrose transform can always be absorbed via an appropriate field redefinition.}
\begin{align}
    \frac{i}{2}\int_{\Kcurl}dX\,\tK\,( B_{\alpha\alpha}B^{\alpha\alpha})\quad \mapsto\quad -\frac{1}{2}\int dX\,B_{\alpha(2s)}B^{\alpha(2s)}\,,
\end{align}
on spacetime. It can be understood as a perturbation around the self-dual sector that contributes to the full Yang-Mills structure in the HS-IKKT model.  

Now we move on to the cubic interaction term for the Yang-Mills sector of the HS-IKKT.\footnote{We will study the other interaction terms in the action \eqref{eq:IKKTN=4SCFspacetime} in a later work.} Recall that the Poisson bracket of $\{A^{\alpha}_{\ \alpha'},A^{\alpha,\alpha'}\}$ involves both the fibre and spacetime derivatives on $\Kcurl$. Therefore, there will be two types of contributions in the $B\{A,A\}$ vertex. The leading contribution is the  one which arises from pure spacetime derivatives, where all fibre coordinates factor out from the $\{,\}$ bracket. Explicitly, 
\begin{align}
    \begin{split}
    &(\lambda^{\beta})^s(\hat{\lambda}^{\rho})^sB_{\alpha\alpha\beta(s)\rho(s)}\lambda_{\gamma}^m\hat{\lambda}_{\delta}^m\lambda_{\kappa}^n\hat{\lambda}_{\tau}^n\{A^{\gamma(m)\delta(m)\alpha,}_{\qquad \qquad \alpha'}(\tx),A^{\kappa(n)\tau(n)\alpha,\alpha'}(\tx)\}\\
    &\sim(\lambda^{\beta})^s(\hat{\lambda}^{\rho})^sB_{\alpha\alpha\beta(s)\rho(s)}\lambda_{\gamma}^m\hat{\lambda}_{\delta}^m\lambda_{\kappa}^n\hat{\lambda}_{\tau}^n\Ecal^{\circ\circ'\bullet\bullet'}\pl_{\circ\circ'}A^{\gamma(m)\delta(m)\alpha,}_{\qquad \qquad \alpha'}\,\pl _{\bullet\bullet'}A^{\kappa(n)\tau(n)\alpha,\alpha'}\,.
    \end{split}
\end{align}
We note that the spacetime fields are traceless with respect to $\epsilon$ tensors since they arise from the first eigenmodes \eqref{eq:positivehelicityansatz} and \eqref{eq:negativehelicityansatz} of the twistor field $A$ and $B$. Hence, using the Penrose transform \eqref{eq:bridge} i.e. integrating over the $\CP^1$ fibre space, we obtain the following spacetime vertex
\begin{align}\label{eq:thevertices}
   V_3^{\lead}=\sum_{m+n=2s-2} B_{\alpha(2s)}\pl_{\alpha \bullet'}A^{\alpha(m),}_{\qquad  \alpha'}\pl_{\alpha}^{\ \bullet'}A^{\alpha(n),\alpha'}
   +V^{\text{irrelevant}}\,.
\end{align}
Due to the restriction on the contraction between unprimed indices, the $\alpha$ indices in the derivatives will be contracted with the $\alpha$ indices of the fields $A$s in all possible way. However, because of the plane-wave solution \eqref{eq:positivehelicityansatz} we will receive only the contribution of type $\langle \zeta i\rangle\langle \zeta j\rangle$ for $i \rangle
\equiv \upsilon_i$ being the external spinors associated to the external fields $A_i$ (see section \ref{sec:5}). For this reason, we do not need to specify how the $\alpha$ indices of the derivatives contract to the ones of the $A$ fields. 
In \eqref{eq:thevertices}, $V^{\text{irrelevant}}$ denotes other contributions of $V_3^{\lead}$ that vanish when we compute the 3pt-scattering amplitudes using the asymptotic states \eqref{eq:positivehelicityansatz} and \eqref{eq:negativehelicityansatz}. It is remarkable that the cubic vertex \eqref{eq:thevertices} is closely related to the one of the higher-spin extension of self-dual gravity (HS-SDGRA) in \cite{Krasnov:2021nsq}. Hence,  
even though the HS-IKKT looks like a higher-spin extension of $\Ncal=4$ SYM, it behaves like a gravitational  theory of higher spins!

Let us now consider the subleading terms that result from  $\{\lambda_{\alpha},\hat{\lambda}_{\beta}\}= -i\epsilon_{\alpha\beta}$ in \eqref{spinor-Poisson}. Observe that we will get contributions that have a pair of unprimed spinorial indices contracted between the gauge potentials $A^{\alpha(2s-1),\alpha'}$. We first compute
\begin{align}\label{eq:subleadingBAA}
    \{\LaTeXunderbrace{\lambda_{\alpha}...\lambda_{\alpha}}_{n\ \text{times}}\LaTeXunderbrace{\hat{\lambda}_{\beta}...\hat{\lambda}_{\beta}}_{n\ \text{times}},\LaTeXunderbrace{\lambda_{\gamma}...\lambda_{\gamma}}_{m\ \text{times}}\LaTeXunderbrace{\hat{\lambda}_{\delta}...\hat{\lambda}_{\delta}}_{m\ \text{times}}\}=imn\,\epsilon_{\gamma\beta}\lambda^{n}_{\alpha}\lambda^{m-1}_{\gamma}\hat{\lambda}^{n-1}_{\beta}\hat{\lambda}^m_{\delta}-imn\,\epsilon_{\alpha\delta}\lambda_{\alpha}^{n-1}\lambda^m_{\gamma}\hat{\lambda}^{m-1}_{\delta}\hat{\lambda}^n_{\beta}\,.
\end{align}
In spacetime, the subleading vertex read
\begin{align}\label{eq:V3sub}
    V_3^{\sub}=2\sum_{m+n=2s-2}mn\,B_{\alpha(2s-2)}A^{\alpha(m)}_{\quad \ \ \bullet,\bullet'}A^{\alpha(n)\bullet,\bullet'}\,.
\end{align}
Their contributions to the scattering amplitudes  vanish due to the plane-wave solutions \eqref{eq:positivehelicityansatz}. 
Note also that the vertex \eqref{eq:V3sub} vanishes identically in the light-cone gauge. Hence, the only non-vanishing contribution of the Poisson bracket between two higher-spin gauge potentials $A$ is the one that has all fibre coordinates outside of the $\{,\}$ bracket. As a remark, although the Poisson brackets will provide more structures compared to higher-spin interactions in field theory approaches on twistor space, the spacetime action is simple thanks to the Penrose transform \eqref{eq:bridge}.

\section{Scattering of the HS-IKKT in the flat limit}\label{sec:5}

We are now in a position to study the 3-pt tree-level amplitude of the SDYM sector of the HS-IKKT in the flat limit where $S^4\rightarrow \RR^4$. The action for this sector is
\begin{align}\label{eq:gravitationalHS-SDYM}
    S=\sum_s\int d^4x\, B_{\alpha(2s)}G^{\alpha(2s)}\,,
\end{align}
where 
\begin{align}
    G^{\alpha(2s)}=\pl^{\alpha}_{\ \alpha'}A^{\alpha(2s-1),\alpha'}+\sum_{m+n=2s+2}\pl_{\alpha \bullet'}A^{\alpha(m),}_{\qquad \kappa'}\pl_{\alpha}^{\ \bullet'}A^{\alpha(n),\kappa'}\,.
\end{align}
The above action is closely related to the action of HS-SDGRA in \cite{Krasnov:2021nsq}. Indeed, the light-cone action for the above is identical with the one of HS-SDGRA. It can be seen as follow. First of all, we can impose $A^{\alpha(2s-2)0,0'}=0$ in the light-cone gauge. Then, the physical component of the higher-spin gauge potential is $A^{1(2s-1),0'}=\Phi_{+s}$. We recall that $\alpha=0,1$ and $\alpha'=0',1'$. Similarly, the physical component of the $B$ field is $B^{0(2s)}=\pl^+\Phi_{-s}$. We will split the partial derivative in spacetime as $\pl^{\mu}=(\pl^+,\pl^-,\pl,\bar{\pl})$. Upon integrating out auxiliary components of the gauge fields
, we end up with the following action
\begin{align}
    S=\sum_s\int d^4x \frac{1}{2}(\Phi_{-s}\Box \Phi_{+s})-\sum_{s_2,s_3}\Big(\Phi_{-(s_2+s_3-2)}\pl^0_{\ \alpha'}\pl^0_{\ \beta'}\Phi_{+s_2}\pl^{0\alpha'}\pl^{0\beta'}\Phi_{+s_3}\Big)\,.
\end{align}
In momentum space the above reduces to
\begin{align}
    S=-\frac{1}{2}\sum_s\int d^4\pvec (\Phi_{-s}\Phi_{+s})\pvec^2+\sum_{s_2,s_3}\int d^4\pvec_{1,2,3}\delta^4(\sum \pvec_i)\, \PPb^2 (\Phi_{-(s_2+s_3-2)}\Phi_{+s_2}\Phi_{+s_3})\,,
\end{align}
where $\pvec:=(\beta,p^-,p,\bar{p})$ and $\PPb_{ij}=\bar{p}_i\beta_j-\bar{p}_j\beta_i$ for $\pvec_i$ being the 4-momenta of the external field $\Phi_{s_i}$. 

Next, it is well-known that the 3-pt amplitude vanishes for real momenta. Hence, we will work in a complexified setting to compute the 3-pt amplitude and we will not discuss reality/positive energy conditions explicitly. To address momentum conservation, we choose a convention such that the sum of momenta entering a vertex is zero. To this end, let us compute the simplest tree-point amplitudes between ($B_1^{-s_1},A_2^{s_2},A_3^{s_3}$). Using the plane-wave solutions \eqref{eq:positivehelicityansatz} and \eqref{eq:negativehelicityansatz}, we obtain
\begin{align}
    \Mcal_{-s_1|s_2,s_3}=\delta(2-(s_2+s_3-s_1))[23]^2\frac{\langle \zeta 1\rangle^{2s_1+2s_2-4}}{\langle  \zeta 2\rangle^{2s_2-2}\langle \zeta 3  \rangle^{2s_3-2}}\,.
\end{align}
Due to conservation of momentum $\pvec_1+\pvec_2+\pvec_3=0$, we have
\begin{align}
    \frac{\langle \zeta 1 \rangle }{\langle \zeta 2\rangle }=\frac{[23]}{[31]}\,,\qquad \frac{\langle \zeta 1 \rangle }{\langle \zeta 3\rangle }=\frac{[23]}{[12]}\,.
\end{align}
Therefore, the three-point amplitude reads
\begin{align}\label{eq:3pamplitudes}
    \Mcal_{-s_1|s_2,s_3}=\delta(2-(s_2+s_3-s_1))\frac{[23]^{2s_2+2s_3-2}}{[31]^{2s_2-2}[12]^{2s_3-2}}\,.
\end{align}
We obtain the usual result for gravity at $s=2$. Note that the action \eqref{eq:gravitationalHS-SDYM} can be viewed as a closed subsector of the HS-IKKT model. We shall refer this theory as gravitational HS-SDYM or gravHS-SDYM for short. In the next section, we will reconstruct this theory from a different point of view.


\section{Twistor construction for gravHS-SDYM}\label{sec:6}

Let us summarize what we have done so far. We started with the IKKT model on a background given by fuzzy $\CP^3$, which is recognized as quantized twistor space. In the semi-classical limit,  higher-spin fields emerge naturally in the BWR. 
After integrating out the fibres using the Penrose transform, we end up with the spacetime description of the HS-IKKT, with fields in the MUR. Since the resulting system  is a field theory on spacetime, it is natural to ask whether we can arrive at the same spacetime action in a different, more commutative manner. We will propose how this can be achieved using the twistor construction for higher-spin fields in \cite{Tran:2021ukl,Haehnel:2016mlb,Adamo:2016ple}. For the twistor constructions of lower-spin theories, we shall refer the readers to e.g. \cite{Mason:2005zm,Boels:2007qn,Mason:2007ct} for further details.

Our goal is to arrive at the gravHS-SDYM action that gives the same scattering amplitudes as in \eqref{eq:3pamplitudes}. Hence, it is appropriate to consider a curved/deformed projective twistor space $\PTc$ \cite{Penrose:1976js} associated to flat Eulidean space $\RR^4$. Since no confusion can arise, we will simply call $\PTc$ as twistor space. The twistor action has the form of ``gravitational'' $\Bcal\Fcal$ action on the uncompactified twistor space and is closely related to the twistor action of SDGRA studied in \cite{Mason:2007ct}. Our derivation follows closely to the references \cite{Mason:2005zm,Boels:2007qn,Tran:2021ukl}.

To begin, recall that the twistor space corresponding to $\RR^4$ is isomorphic to the projectivisation of the unprimed spinor bundle denoted as $\PS\cong\CP^1\times \RR^4$. Hence, as before, we can use $\lambda^{\alpha}$ as coordinates on the $\CP^1$ fiber while the spinors $\mu^{\alpha'}$ are coordinates up the fibers of the normal bundle\footnote{See e.g. 
 \cite{Penrose:1968me,Penrose:1976js,Penrose:1967wn,Huggett:1986fs} for an explanation.}
\begin{align}
    T(\PTc)|_{\CP^1}/T(\CP^1)=\Ocal(1)\oplus \Ocal(1)\,,
\end{align}
in the view of the fibration $\pi:\PTc\rightarrow \CP^1$. It is convenient to define the following basis for the $(0, 1)$-vector
fields on the projectivisation of the unprimed spinor bundle $\PS$ \cite{Mason:2005zm}
\begin{align}\label{eq:basis}
    \bar{\pl}_0=-\langle \hat{\lambda}\lambda\rangle \lambda_{\alpha}\frac{\pl}{\pl \hat{\lambda}_{\alpha}}\,,\qquad \qquad \qquad \qquad  \bar{\pl}_{\alpha'}=-\lambda^{\alpha}\frac{\pl}{\pl x^{\alpha\alpha'}}\,.
\end{align}
Their dual $(0,1)$-forms read
\begin{align}\label{eq:dualbasis}
    \bar{e}^0=\frac{\langle \hat{\lambda}d\hat{\lambda}\rangle}{\langle \hat{\lambda}\lambda\rangle^2}\,,\qquad \qquad \qquad \qquad  \bar{e}^{\alpha'}=\frac{\hat{\lambda}_{\alpha}dx^{\alpha\alpha'}}{\langle \hat{\lambda}\lambda\rangle}\,.
\end{align}
It is easy to check that $\bar{\pl}=\bar{e}^0\bar{\pl}_0+\bar{e}^{\alpha'}\bar{\pl}_{\alpha'}=d\hat Z^A\frac{\pl}{\pl \hat Z^A}$. Essentially, the above basis is based on the fact that $T(\CP^1)\sim \Ocal(2)$ and $T^*(\CP^1)\sim \Ocal(-2)$, where $\Ocal(n)$ is the line bundles whose sections are polynomials of homogeneity $n$ in $\lambda$. The above basis provides a convenient way to work with Penrose transform in the Woodhouse's gauge (also known as harmonic gauge). 

Since the vertices of gravHS-SDYM involve derivatives, it is convenient to define the following $(1,0)$-vector fields and their dual $(1,0)$-forms on $\PS$:
\begin{subequations}\label{eq:nicebasisfordeformation}
\begin{align}
   \partial_0&:=-\frac{\hat{\lambda}_{\alpha}}{\langle \hat \lambda \lambda  \rangle}\frac{\partial}{\partial \lambda_{\alpha}}\,,  \qquad  &\partial_{\alpha'}&:=\frac{\hat{\lambda}^{\alpha}}{\langle \hat\lambda \lambda\rangle}\partial_{\alpha\alpha'}\,,\\
   e^0&:=\langle \lambda d\lambda\rangle\,,  \qquad  &e^{\alpha'}&:=\lambda_{\alpha}dx^{\alpha\alpha'}\,.
\end{align}
\end{subequations}
Notice that $e^0$ is the holomorphic top-form of the fiber $\CP^1$.

Now, let us consider the following action on deformed projective twistor space $\PTc$
\begin{align}\label{eq:BF}
    S_{SD}=\int_{\PTc} D^3Z\, \Bcal\Fcal+S_c\,,
\end{align}
where $\Bcal$ is a $(0,1)$-form Lagrangian multiplier and $\Fcal$ is a curvature $(0,2)$-form. Here, $S_c$ is the correction part of the action that addresses the non-gauge-invariance of the measure $D^3Z$ under higher-spin diffeomorphism on $\PTc$ viz.
\begin{align}
    \delta Z=\{Z,\xi\}_h\,,
\end{align}
where $\xi$ is some $\hs$-valued (gauge parameter) section on $\PTc$. Furthermore, $\{\,,\}_h$ with the subscript $h$ is the holomorphic Poisson bracket on $\PTc$ that has the following property
\begin{align}\label{eq:propertyofholoPoisson}
    &\{a,b\}_h=-\{b,a\}_h\,.
\end{align}
The gauge transformations for $\Bcal$ and $\Fcal$ read
\begin{align}\label{holotransform}
    \delta \Fcal=\{\Fcal,\xi\}_h\,,\qquad\qquad \delta \Bcal=\{\Bcal,\xi\}_h\,.
\end{align}
We observe that if we work with $\PS\cong \CP^1\times \RR^4$ instead of $\PTc$, the correction $S_c$ to the higher-spin diffeomorphism can be dropped since we can write the anti-holomorphic measure $D^3\bar{Z}$ as
\begin{align}
   D^3\bar{Z}=\bar{e}^0\wedge [\bar{e}^{\alpha'}\wedge \bar{e}_{\alpha'}]\,,
\end{align}
using the basis  \eqref{eq:dualbasis}. We, then, have
\begin{align}
    D^3Z\wedge D^3\bar{Z}=d^4x\frac{\langle \hat\lambda d\hat\lambda\rangle \wedge \langle \lambda d\lambda\rangle}{\langle \hat\lambda\lambda\rangle^2}\,,
\end{align}
when the base manifold is $\RR^4$. The holomorphic Poisson bracket on $\PS$ is defined as
\begin{align}
\label{holo-Poisson}
    \{f,g\}_h:=\pl_{\alpha'}\,f\wedge \pl^{\alpha'}\,g=\frac{\hat\lambda^{\alpha}\hat\lambda^{\beta}}{\langle \hat\lambda\lambda\rangle^2}\pl_{\alpha\alpha'}f\wedge \pl_{\beta}{}^{\alpha'}g\,,
\end{align}
where $\pl_{\alpha'}$ is defined in \eqref{eq:nicebasisfordeformation}.\footnote{This deformation was previously used in the context of non-linear graviton construction, see e.g. \cite{Penrose:2015lla,Lukierski:2021zta}.} Here, $f$ and $g$ are holomorphic forms/sections on $\PS\cong\PTc$. Hence, the holomorphic Poisson bracket preserves the holomorphicity of the wedge product between $f$ and $g$. One can check that \eqref{holo-Poisson} satisfies the property \eqref{eq:propertyofholoPoisson}. Moreover, the action \eqref{eq:BF} on $\PS$ is now gauge invariant under \eqref{holotransform} since we can freely move the $\pl_{\alpha'}$ derivatives around without creating additional terms when $\pl_{\alpha'}$ acting on the measure.

Let us now define our fundamental fields on $\PTc$ by recalling that a free massless spin-$s$ field on $4d$ spacetime corresponds to a twistor cohomology representative in the Dolbeault cohomology group $H^{0,1}(\PT,\Ocal( 2s-2))$ if it has positive helicity, and $H^{0,1}(\PT,\Ocal(-2s-2))$ otherwise \cite{Eastwood:1981jy,Sparling,Woodhouse:1985id,Hitchin:1980hp}, where\footnote{See \cite{Adamo:2013cra,Jiang:2008xw} for a nice review.}
\begin{align}
    H^{0,1}(\PT,\Ocal(n)):=\frac{\{\omega \in \Omega^{0,1}(\PT)(n)|\bar{\pl}\omega=0\}}{\{\omega|\omega=\bar{\pl}\beta\}}\,.
\end{align}
Here, $\Omega^{0,1}(\PT)(n)$ is the space of $(0,1)$-forms on $\PT$ of weight $n$, i.e. $f(tZ) = t^nf(Z)$. Note that $\PT$ is the twistor space where there is no deformation of the complex structures, and we will take the spinors $\lambda$ to be our reference of weight.

Now, to address interactions between higher-spin fields, we can consider the following twistor representatives \cite{Tran:2021ukl} in the Woodhouse gauge 
\begin{align}\label{eq:HSconnection}
     \omega&\equiv\bar{e}^{\alpha'}\omega_{\alpha'}=\sum_s\omega_{\alpha'}^s\bar{e}^{\alpha'}=\sum_s \LaTeXunderbrace{\lambda^{\alpha}\cdots \lambda^{\alpha}}_{2s-1\  \text{times}} A_{\alpha(2s-1),\alpha'}\bar{e}^{\alpha'}\,,
\end{align}
and
\begin{align}\label{eq:HSmultiplier}
     \Bcal&=\sum_s\big(\Bcal_0^s\bar{e}^0+\Bcal_{\alpha'}^s\bar{e}^{
\alpha'}\big)=\sum_s(2s+1)\,\LaTeXunderbrace{\hat{\lambda}^{\alpha}\cdots \hat{\lambda}^{\alpha}}_{2s\  \text{times}}B_{\alpha(2s)}\frac{\langle \hat{\lambda}
d\hat{\lambda}\rangle }{\langle \hat{\lambda}\lambda\rangle^{2s+2} }+\sum_s\Bcal_{\alpha'}^s\bar{e}^{\alpha'}\,.
\end{align}
Note that $\bar{e}^{\alpha'}$ and $\bar{e}^0$ are defined in \eqref{eq:dualbasis}. We will refer $\omega$ to as the generalized connection that is a twistor representative of $\bigoplus_s \Omega^{0,1}(\PTc,\Ocal(2s-2))$. On the other hand, the $\Bcal$ field is referred to as the generalized Lagrangian multiplier twistorial field that belongs to the following group $\bigoplus_s \Omega^{0,1}(\PTc,\Ocal(-2s-2))\,$. Note that the above twistor representatives of $\omega$ and $\Bcal$ allows us to get HS-SDYM in spacetime \cite{Tran:2021ukl} through the Penrose transform \eqref{eq:bridge}. Furthermore, we will assume that all deformations are sufficiently small so that they will not affect the complex structures on twistor space to avoid the complication arise from Kodaira's theory \cite{kodaira2006complex}.

Next, the field strength $\Fcal$ is defined as\footnote{A derivation for this field strength will appear elsewhere.}
\begin{align}\label{eq:Fdef}
    0=\Fcal:=\bar{\pl}\omega+\frac{1}{2}\{\omega,\omega\}_h\,.
\end{align}
By varying the action \begin{align}
    S=\int_{\PS} e^0\wedge [e^{\alpha'}\wedge e_{\alpha'}]\, \Bcal (\bar{\pl}\omega +\frac{1}{2}\{\omega,\omega\}_h)\,,
\end{align}
with respect to $\omega$, we obtain the equation of motion for $\Bcal$ as
\begin{align}
    \bar{\Dcal}\Bcal=\bar{\pl}+\{\omega,\Bcal\}_h=0\,.
\end{align}
The above equation is invariant under $\Bcal\rightarrow \Bcal+\bar{\Dcal}\chi$ due to Bianchi identity. Therefore, the gauge transformations for $\omega$ and $\Bcal$ read
\begin{align}\label{eq:twistorgaugetransformation}
   \delta \omega=\bar{\pl}\xi+\{\omega,\xi\}\,,\qquad \qquad  \delta \Bcal =\{\Bcal,\xi\}_h+\bar{\pl}\chi+\{\omega,\chi\}\,.
\end{align}
To obtain spacetime description of the above $\Bcal\Fcal$ theory, we will rewrite the field strength $\Fcal$ as 
\begin{align}
    \Fcal=(\bar{\pl}_0\omega_{\alpha'}-\bar{\Dcal}_{\alpha'}\omega_0)\bar{e}^0\wedge \bar{e}^{\alpha'}+\bar{\Dcal}_{\alpha'}\omega_{\beta'}\bar{e}^{\alpha'}\wedge \bar{e}^{\beta'}=0\,,
\end{align}
where we have used the basis \eqref{eq:dualbasis} to decomposed $\bar{\Dcal}$, $\Bcal$ and $\omega$ in terms of coefficients of $\bar{e}^0$ and $\bar{e}^{\alpha'}$. The above can be understood as the integrability condition with the deformed complex structure $\bar{\Dcal}$. We note that
\begin{align}
    \bar{\Dcal}_{\alpha'}\omega_0=\bar{\pl}_{\alpha'}\omega_0+\{\omega_{\alpha'},\omega_0\}_h\,,\qquad \qquad \bar{\Dcal}_{\alpha'}\omega_{\beta'}=\bar{\pl}_{\alpha'}\omega_{\beta'}+\{\omega_{\alpha'},\omega_{\beta'}\}_h\,,
\end{align}
according to \eqref{eq:Fdef}. Since, $\bar{\Dcal}_{\alpha'}\omega_0\bar{e}^0\wedge \bar{e}^{\alpha'}=\bar{\Dcal}(\omega_0\bar{e}^0)$, we can remove this contribution by the gauge transformation \eqref{eq:twistorgaugetransformation}. Namely, we can set $\omega_0=0$ as a natural gauge fixing condition on commutative twistor space. This particular gauge is the aforementioned Woodhouse gauge or harmonic gauge. It indeed agrees with various theorems \cite{Eastwood:1981jy,Sparling,Woodhouse:1985id,Hitchin:1980hp} that tightly constrain twistor cohomology representatives which are harmonic on the fibres. In the end, we only have to consider the following decomposition of $\Bcal$ and $\omega$:
\begin{align}
     \Bcal=\sum_s\big(\bar{e}^0\Bcal_0^s+\bar{e}^{\alpha'}\Bcal_{\alpha'}^s\big)\,,\qquad \qquad  \omega=\sum_s\bar{e}^{\alpha'}\omega_{\alpha'}^s\,,
\end{align}
where the letter $s$ indicates the spins of fields that take value in $\hs$.

 The integrability condition \eqref{eq:Fdef} contains two sub-equations
\begin{align}
    \sum_s\bar{\pl}_0\omega_{\alpha'}^s=0\,,\qquad \qquad \sum_s\bar{\pl}_{\alpha'}\omega^s_{\beta'}+\sum_{m+n=s}\{\omega_{\alpha'}^{m},\omega_{\beta'}^n\}_h\,.
\end{align}
The first equation does not contain spacetime derivative and is solved by \eqref{eq:HSconnection}. It tells us that $\omega^s_{\alpha'}$ are holomorphic in $\lambda$. The second equation, which contain spacetime derivatives, addresses interactions and can be written as
\begin{align}
    0=\epsilon_{\alpha'\beta'}\sum_s\lambda^{\alpha}...\lambda^{\alpha}\Big[\pl_{\alpha\bullet'}A_{\alpha(2s-1),}^{\qquad \quad  \bullet'}+\sum_{m+n=2s-2}\{A_{\alpha(m),\bullet'},A_{\alpha(n),}^{\quad \ \ \bullet'}\}_h\Big]\,.
\end{align}
After integrating out the fibre, we will obtain the two-derivative vertices for higher spins as in \eqref{eq:thevertices} without the $V^{\text{irrelevant}}$ terms. On spacetime, it is interesting to note that the gauge transformation of higher-spin fields contain two derivatives in the holomorphic Poisson bracket $\{\,,\}_h$ after the Penrose transform. It is somewhat similar to the case of Moyal-like higher-spin theories studied in \cite{Bonora:2018ggh}.



\paragraph{Remarks.} We have just shown that the twistor action for the gravitational self-dual Yang-Mills sector of the HS-IKKT model on commutative twistor space is a gravitational $\Bcal\Fcal$ one. Therefore, from the geometrical point of view, it should be integrable in accordance with the light-cone result in \cite{Ponomarev:2017nrr}. It is then reasonable to conjecture that the twistor action of the $\Ncal=4$ self-dual HS-IKKT is a deformed Chern-Simons action on super twistor space $\CP^{3|4}$. It would be interesting to check whether the non-self-dual (n.SD) part of the action \eqref{eq:nonselfdualHSIKKT} can have a twistor analogue like the one in \cite{Boels:2006ir,Cachazo:2004kj}. Namely,
\begin{align}\label{eq:nonselfdual}
    \begin{split}
    \text{n.SD}&=\int_X D\mathcal{X}\, \log\det\Big((\bar{\pl}+\AB)|_{(\projectivespace^1)^{\otimes n}}\Big)\\
    &=\int_X D\Xcal\, \Tr\Bigg[\log \bar{\pl}+\sum_{n=1}^{\infty}\frac{1}{n}\Big(\frac{i}{2\pi}\Big)^n\int\frac{\langle \lambda_1 d\lambda_1\rangle}{\langle \lambda_n\lambda_1\rangle }\AB_1...\frac{\langle \lambda_n d\lambda_n\rangle}{\langle \lambda_{n-1}\lambda_n\rangle}\AB_n\Bigg]\,,
    \end{split}
\end{align}
where $\AB$ is some super field that contains the gauge fields $(\omega,\Bcal)$, the fermionic fields $(\chi^{\Ical},\widetilde{\chi}^{\Ical})$ and the scalar fields $\phi^{\Ical\Jcal}$.  
The above action exhibits an interesting feature of twistor space: the interactions on twistor space can be non-local as long as the effective spacetime vertices are local. We shall refer the readers to \cite{Witten:2003nn,Cachazo:2004kj,Boels:2006ir} for further discussions.

Recall that by averaging out the fibre on fuzzy twistor space, we obtain a term $\sum_sB_{\alpha(2s)}B^{\alpha(2s)}$ in the action of the HS-IKKT on spacetime. This term can be understood as a deformation away from the self-dual gauge sector of the HS-IKKT model. Therefore, it is natural to consider the following action 
\begin{align}
    S=\sum_s\int d^4x\, B_{\alpha(2s)}G^{\alpha(2s)}-\frac{1}{2}\sum_s\int d^4x\,B_{\alpha(2s)}B^{\alpha(2s)}\,.
\end{align}
By integrating out the $B$ fields, we end up with a gravitational-type HS-YM (gravHS-YM) action
\begin{align}
    S=\frac{1}{2}\sum_s\int d^4x\, G_{\alpha(2s)}G^{\alpha(2s)}\,.
\end{align}
It is easy to see that the above action contains at most 4 derivatives in the interactions. Hence, it is a local gravitational higher-spin theory. This theory is can be thought of as the higher-derivative extension version of the HS-YM in \cite{Tran:2021ukl}. Note that due to the expansion purely in terms of unprimed indices, the above theory is intrinsically chiral. 



\section{Discussion}\label{sec:7}

In this work, we established a relation between two different approaches to higher-spin theories: one based on the IKKT matrix model on covariant quantum spaces \cite{Steinacker:2016vgf,Sperling:2017gmy,Sperling:2019xar}, and another based on a Poisson deformation of twistor space with higher-spin symmetry encoded in the twistor cohomology representatives \cite{Tran:2021ukl,Haehnel:2016mlb,Adamo:2016ple}. 

More specifically, we clarified the relation between the fuzzy 4-sphere $S^4_N$ and (fuzzy) twistor space, and studied the $\hs$-valued fields which naturally arise in the balanced weight representation (BWR).  This allows for a transparent analysis of higher-spin modes in the IKKT matrix model on such a background. We have also spelled out the effective vielbein and metric in spinorial language. This provides us a novel twistorial description for the IKKT-matrix model on fuzzy twistor space in full non-linearity. We then studied in detail the higher-spin gauge theory induced by the IKKT model in the large $N$ (semi-classical) the flat limit. Upon performing the Penrose transform, we obtained the spacetime action for the HS-IKKT. We also studied the simplest 3-pt scattering tree-level amplitude of the gravitational self-dual Yang-Mills sector of the HS-IKKT (gravHS-SDYM). The result matches with the 3-pt amplitude of the HS-SDGRA considered in \cite{Krasnov:2021nsq}. 

Having the spacetime action of gravHS-SDYM, we also show that its twistor action can also be formulated as a gravitational $\Bcal\Fcal$ theory on deformed projective twistor space $\PTc$. Therefore, the gravHS-SDYM is integrable. This strongly suggests
that the self-dual sector of the HS-IKKT model is also integrable along the line of \cite{Witten:2003nn,Boels:2006ir}.



The remarkable feature of the IKKT model is that it provides a simple non-perturbative definition of a higher-spin gauge theory, which appears to have good locality properties, and is well suited for quantization.
Even though the action is different from the more standard definitions of gravity, quantum effects are expected to bridge this gap \cite{Steinacker:2021yxt}. Due to this non-standard formulation, the significance of the resulting  higher-spin theory is not evident. The present re-formulation in terms of spinors and the relation to twistor space should be very useful to clarify its physical significance, as illustrated by the computation of a simple scattering amplitude.

Some further remarks are in order. First of all, the $\hs$-valued fields in the  fuzzy twistor construction naturally arise in the the BWR, while the twistor construction uses MUR. Nevertheless, the  higher-spin fields   on spacetime always belong to MUR. The main advantage of the fuzzy twistor construction is that we do not need to rely on twistor cohomology, and it is more natural to consider higher-spin interactions on twistor space compared to the twistor construction using MUR. Indeed, one can straightforwardly write down the non-self-dual interactions involving higher-spin fields in the fuzzy twistor construction. Moreover, we can maintain locality all the way from twistor space to spacetime in this construction compared to the usual twistor construction in \cite{Mason:2005zm,Boels:2006ir}. However, it does not give us a clear interpretation of why the self-dual sector of HS-IKKT is integrable.\footnote{The spacetime theory of self-dual HS-IKKT matrix model should be a quasi-topological theory. Namely, it has propagating degrees of freedom and interactions, but has trivial scattering amplitudes. Moreover, it should be a gravitational supersymmetric extension of HS-SDYM \cite{Ponomarev:2017nrr,Krasnov:2021nsq}. See also  \cite{Metsaev:2019aig} for a recent development of supersymmetric extension of chiral higher-spin gravities in the light-cone gauge.} We believe that the fuzzy twistor construction considered in this paper and usual the twistor construction in \cite{Mason:2005zm,Boels:2006ir,Mason:2007ct,Haehnel:2016mlb,Adamo:2016ple,Tran:2021ukl} can complement each other in finding consistent twistorial higher-spin theories in spacetime.

In the past, it was relatively simple to rule out the existence of higher-spin theories under the requirement that the S-matrix should be Poincare invariant, analytic and local \cite{Weinberg:1964ew,Coleman:1967ad,Benincasa:2007xk,Benincasa:2011kn,Benincasa:2011pg}. It turned out that the assumption about higher-spin symmetries \cite{Berends:1984rq,Boulanger:2013zza,Fronsdal:1978vb,Fradkin:1986ka}, i.e. interactions between higher-spin fields, is crucial for higher-spin theories to exist. However, it is not always possible. For example, if we use the Fronsdal approach to tackle the problem of higher-spin interactions, sooner or later we will once again run into No-go results \cite{Maldacena:2011jn,Sleight:2017pcz,Bekaert:2015tva,Boulanger:2015ova}. The reason is that most of the parity-invariant higher-spin theories using Fronsdal fields as the main objects in the literature exhibit non-local features starting from the quartic interactions. Specifically, in the Fronsdal's approach, spins and the number of derivatives in the vertices are closely correlated. In particular, as the spins of the fields increase, so do the numbers of derivatives.

The difficulty of non-locality in the Fronsdal's approach is overcome by the light-cone \cite{Bengtsson:1983pd,Bengtsson:1986kh,Bengtsson:2014qza,Metsaev:1991mt,Metsaev:1991nb,Metsaev:2005ar,Ponomarev:2016lrm,Ponomarev:2017nrr} and twistor/spinor formalisms \cite{Krasnov:2021nsq}. There, one can handle the number of derivatives and spins/helicities almost independently. However, the higher-spin theories constructed by the light-front or twistor/spinor approaches are usually non-unitary and self-dual. Being tailored for mainly constructing chiral theories, twistor theory might be an appropriate framework to formulate higher-spin theories that can avoid No-go theorems/results. 


\section*{Acknowledgement}
We are grateful to Tim Adamo, Mitya Ponomarev, Zhenya Skvortsov for useful discussions. We also thank the anonymous JHEP referee for many valuable suggestions and improvements. The work of TT is partially supported by the Fonds de la Recherche Scientifique - FNRS
under Grants No. F.4503.20 (HighSpinSymm), Grant No. 40003607 (HigherSpinGraWave), T.0022.19 (Fundamental issues
in extended gravitational theories) and the funding from the European Research Council (ERC) under the European Union’s Horizon 2020 research and innovation programme (grant agreement No 101002551). The work of HS is supported by  the Austrian Science Fund (FWF) grant P32086.

\appendix
\section{Spinor algebra}\label{sec:spinoralgebra}
First of all, the $\Sigma$-matrices provide a realization of $\mso(5)$ algebra
\begin{align}\label{eq:Sigmarelations}
    [\Sigma_{ab},\Sigma_{cd}]=(\Sigma_{ad}\delta_{bc}-\Sigma_{ac}\delta_{bd}-\Sigma_{bd}\delta_{ac}+\Sigma_{bc}\delta_{ad})\,.
\end{align}
For $\gamma$-matrices, we have the following useful relations ($\mu=1,2,3,4$)
\begin{subequations}
\begin{align}
    (\gamma^{a})_{\Acal}^{\ \Bcal}(\gamma_a)_{\Ccal}^{\ \Dcal}&=-\delta_{\Acal}^{\ \Bcal}\delta_{\Ccal}^{\ \Dcal}+2(\delta_{\Acal}^{\ \Dcal}\delta_{\Ccal}^{\ \Bcal}+C_{\Acal\Ccal}C^{\Bcal\Dcal})\,,\label{eq:gammaimportant}\\
    (\gamma^{\mu})_{\Acal[\Bcal}(\gamma_{\mu})_{\Ccal\Dcal]}&=0\,,\\
    \gamma^{\mu}\gamma^{\nu_1}...\gamma^{\nu_s}\gamma_{\mu}&=(-)^s(4-2s)\gamma^{\nu_1}...\gamma^{\nu_s}\,.
\end{align}
\end{subequations}
We use $C^{\Acal\Bcal}$ to raise and lower indices as
\begin{align}
    P^{\Acal}=P_{\Bcal}C^{\Acal\Bcal}\,,\qquad P_{\Acal}=P^{\Bcal}C_{\Bcal\Acal}\,,\qquad C_{\Acal\Mcal}C^{\Bcal\Mcal}=\delta_{\Acal}^{\ \Bcal}\,.
\end{align}
We also have the following relations $\epsilon$ matrices
\begin{align}
    u^{\beta}=u_{\gamma}\epsilon^{\beta\gamma}\,,\qquad u_{\beta}=u^{\gamma}\epsilon_{\gamma\beta}\,,\qquad \epsilon_{\alpha\gamma'}\epsilon^{\beta\gamma'}=\delta_{\alpha}^{\ \beta}\,,
\end{align}
from which we can define the angle and square brackets often used in the literature
\begin{align}
    \langle ab\rangle =a^{\alpha}b_{\alpha}\,,\qquad \qquad [ab]=a^{\alpha'}b_{\alpha'}\,.
\end{align}
We note that for a generic tensor $T_{\alpha\beta}$, it can be decomposed as 
\begin{align}
    T_{\alpha\beta}=T_{(\alpha\beta)}+\frac{1}{2}\epsilon_{\alpha\beta}T_{\gamma}^{\ \gamma}\,.
\end{align}
As a consequence, one can write an $\msp(4)$-tensor $T^{\Acal_1\Bcal_1,...,\Acal_s\Bcal_s}$ as
\begin{align}
    T^{\alpha_1\alpha_1'\beta_1\beta_1',...,\alpha_s\alpha_s'\beta_s\beta_s'}=T^{\alpha(s)\beta(s)}\epsilon^{\alpha_1'\beta_1'}...\epsilon^{\alpha_s'\beta_s'}+T^{\alpha'(s)\beta'(s)}\epsilon^{\alpha_1\beta_1}...\epsilon^{\alpha_s\beta_s}\,,
\end{align}
where
\begin{align}
    T^{\alpha(2s)}=\frac{(-)^s}{2^s}T^{\Acal_1\Bcal_1,...,\Acal_s\Bcal_s}(\sigma_{\Acal_1})^{\alpha}_{\ \gamma_1'}(\sigma_{\Bcal_1})^{\alpha\gamma_1'}...(\sigma_{\Acal_s})^{\alpha}_{\ \gamma_s'}(\sigma_{\Bcal_s})^{\alpha\gamma_s'}\,.
\end{align}
The tensor $T^{\alpha(2s)}$ and $T^{\alpha'(2s)}$ are referred to as the self-dual and anti-self-dual components of the tensor $T^{\Acal_1\Bcal_1,...,\Acal_s\Bcal_s}$.\footnote{We note that in Lorentzian signature, $T^{\alpha(2s)}$ are $T^{\alpha'(2s)}$ are complex conjugate of each other otherwise the corresponding field strength of $T$ will be complex.} There are also useful relations of $\sigma$-matrices 
\begin{subequations}
\begin{align}\label{eq:sigmaidentities}
    (\sigma^{\ta})^{\alpha}_{\ \alpha'}(\sigma^{\tb})^{\beta \alpha'}&=\frac{1}{2}\eta^{\ta\tb}\epsilon^{\alpha\beta}+\frac{1}{2}[\sigma^{\ta},\sigma^{\tb}]^{\alpha\beta}\,,\\
    (\sigma_\ta)^{\alpha\alpha'}(\sigma^{\tb})_{\alpha\alpha'}&=\delta_{\ta}^{\ \tb}\,,\\
    (\sigma_{\ta})^{\alpha\alpha'}(\sigma^{\ta})_{\beta\beta'}&=\epsilon_{\beta}^{\ \alpha}\epsilon_{\beta'}^{\ \alpha'}\,,\\
    (\sigma_{[\ta})^{\alpha\alpha'}(\sigma_{\tb]})_{\alpha}^{\ \beta'}&=-\frac{1}{2}\epsilon_{\ta\tb\tc\td}(\sigma^{\tc})^{\alpha\alpha'}(\sigma^{\td})_{\alpha}^{\ \beta'}\,,
\end{align}
\end{subequations}
for $\ta,\tb=1,2,3,4$. 
\section{Relations of \texorpdfstring{$S^4_N$}{sphere}}\label{app:S4Nbasic}

\paragraph{$\mso(5)$ relations.} Beside \eqref{eq:highestweightS4N}, we also have the following identities \cite{Sperling:2017dts,Sperling:2018xrm}:
\begin{subequations}\label{eq:so(5)relations1}
\begin{align}
    \{M_{ab},Y^b\}_+&=0\,, \label{M-Y-id}\\
    \frac{1}{2}\{M_{ac},M_{b}^{\ c}\}_+&= \frac{R_N^2}{r^2}\Big(\delta_{ab}-\frac{1}{2R_N^2}\{Y_a,Y_b\}_+\Big) \ ,  \label{M-M-id}\,\\
     \epsilon^{abcde} M_{cd} Y_e &= r(N+2) M^{ab} \ .
\end{align}
\end{subequations}
Note that the relation with the $\epsilon_{abcde}$ tensor is a self-duality constraint that makes all possible Young diagrams associated to higher-spin modules in the space of functions on $S^4_N$ into two-row ones \cite{Sperling:2017dts,Sperling:2017gmy}.

The identities \eqref{eq:so(5)relations1} follow directly from $Y_aY^a=R^2$ by acting with $[Y^b,\bullet]$. The identities in eq. \eqref{eq:highestweightS4N} which are related to the highest weight $\Xi=(N,0,0)$ of $\mso(6)\simeq \msu(4)$ can be obtained as follow. First of all, the oscillator relations of $Y_a$ that satisfy \eqref{eq:noncom} reads
\begin{align}
\label{fuzzy-S4-osc}
   Y_a=\frac{r}{2}\,(Z_{\Acal})^{\dagger}(\gamma_a)^{\Acal}_{\ \Bcal}Z^{\Bcal}\,. 
\end{align}
By putting the dagger, we want to emphasize that $Z_{\Acal}^{\dagger}$ and $Z^{\Acal}$ are non-commutative variables that obey \eqref{eq:fuzzytwistor}.

For convenience we reproduce the computation of the radius:
\begin{align}
    \begin{split}
 Y^a Y_a &=  \frac{r^2}4 Z_\Acal^\dagger (\g^a)^\Acal_{\ \Bcal} Z^{\Bcal}
 Z^\dagger_\Ccal (\g_a)^\Ccal_{\ \Dcal} Z^\Dcal  \\ 
  &= \frac{r^2}4 Z_\Acal^\dagger (\g^a)^\Acal_{\ \Bcal} 
  ([Z^{\Bcal},Z^\dagger_\Ccal] + Z^\dagger_\Ccal Z^{\Bcal})
  (\g_a)^\Ccal_{\ \Dcal} Z^\Dcal  \\ 
   &=  \frac{r^2}4 Z_\Acal^\dagger (\g^a)^\Acal_{\ \Bcal} (\g_b)^\Bcal_{\ \Dcal} Z^\Dcal 
  +  \frac{r^2}4 Z_\Acal^\dagger Z^\dagger_\Ccal (\g^a)^\Acal_{\ \Bcal}  
  (\g_a)^\Ccal_{\ \Dcal} Z^{\Bcal} Z^\Dcal   \\ 
   &=  \frac{r^2}4 Z_\Acal^\dagger (\g^a)^\Acal_{\ \Bcal} (\g_a)^\Bcal_{\ \Ccal} Z^\Ccal 
  +  \frac{r^2}8 Z_\Acal^\dagger Z^\dagger_\Ccal (Z^{\Acal} Z^\Ccal + Z^{\Ccal} Z^\Acal )  \\ 
   &=  \frac{r^2}4 \hat N(\hat N +4)\,,
   \end{split}
\end{align}
where we have used \eqref{eq:fuzzytwistor} and \eqref{eq:gammaimportant}. Similarly, using the fact that $M_{ab}=(Z_{\Acal})^{\dagger}(\Sigma_{ab})^{\Acal}_{\ \Bcal}Z^{\Bcal}$ together with \eqref{eq:Sigmarelations} and \eqref{eq:fuzzytwistor}, we arrive at
\begin{align}
    \epsilon_{abcde}M^{bc}M^{de}=12Z^{\dagger}\Sigma_{a6}Z+4(NZ^{\dagger}\Sigma_{a6}Z-Z^{\dagger}\Sigma_{a6}Z)=\frac{4}{r}(N+2)Y_a\,,
\end{align}
where $\Sigma^{a6}=\frac{1}{2}\gamma^a$. This is the self-duality constraint \eqref{M-M-id}.

\paragraph{$\msp(4)$ and $\msu(4)$ relations.} The relations \eqref{eq:so(5)relations1}, which hold on $\Hcal_N$, take the following form in terms of the new generators $L$ and $Y$:
\begin{subequations}\label{eq:sp(4)relations}
\begin{align}
    Y_{\Acal\Bcal}Y^{\Acal\Bcal}&=  L_{\Acal\Bcal}L^{\Acal\Bcal} = 4R_N^2=N(N+4)\,,\\
    \{L_{[\Acal \Mcal},Y_{\Bcal]}^{\  \Mcal}\}_+&=0\,,\\
    \{L_{[\Acal}^{\ \  [\Bcal},L_{\Ccal]}^{\  \Dcal]}\}_+&=-\{Y_{[\Acal}^{\  [\Bcal},
    Y_{\Ccal]}^{\  \Dcal]}\}_+\,,\\
    \epsilon_{\Acal\Bcal\Ccal\Dcal}Y^{\Acal\Bcal} &= -Y_{\Ccal\Dcal}\,.
    \label{self-duality-sp4} 
\end{align}
\end{subequations}
In obtaining the above, we have used some useful relations in the appendix \ref{sec:spinoralgebra}.

The first relation in \eqref{eq:sp(4)relations} can be obtained directly by the map $Y^a\gamma_a^{\Acal\Bcal}=Y^{\Acal\Bcal}$ while others can be obtained by acting with $[Y^{\Acal\Bcal},\bullet]$ on $Y_{\Acal\Bcal}Y^{\Acal\Bcal}=4R^2$. To derive the self-duality constraint \eqref{self-duality-sp4}, we note the following identification:
\begin{align}
Y^a &= \frac{r}{4}   Y^{\Acal\Bcal}\g^a_{\Acal\Bcal}\,. 
\end{align}
Now recall the following $\mso(5)$ Fierz identity from \cite{Steinacker:2016vgf} 
\begin{align}
  \g_a \otimes \g^a 
  &= \frac 12 (\one+P) -\frac 32 (\one-P) + 8 P_1 \,,
  \label{gamma-tens-id} 
\end{align} 
where $P$ is the permutation operator, and $P_1 = \frac 14 C\otimes C$ is the projector on the $\mso(5)$ singlet in 
$(4)\otimes (4) = ((10)_S \oplus (5)_{AS} \oplus (1)_{AS}\big)_{\mso(5)}$. In terms of indices, the identity \eqref{gamma-tens-id} reads
\begin{align}
    (\gamma^a)_{\Acal\Bcal}(\gamma_a)_{\Ccal\Dcal}=- C_{\Acal\Bcal} C_{\Ccal\Dcal} + 2C_{\Acal\Dcal} C_{\Ccal\Bcal} 
   + 2 C_{\Acal\Ccal}C_{\Bcal\Dcal} \,,
\end{align}
which is nothing but the identity \eqref{eq:gammaimportant}. This allows us to compute
\begin{align}
  \epsilon_{ABCD}(\g_a)^{\Acal\Bcal} (\g_b)^{\Ccal\Dcal}
   = c\, \d_{ab}\qquad \Rightarrow \qquad 
    \epsilon_{ABCD}(\g_a)^{\Acal\Bcal} (\g^a)^{\Ccal\Dcal}
   = 5 c\,,
\end{align}
which gives
\begin{align}
    -5\epsilon_{ABCD}C^{AB}C^{CD}=5c\,.
\end{align}
Therefore, we can fix the normalization constant
\begin{align}
 c=-4\,.
\end{align}
The self-dual constraint for $\msp(4)$ reads
\begin{align}
    \begin{split}
   \epsilon_{\Acal\Bcal\Ccal\Dcal}Y^{\Acal\Bcal}Y^{\Ccal\Dcal}
   &= r^{-2}  \epsilon_{\Acal\Bcal\Ccal\Dcal} (\g_a)^{\Acal\Bcal} (\g_b)^{\Ccal\Dcal} Y^a Y^b \\
    &= c\, r^{-2} Y^a Y_a = - N(N+4)
    \end{split}
\end{align}
or equivalently
\begin{align}
\label{selfdual-YAB}
    \epsilon_{\Acal\Bcal\Ccal\Dcal}Y^{\Acal\Bcal} &=  \frac{c}{4} \,Y_{\Ccal\Dcal}=-Y_{\Ccal\Dcal}\,.
\end{align}    
Further identities are obtained by noting that 
\begin{align}
    T^{\Acal\Bcal} = \frac 12 Y^{\Acal\Bcal} + L^{\Acal\Bcal}
\end{align}
satisfies the relations of $\msu(4)$, as well as the characteristic equation \cite{Carow-Watamura:2004mug}
\begin{align}
(T-3)(T+1+\frac{4}{N}) = 0
\end{align}
on $\Hcal_N$.
This is the origin of \eqref{M-M-id}.




\section{Derivation of effective metric}\label{app:C}
Let $g^{\alpha\alpha'\beta\beta'}$ is the effective metric in the tangential direction while $\varrho^{\alpha\alpha'\beta\beta'}$ is the effective metric in the transversal one. By using the following identities
\begin{align}\label{eq:magic}
    \lambda^{[\alpha}\hat{\lambda}^{\beta]}=\frac{1}{2}\epsilon^{\alpha\beta}\langle \hat{\lambda}\l\rangle\,,\qquad \mu^{[\alpha'}\hat{\mu}^{\beta']}=\frac{1}{2}\epsilon^{\alpha'\beta'}[\hat{\mu}\mu]\,,
\end{align}
we can write the first term $g^{\alpha\alpha'\beta\beta'}$ in \eqref{eq:preeffectivemetric} as 
\begin{align}\label{eq:tensoreffectivemetric1}
    \begin{split}
    g^{\alpha\alpha'\beta\beta'}&=\epsilon^{\alpha\beta}\epsilon^{\alpha'\beta'}(\langle\hat{\lambda}\lambda\rangle^2+[\hat{\mu}\mu]^2)+8\lambda^{(\alpha}\hat{\lambda}^{\beta)}\mu^{(\alpha'}\hat{\mu}^{\beta')}\\
    &=N^2\epsilon^{\alpha\beta}\epsilon^{\alpha'\beta'}-2[\hat\mu\mu]\langle \hat\l\l\rangle\epsilon^{\alpha\beta}\epsilon^{\alpha'\beta'}+8\lambda^{(\alpha}\hat{\lambda}^{\beta)}\mu^{(\alpha'}\hat{\mu}^{\beta')}\\
    &=N^2\epsilon^{\alpha\beta}\epsilon^{\alpha'\beta'}+2(\hat{\lambda}^{\alpha}\lambda^{\beta}\hat{\mu}^{\beta'}\mu^{\alpha'}+\lambda^{\alpha}\hat{\lambda}^{\beta}\mu^{\beta'}\hat{\mu}^{\alpha'})\,.
    \end{split}
\end{align}
The second term $\varrho^{\alpha\alpha'\beta\beta'}$ in \eqref{eq:preeffectivemetric} reads
\begin{align}
    \varrho^{\alpha\alpha'\beta\beta'}=-(\hat{\lambda}^{\alpha}\hat{\lambda}^{\beta}\mu^{\alpha'}\mu^{\beta'}+\hat{\lambda}^{\alpha}\lambda^{\beta}\mu^{\alpha'}\hat{\mu}^{\beta'}+\lambda^{\alpha}\hat{\lambda}^{\beta}\hat{\mu}^{\alpha'}\mu^{\beta'}+\lambda^{\alpha}\lambda^{\beta}\hat{\mu}^{\alpha'}\hat{\mu}^{\beta'})\,.
\end{align}
Combine them together, we obtain the full metric in \eqref{fullmetric}.











\footnotesize
\setstretch{1.0}
\bibliographystyle{JHEP-2}
\bibliography{IKKT.bib}

\end{document}